\pdfoutput=1
\documentclass[aps,prd,amsmath,floats,floatfix, twocolumn,
superscriptaddress,nofootinbib,showpacs]{revtex4-1}

\usepackage[T1]{fontenc}
\usepackage[utf8]{inputenc}
\usepackage{lmodern}

\usepackage{verbatim}

\usepackage[dvipsnames, usenames]{xcolor}
\definecolor{linkcolor}{rgb}{0.0,0.3,0.5}
\usepackage[hypertexnames=false, unicode, colorlinks=true, linkcolor=linkcolor,
citecolor=linkcolor, filecolor=linkcolor,urlcolor=linkcolor,
pdfusetitle]{hyperref}
\usepackage[all]{hypcap}
\usepackage{xspace}
\usepackage{amssymb}
\usepackage{microtype}
\usepackage{enumitem}
\usepackage{cancel}

\usepackage{mathrsfs}
\usepackage{graphicx}
\usepackage{color}
\usepackage{dcolumn}
\usepackage{bm}
\usepackage{float}
\usepackage{subfigure}
\usepackage{multirow}
\usepackage{physics}
\usepackage{soul}
\usepackage[normalem]{ulem} 

\newcolumntype{L}[1]{>{\raggedright\let\newline\\\arraybackslash\hspace{0pt}}m{#1}}
\newcolumntype{C}[1]{>{\centering\let\newline\\\arraybackslash\hspace{0pt}}m{#1}}
\newcolumntype{R}[1]{>{\raggedleft\let\newline\\\arraybackslash\hspace{0pt}}m{#1}}

\newcommand{\mnras}{Mon. Not. R. Astron Soc.}

\newcommand{\eff}{\text{eff}}
\newcommand{\FS}{\text{FS}}
\newcommand{\FC}{\text{FC}}

\newcommand{\omd}{\dot{\Omega}_r}
\newcommand{\lol}{\tilde{\lambda}}
\newcommand{\chirp}{\mathcal{M}}
\newcommand{\tha}{\hat{t}}
\newcommand{\etal}{\textit{et al.\ }}

\newcommand{\ra}{R_{\rm NS}}

\newcommand\aap{\ref@jnl{A\&A}}%

\newlist{todolist}{itemize}{2}
\setlist[todolist]{label=$\square$}
\usepackage{pifont}

\allowdisplaybreaks
\begin{document}
\title{Excitation of $f$-modes during mergers of spinning binary neutron star}
\author{Sizheng Ma}
\email{sma@caltech.edu}
\author{Hang Yu}
\email{hangyu@caltech.edu}
\author{Yanbei Chen}
\email{yanbei@caltech.edu}
\affiliation{TAPIR, Walter Burke Institute for Theoretical Physics, California Institute of Technology, Pasadena, California 91125, USA}
\date{\today}

\begin{abstract}
Tidal effects have important imprints on gravitational waves (GWs) emitted during the final stage of the coalescence of binaries that involve neutron stars (NSs). 
Dynamical tides can be significant when NS oscillations become resonant with orbital motion; understanding this process is important for accurately modeling GW emission from these binaries, and for extracting NS information from GW data. 
In this paper, we use semi-analytic methods to carry out a systematic study on the tidal excitation of fundamental modes ($f$-modes) of spinning NSs in coalescencing binaries, focusing on the case when the NS spin is anti-aligned with the orbital angular momentum -- where the tidal resonance is most likely to take place.  We first expand NS oscillations into stellar eigenmodes, and then obtain a Hamiltonian that governs the tidally coupled orbit-mode evolution. (Our treatment is at Newtonian order, including gravitational radiation reaction at quadrupole order.) We then find a new approximation that can lead to analytic expressions of tidal excitations to a high accuracy, and are valid in all regimes of the binary evolution: adiabatic, resonant, and post-resonance.
Using the method of osculating orbits, we obtain semi-analytic approximations to the orbital evolution and GW emission; their agreements with numerical results give us confidence in on our understanding of the system's dynamics.  In particular, we recover both the averaged post-resonance evolution, which differs from the pre-resonance point-particle orbit by shifts in orbital energy and angular momentum, as well as instantaneous perturbations driven by the tidal motion. 
%
%
%
Finally, we use the Fisher matrix technique to study the effect of dynamical tides on parameter estimation. We find that, for a system with component masses of $(1.4,1.4)M_\odot$ at 100Mpc, the constraints on the effective Love number of the $(2,2)$ mode at Newtonian order can be improved by factor of $3\sim4$ if spin frequency is as high as 500Hz. The relative errors are $0.7\sim0.8$ in the Cosmic Explorer (CE) case, and they might be further improved by post-Newtonian effects. The constraints on the $f$-mode frequency and the spin frequency are improved by factors of $5\sim6$ and $19\sim27$, respectively. In the CE case, the relative errors are $0.2\sim0.4$ and $0.7\sim1.0$, respectively. Hence the dynamical tides  may potentially provide an additional channel to study the physics of NSs. 
The method presented in this paper is generic and not restricted to $f$-mode; it can also be applied to other types of tide.

\end{abstract}

\maketitle

\section{Introduction}

The detection of gravitational waves (GWs) and its electromagnetic counterparts from binary neutron star (BNS) coalescence GW170817 \cite{Abbott+BNS+17,abbott2017gravitational,goldstein2017ordinary,savchenko2017integral}, as well as the recent event GW190425 \cite{Abbott:2020uma}, has started a new approach to study the physics of NSs. The observations have already provided new constraints on tidal deformabilities \cite{Abbott:2018exr,LIGOScientific:2019eut,Annala+Gorda+18,Most+Weih+18}, the maximum mass \cite{LIGOScientific:2019eut,margalit2017constraining,rezzolla2018using,Ruiz+Shapiro+18,Shibata+Fujibayashi+17}, radii \cite{Abbott:2018exr,Most+Weih+18} and $f$-mode frequencies \cite{pratten2019gravitational} of NSs. With the improvement of detector sensitivity, more BNS coalescence detections are expected for the near future \cite{abadie2010predictions,kalogera2004cosmic,kim2006effect,o2008constraining}. 
Furthermore, 3G detectors, like the Einstein Telescope (ET) \cite{hild2009xylophone,sathyaprakash2012scientific} and the Cosmic Explorer (CE) \cite{abbott2017exploring}, are being planned for operation in the 2030s. These 3G detectors may increase neutron star black-hole (NSBH)  and BNS detection rates by 3-4 orders of magnitude \cite{baibhav2019gravitational}. As a result, accurately modeling NSs in  binary systems is necessary and timely.

During the inspiral process, NSs in binaries are distorted due to the tidal field of their companions. Tidal coupling between compact objects in binaries allows the equation of state (EoS) of these objects to leave an imprint on GW signals, both during the early inspiral stage \cite{Flanagan+Hinderer+08} and during the late inspiral stage \cite{Faber+02,bejger2005impact}. 
Under the equilibrium-tide approximation, the effect of tidal interaction can be characterized by the relativistic tidal Love number.  Hinderer \etal  studied the effect of equilibrium tide on gravitaitonal waveforms, both using polytropic~\cite{hinderer2008tidal,Flanagan+Hinderer+08}  and more realstic EoS~\cite{Hinderer+10}. They found that 3G detectors are likely able to probe the clean tidal signatures from the early stage of inspirals. 
The post-Newtonian (PN) \cite{Damour+Soffel+91,Damour+Soffel+92,Damour+Soffel+93,Damour+Soffel+94,Racine+Flanagan+05} tidal effects were studied by Vines and Flanagan \cite{Vines+Flanagan+13}, who explicitly obtained equations of motion with quadrupolar tidal interactions up to 1PN order. They pointed out that spin-orbit coupling must be included at this order in order to conserve angular momentum. The spin-tidal couplings and higher PN orders were studied later by Abdelsalhin \etal ~\cite{Abdelsalhin+Gualtieri+18}. 
%


In the late stage of an inspiral, the binary's orbital frequency sweeps through from hundreds of Hz to thousands of Hz. As the tidal driving frequency comes close to a normal mode frequency of the NS, internal stellar oscillations can be excited --- giving rise to dynamical tide (DT). Exchanges of energy and angular momentum between orbital motion and stellar oscillations cause changes in orbital motion,  leading to additional features in the  the gravitational waveform. 

The tidal excitation of $f$-modes of stars was first investigated by Cowling \cite{cowling1941non}. Later, several authors studied the DTs of non-spinning stars in the context of Newtonian physics \cite{Lai+94,reisenegger1994excitation,Kokkotas:1995xe}, and in the context of general relativity (incorporating gravitational radiation reaction and treating the NS relativistically) \cite{Gualtieri:2001cm,Pons:2001xs,Miniutti:2002bh,Berti:2002ry}. In particular, Lai (hereafter L94) \cite{Lai+94} split the whole process into three regimes: the adiabatic, resonant and post-resonance regimes. The first one is described by the the well-known adiabatic approximation to a high accuracy. At the post-resonance stage, they assumed that each stellar mode oscillates mainly at its own eigenfrequency; by factoring out the eigenfrequency, the motion can then be described by a slowly varying amplitude.  This allowed them to obtain a simple form of post-resonance tidal amplitude with the stationary-phase approximation (SPA), which further leads to changes in the orbital separation, energy, angular momentum and the phase of GWs. They found that the amount of energy transfer due to resonance and the induced GW phase shift are negligible, since the coupling between $g$-mode and tidal potential is weak. They also pointed out that $f$-mode frequency is too high for resonance to take place.

As it turns out, the effect of DT can be strengthened by stellar rotation\footnote{In the inertial frame, mode frequencies are shifted by the spin frequency to a lower value. As a result, those modes become easier to be excited. See Fig.~\ref{fig:reson-fre} for more details.} \cite{Lai_1997,Ho+Lai+99} and orbital eccentricity \cite{Chirenti:2016xys,Yang:2018bzx,Yang:2019kmf,Vick:2019cun}. In this paper, we mainly focus on the significance of stellar rotation. It is conventionally believed that high rotation rate is unlikely when binaries which enter the LIGO band, since such systems usually have had enough time to evolve and spin down. For example, recent events GW170817 \cite{Abbott+BNS+17} and GW190425 \cite{Abbott:2020uma} are all consistent with low spin configurations. The fastest spinning pulsar observed in BNS is PSR J0737-3039A, which spins at 44Hz \cite{Lyne1153}. Andersson \etal ~\cite{Andersson+Ho+18} estimated that it will spin down to 35 Hz as it enters the LIGO band. However, high spin rate is still physically allowed. In such systems, retrograde rotation (respect to the orbit) drags the mode frequency to a lower value in the inertial frame. This makes the tidal resonance take place earlier. The energy and angular momentum transfers due to DTs in spinning stars were calculated in Ref. \cite{Lai_1997}. Ho and Lai \cite{Ho+Lai+99} found that the resonance of the dominant $g$-mode is enhanced by spin if the star rotates faster than 100Hz, it can induce a phase shift of $\sim$0.05 rad in the waveform. Additionally, $f$-mode resonance can produce a significant phase shift if the spin frequency is higher than 500Hz (depending on the EoS).

However, Ref. \cite{Lai_1997} was based on a configuration-space decomposition of the stellar oscillation, which does not use an orthonormal basis for a spinning star. This problem can be fixed by a phase-space mode expansion method \cite{Schenk+01}. Within this formalism, Lai and Wu \cite{Lai+Wu+06} investigated the effect of the inertial modes\footnote{Inertial modes, or generalized $r$-modes, are a class of modes in spinning NSs who are not purely axial when spin frequency goes to 0, whereas $r$-modes are axial in this limit.}, and found that the phase shift is of order 0.1 rad when the spin frequency is lower than 100Hz. The exception is $m=1$ mode, which can be excited at tens of Hz for nonvanishing spin-orbit inclinations, and hence, generate a large phase shift in GW phase.

Accurate theoretical templates are needed in order to extract tidal information from GWs. Although extensive work has been done on adiabatic tide (AT), the study on DTs still requires improvements. For example, L94 \cite{Lai+94} only estimated the changes of several parameters due to DTs. Their work did not explicitly treat the effect of tidal back-reaction on the orbit. The treatment did not provide detailed time evolution near the resonance, either. One approximate model was provided by Flanagan and Racine (hereafter FR07) \cite{Flanagan+Racine+07}. They approximated the post-resonance orbit by a point particle (PP) trajectory, since the energy and angular momentum transfers only take place near the resonance. After that, the NS is treated as freely oscillating without interacting with the orbit. This model averages the dynamics over the tide-oscillation timescale, therefore does not describe the tidal perturbation at shorter timescales. 

More recently, Hinderer \etal (hereafter, H+16)~\cite{Steinhoff+Hinderer+16,Hinderer+Taracchini+16} incorporated DT, in particular, the resonance of the $f$-mode in non-spinning NS, into the Effective-One-Body (EOB) formalism. A frequency domain model was developed later in Ref. \citep{Schmidt+Hinderer+19}. In these works, DT is described by effective Love number
\begin{align}
\lambda_\eff=-\frac{E_{ij}Q^{ij}}{E_{kl}E^{kl}}, \label{eff-love-h}
\end{align}
where $E_{ij}$ is the tidal field induced by the companion and $Q^{ij}$ is the quadrupole moment of the NS. To evaluate this quantity, they expanded the NS's response function near resonance; and described the evolution of DT by Fresnel functions in the resonant regime. They then used asymptotic analyses to piece adiabatic expressions and Fresnel functions together to obtain a single formula. The formula is precise prior to resonance. But it does not describe the phasing of the post-resonance regime. This is not a big issue for slowly spinning NSs, since in this case the mode is not excited until the end of coalescence, and post-resonance dynamics is extremely short. Furthermore, because current detections are all consistent with low spin configuration \cite{Abbott+BNS+17,Abbott:2020uma}, this model is accurate enough for current data analysis. However, this method cannot describe rapidly spinning NSs \cite{PhysRevD.99.044008}; given the fact that rapidly spinning NSs are physically allowed, an accurate GW model for these systems is still necessary. In this paper, we extend H+16 \cite{Steinhoff+Hinderer+16,Hinderer+Taracchini+16} to arbitrary spin, by deriving new analytic formulae to describe the entire process of DT, accurate throughout the adiabatic, resonant and post-resonance regimes. The formulae agree with numerical integrations to high accuracies. We then carry out a systematic study on the post-resonance dynamics, by using the tidal response formulae and the method of osculating orbits. Finally, we analyze the impact of DT on parameter estimations by Fisher information matrices formalism. In order to more optimistically illustrate a best-case scenario in which $f$-mode DT might bring more information, we will be assuming high NS spin frequencies and stiff EoS. However, as we will discuss later, the qualitative features of DT shown in this paper do not depend on the specific properties of NSs.

This paper is organized as follows. In Sec.\ \ref{sec:EoS}, we introduce the EoS used in this paper, and construct approximations for the spin's effect on mode frequencies using the Maclaurin spheroid model. In Sec.\ \ref{sec:eq} we derive equations of motion using the phase-space mode expansion method and a Hamiltonian approach. With these at hand, we give a comprehensive discussion on DT in Secs.\ \ref{sec:new-dyn-tides} and \ref{sec:post-resonance orbit dynamics}. In Sec.\ \ref{sec:new-dyn-tides}, we mainly work on the stellar part. We first review previous studies on DT in Sec.\ \ref{sec:DT-review} and propose our new approach in Sec.\ \ref{sec:new_DT}, where we also compare these models with numerical integrations. In Sec.\ \ref{sec:post-resonance orbit dynamics}, we use our new formulae and the method of osculating orbits to study post-resonance orbital dynamics. We get a set of first order differential equations to describe the time evolution of osculating variables (e.g. Runge-Lenz vector, angular momentum, and orbital phase). These equations can provide rich information of the orbit near resonance, as discussed in Sec.\ \ref{sec:init}. Then in Sec.\ \ref{sec:orbit-num-comp}, we compare our osculating equations with numerical integrations and provide a new way to obtain the post-resonance averaged orbit over the tide-oscillation timescale, which agrees with FR07 \cite{Flanagan+Racine+07} to the leading order in of tidal interaction. By combining our new method and FR07 \cite{Flanagan+Racine+07}, we obtain an analytic expression for the time of resonance. Sec.\ \ref{sec:GW} mainly focuses on GWs. We first quantify the accuracy of several models by mismatch between waveforms. Then in Sec.\ \ref{sec:fim} we use the Fisher information matrix formalism to discuss the influence of DT on parameter estimation. Finally, in Sec.\ \ref{sec:con} we summarize our results. 

Throughout this paper we use the following conventions unless stated otherwise. We use the geometric units with $G=c=1$. We use Einstein summation notation, i.e. summation over repeated indexes.



\section{Basic equations of dynamical tides}
\label{sec:mode-orbit-eq}
This section will provide equations of motion of the system undergoing DT. In Sec. \ref{sec:EoS}, we construct approximations on spin's influence on $f$-mode frequencies, based on the Maclaurin spheroid model. In Sec. \ref{sec:eq} we use the phase-space mode expansion method and a Hamiltonian approach to obtain the coupled equations of motion.

\subsection{Neutron star equations of state and properties}
\label{sec:EoS}

In this paper, we shall use, as input for our studies, properties of spinning neutron stars such as values of $f$-mode frequencies and tidal Love numbers.  

Properties of non-spinning NSs have been studied extensively. In this paper we use two of them for comparison purposes. One is the H4 model \citep{Hinderer+10}, which gives dimensionless Love number $k_2=0.104$ for a NS with mass $M_{\rm NS}=1.4M_\odot$ and radius $\ra=13.76$km. Here $k_2$ is defined as \cite{Flanagan+Hinderer+08}
\begin{align}
k_2=\frac{3}{2}\frac{\lambda}{R^5_{\rm NS}}, \label{lambda-k}
\end{align}
where $\lambda$ is the value of $\lambda_{\rm eff}$ in the equilibrium limit [Eq.\ (\ref{eff-love-h})].
The other one is a $\Gamma=2$ polytrope with $M_{\rm NS}=1.4M_\odot$ and $\ra=14.4$km, which has $k_2=0.07524$. The latter model is the same as the one used in H+16 \cite{Steinhoff+Hinderer+16,Hinderer+Taracchini+16}. Their $f$-mode frequencies are $2\pi\times1.51$kHz and $2\pi\times1.55$kHz, respectively, consistent with the universal relation of NS properties \cite{Chan+Sham+14,Yagi+Yunes+13+2,Yagi+Yunes+13,Yagi+Yunes+17}. We want to note that H4 is a stiff EoS that is not favored by GW170817 \cite{LIGOScientific:2019eut}, yet our focus is on exploring what information DT might bring,  hence  the H4 EoS will be more ``optimistic'', since it  leads to stronger tidal features than the softer, more compact EoS.

For spinning NSs, $f$-mode frequencies will split, and the Love number will also change. Unlike the non-spinning case, there is not yet a systematic parameterization of spinning NS properties, depending on EoS. For Love number, we shall simply use their non-spinning values; we will justify the validity of this treatment later [below Eq.~(\ref{eq-lambda0})]. On the other hand, since the $f$-mode frequency split is important for bringing down the orbital frequency required for resonance, we will need more accurate input. Oscillation of spinning NS has been studied extensively in different limits, such as the (post-)Newtonian limit \cite{1986ApJ...309..598M,1990ApJ...355..226I,1991ApJ...373..213I,1995ApJ...438..830Y,Passamonti:2008tt,1991ApJ...374..248C,1992ApJ...385..630C}, the slow-rotation limit \cite{1993ApJ...414..247K,10.1143/ptp/90.5.977,1997MNRAS.289..117Y,PhysRevD.75.064019,PhysRevD.77.024029} and the Cowling approximation \cite{Yoshida:1998mu,1992ApJ...389..392I,PhysRevD.81.084019,PhysRevD.78.064063} (see Sec.~8.6.1 of Ref.\ \cite{rotat_rel_star} and references therein). The case of full relativistic NS with arbitrary high rotation rate has also been studied, for example, by Zink \etal \cite{Zink:2010bq}, by using nonlinear time-evolution code. In this paper, for simplicity, we shall use the features of Maclaurin spheroid to construct an approximation on how spin influences $f$-mode frequencies. 

The Maclaurin spheroid describes a self-gravitating, rigidly rotating body of uniform density in Newton's theory. In the coordinate system $(x^\prime,y^\prime,z^\prime)$ which co-rotates with the NS, the NS surface in hydrostatic equilibrium is described by \cite{chandrasekhar1969ellipsoidal}
\begin{align}
\frac{x^{\prime 2}+y^{\prime 2}}{a_1^2}+\frac{z^{\prime 2}}{a_3^2}=1,
\end{align}
where we assume that the spin vector is along the $z^\prime$-axis. The spheroid's semi-axes in the $x^\prime(y^\prime)$- and $z^\prime$-directions are denoted by $a_1$ and $a_3$, respectively. They are related to the eccentricity $e_s$ of the star by
\begin{align}
e_s=\sqrt{1-\frac{a_3^2}{a_1^2}}.
\end{align}
Note that the NS surface is oblate due to the spin ($a_3<a_1$), so the stellar eccentricity is always smaller than 1. Hydrostatic equilibrium lead to a one-to-one mapping between the spin angular frequency $\Omega_s$ and the stellar eccentricity $e_s$ \cite{chandrasekhar1969ellipsoidal}
\begin{align}
\Omega^2_s=\frac{2\pi \rho}{e^3_s}\left[(1-e^2_s)^{1/2}(3-2e^2_s)\sin^{-1}e_s-3e_s(1-e^2_s)\right], \label{spin-e}
\end{align}
where $\rho$ is the mass density of the star. 
For a Maclaurin spheroid, $f$-mode frequencies are specified in terms of the stellar eccentricity $e_s$, which is further determined by $\rho$ and $\Omega_s$. In this paper we mainly focus on the $(j=2,k=\pm2)$ and $(j=2,k=0)$ modes. Here $(j,k)$ are the angular quantum numbers of multipole expansion, see Sec.\ \ref{sec:ste-osc} for more details. Their mode frequencies (in the co-rotating frame) are given by [see Eq. (32) of Ref.\ \cite{Braviner+Ogilvie+14} and Eqs.\ (12)--(13) of Ref.\ \cite{1979MNRAS.189..255C}] 
\begin{subequations}
\begin{align}
&\frac{\omega_0^2}{\Omega_s^2}=\frac{1+\zeta_0^2}{1+3\zeta_0^2}\left[(1-9\zeta_0^2)+\frac{12\zeta_0(1-\zeta_0\arccot\zeta_0)}{(1+3\zeta_0^2)\arccot\zeta_0-3\zeta_0}\right]\label{omega00}, \\
&\frac{\omega_{2\pm}}{\Omega_s}=-1\pm\left[1-\frac{4e_s^2R_2}{(3-2e_s^2)\sin^{-1}e_s-3e_s(1-e_s^2)^{1/2}}\right]^{1/2}, \label{omega22}
\end{align}
\label{omega-total}%
\end{subequations}
where $\zeta_0=\sqrt{1-e_s^2}/e_s$ and 
\begin{align}
R_2&=\frac{3(1-e_s^2)^{1/2}}{8e_s}\sum_{p=3}^\infty\frac{(2p-2)!!}{(2p-1)!!}e_s^{2(p-2)} \notag \\
&+\frac{1-e_s^2}{e_s^2}\left[\arcsin e_s-\frac{e_s}{(1-e_s^2)^{1/2}}\right], \notag \\
&=\frac{10e_s^4-7e_s^2-3}{8e_s^3\sqrt{1-e_s^2}}+\frac{3+8e_s^2-8e_s^4}{8e_s^4}\arcsin e_s. \label{R2-e}
\end{align}
It is straightforward to see that each mode has two frequencies with opposite signs. The positive (negative) one corresponds to the prograde (retrograde) mode. The absolute value of two $(2,2)$ mode eigenfrequencies split due to the spin, this is an analogue to the Zeeman split.

\begin{figure}[htb]
        \includegraphics[width=\columnwidth,height=5.6cm,clip=true]{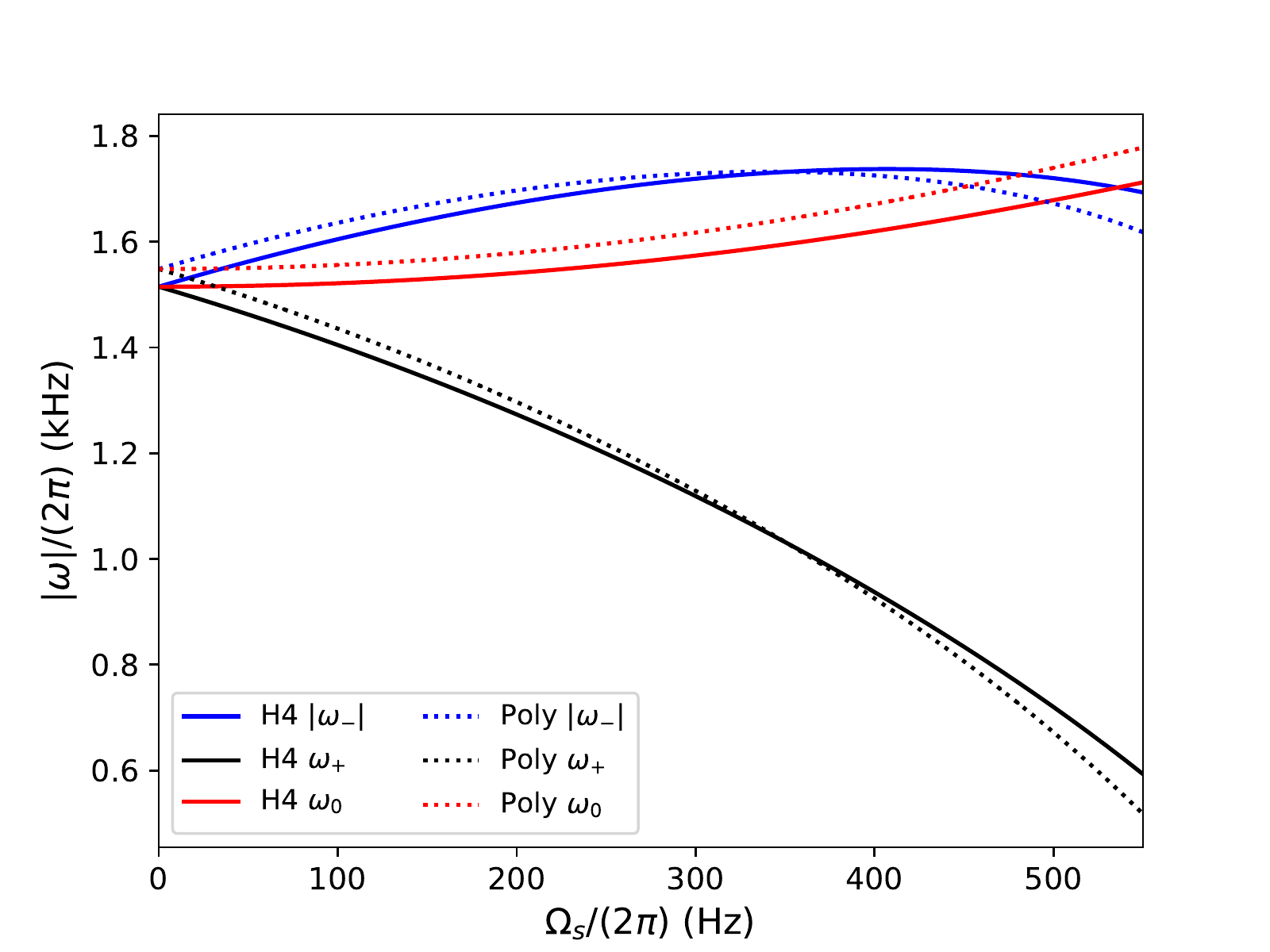}
  \caption{The dependence of $f$-mode frequencies (in the co-rotating frame) on spin for NS with mass $1.4M_\odot$, following our prescription. The H4 EoS, represented by solid lines, gives $\omega_0,|\omega_{2\pm}|=2\pi\times1.51$Hz for non-spinning NS, while $\Gamma=2$ polytrope gives $2\pi\times1.55$Hz. The frequencies of prograde (black line) and retrograde (blue line) modes split due to spin. 
  }
 \label{fig:f-fre}
\end{figure}

Eqs.\ (\ref{spin-e})--(\ref{R2-e}) are valid for any $0\leq e_s<1$\footnote{Maclaurin spheroids become unstable as $e_s>0.813$, corresponds to $\sim 900$ Hz. Such high rotation rate, however, is not of our interest.}. In the small-eccentricity (low-rotation) regime, we have 
\begin{align}
&\Omega_s=\sqrt{\frac{8\pi\rho}{15}}e_s+\mathcal{O}(e_s^3),
&R_2=-\frac{2}{15}e_s+\mathcal{O}(e_s^3).
\end{align}
As a result, $\omega_{0,2\pm}/\Omega_{s}$ in Eqs.\ (\ref{omega-total}) diverge when $e_s\to 0$. However, mode frequencies $\omega_{0,2\pm}$ themselves converge to finite values, which are given by
\begin{subequations}
\begin{align}
&\omega_{2\pm}=\pm\sqrt{\frac{16\pi\rho}{15}}-\sqrt{\frac{8\pi\rho}{15}} e_s+\mathcal{O}(e_s^2), \\
&\omega_{0}=\sqrt{\frac{16\pi\rho}{15}}+\mathcal{O}(e_s^2),
\end{align}%
\end{subequations}
where the leading term $\sqrt{16\pi\rho/15}$ is the mode frequency of a non-spinning NS. But it turns out that this prediction differs from the true $f$-mode frequencies for a realistic EoS, if we use the mean density of the star as $\rho$. This is due to the assumption of homogeneity and incompressibility in the Maclaurin case. We refer the interested readers to Ref.~\cite{1993ApJ...414..247K} for a comprehensive comparison between the Maclaurin spheroid and the relativistic NS in the slow-rotation limit. Therefore, one should not directly use Eqs.\ (\ref{omega-total}). To obtain $f$-mode frequency for a NS with generic spin, we define an effective density $\rho_{\rm eff}$, such that $\sqrt{16\pi\rho_{\rm eff}/15}$ coincides with $f$-mode frequency of a non-spinning NS with realistic EoS (H4 EoS or $\Gamma=2$ polytropic EoS). Meanwhile, we still assume the functional dependence of the mode frequencies $\omega_{0,2\pm}$ on $\Omega_s$ and $\rho_{\rm eff}$ to be the same as Eqs.~(\ref{omega-total}). With such approximation, $f$-mode frequencies for non-spinning NSs can be extended to NSs with generic spins. In Fig. \ref{fig:f-fre}, we plot $\omega_0, |\omega_{2\pm}|$ as functions of $\Omega_s$ with both H4 EoS and $\Gamma=2$ polytropic EoS. Results agree qualitative with previous studies [see Fig.~5 of Ref.\ \cite{Zink:2010bq}].

\subsection{Equations of motion}
\label{sec:eq}

Using the same convention as Ref. \cite{Ho+Lai+99}, we consider a BNS system with individual masses ${M_1}$ and ${M_2}$ moving in the $x-y$ plane, whose orbital angular momentum is along the $z$-axis. For simplicity, we assume that only ${M_1}$ is tidally deformed. We still use $(x^\prime,y^\prime,z^\prime)$ as the body coordinate system that co-rotates with ${M_1}$. Two coordinate planes $x^\prime-y^\prime$ and $x-y$ intersect at the line $\ell$. The angle between the $z$-axis and the $z^\prime$-axis is $\beta$ and the angle between $\ell$ and the $y$-axis is $\alpha$. And let $\gamma$ be the angle that the star rotates about $z^\prime$-axis. Therefore two coordinate systems are related by Euler angles $(\alpha,\beta,\gamma=\Omega_st)$.
\subsubsection{The evolution of stellar oscillation}
\label{sec:ste-osc}
In the co-rotating frame, the oscillation of the rotating star is governed by\footnote{Throughout this paper we ignore the effect of dissipation. For $f$-modes, the most significant dissipation comes from the GW radiation of the mode itself, with a damping timescale of $\sim 0.03\,{\rm s}$~\cite{Ipser:91}, which is much longer than the mode period in the co-rotating frame. Shear and bulk viscosity due to electron scattering~\cite{Lai+94}, as well as Urca reactions~\cite{Arras:19} have even more negligible effects on the dynamics. Therefore, we also assume that the background star's spin is unaffected by the tidal interaction (see also Ref.~\cite{Bildsten+Cutler+92}).} \cite{Schenk+01,Ho+Lai+99}
\begin{align}
\frac{\partial^2 \bm{\xi}}{\partial t^2}+2\bm{\Omega}_s\times\frac{\partial \bm{\xi}}{\partial t}+\bm{C}\cdot\bm{\xi}=-\nabla U,
\end{align}
where $\bm{\xi}$ is the Lagrangian displacement of fluid elements, and $\bm{C}$ is a self-adjoint operator. The external gravitational potential $U$ can be expanded in terms of spherical harmonics
\begin{align}
&U=-G{M_2}\sum_{lm}\frac{W_{lm}r_s^l}{{r}^{l+1}}e^{-im{\phi}(t)}Y_{lm}(\theta,{\iota}), \notag \\
&=-G{M_2}\sum_{lmm^\prime}\frac{W_{lm}r_s^{\prime l}}{{r}^{l+1}}e^{-im{\phi}(t)+im^\prime \Omega_st}Y_{lm^\prime}(\theta^\prime,{\iota}^\prime)\mathcal{D}_{m^\prime m}^{(l)}(\alpha,\beta), \label{U-Y}
\end{align}
where ${r}$ is the separation between the two stars, ${\phi}(t)$ is the orbital phase, and $r_s={r_s^\prime}=\sqrt{x^2+y^2+z^2}$ is the distance of fluid element to the origin. Here $(l,m)$ are the angular quantum numbers of multipole expansion; for example $l=0,1$ are the monopole and dipole pieces, which do not couple to NS internal oscillations, while tidal effects start from $l=2$. Variables $\theta,{\iota}$ are the angular coordinates of fluid elements in the inertial (unprimed) coordinate system; and $\theta^\prime,{\iota}^\prime$ are in the co-rotating (primed) coordinate system. We should note that $\Omega_s$ is always positive in our convention. The quantity $W_{lm}$ is given by \cite{Press+Teukolsky+77}
\begin{align}
&W_{lm}=(-1)^{(l+m)/2}\left[\frac{4\pi}{2l+1}(l+m)!(l-m)!\right]^{1/2} \notag \\
&\times\left[2^l\left(\frac{l+m}{2}\right)!\left(\frac{l-m}{2}\right)!\right]^{-1},
\end{align}
which is non-vanishing only when $l+m=\text{even}$. We have used the Wigner $\mathcal{D}$-functions to transform spherical harmonics between the unprimed and primed coordinate systems.

Using the phase-space mode expansion method developed in Ref. \cite{Schenk+01}, the Lagrangian displacement and its time derivative can be expressed as
\begin{align}
\left(\begin{matrix}
\bm{\xi} \\
\bm{\dot{\xi}}
\end{matrix}\right)=\sum_{\sigma}c_\sigma(t)
\left(\begin{matrix}
\bm{\xi}_{\sigma} \\
-i\omega_{\sigma}\bm{\xi}_{\sigma}
\end{matrix}\right), \label{mode-exp}
\end{align}
where modes are labeled by $\sigma=(j,k,\nu=\pm)$. The angular quantum numbers $j$ and $k$ are integers with $k=\pm j,\pm (j-1)\ldots0$. In our case the mode functions with negative $k$ are related to the positive ones by complex conjugate (up to a constant), therefore we restrict ourselves to $k\geq0$. The label $\nu$ stands for the propagation direction of modes, as mentioned in Sec. \ref{sec:EoS}.

The modes in Eq. (\ref{mode-exp}) are normalized by the condition
\begin{align}
\left<\bm{\xi}_\alpha,\bm{\xi}_\alpha\right>=1. \label{norm-con}
\end{align}
where the inner product is defined by
\begin{align}
\left<\bm{\xi}_\sigma,\bm{\zeta}_\tau\right>=\int d^3\textit{x}^\prime\rho(x^\prime) \bm{\xi}_\sigma^*\cdot\bm{\zeta}_\tau.
\end{align}
The amplitudes $c_\sigma(t)$ satisfy the equation
\begin{align}
\dot{c}_\sigma(t)+i\omega_\sigma c_\sigma(t)=-\frac{i}{b_\sigma}\left<\bm{\xi}_\sigma,\nabla U\right>, \label{c-evo}
\end{align}
where $b_\sigma$ depends on the structure of the star
\begin{align}
b_\sigma=\left<\bm{\xi}_\sigma,2i\bm{\Omega}_s\times\bm{\xi}_\sigma\right>+2\omega_\sigma\left<\bm{\xi}_\sigma,\bm{\xi}_\sigma\right>. \label{balpha}
\end{align}
Henceforth we restrict our discussions to systems where the spin is anti-aligned with the orbital angular momentum, with $(\alpha,\beta)=(0,\pi)$. In this case, the Wigner $\mathcal{D}$-functions reduce to $\mathcal{D}^{(2)}_{m^\prime m}=\delta_{m^\prime, -m}$, and Eq.\ (\ref{U-Y})
becomes
\begin{align}
U=-G{M_2}\sum_{lm}\frac{W_{lm}r_s^l}{{r}^{l+1}}e^{-im({\phi}+\Omega_st)}Y_{l,-m}(\theta^\prime,{\iota}^\prime).  
\end{align}
Here we focus on $(j=2,k=2,0)$ modes coupled to the gravitational fields labeled by $(l=2,m=-2,0)$, since they are the leading order terms in $\ra/r$, and give the strongest effects

The amplitudes of these modes are denoted by $c_0, c_{2,+}$ and $c_{2,-}$, where we have suppressed the mode index $j$. The equations of motion of these amplitudes are given by
\begin{subequations}
\begin{align}
&\dot{c}_0+i\omega_0c_0=f_0,  \label{c0-eq}\\
&\dot{c}_{2,\nu}+i\omega_{2,\nu} c_{2,\nu}=f_{2,\nu}, \label{c2m-eq} 
\end{align}
\label{c-eq-full}%
\end{subequations}
with the driving force $f_{2,\nu}$ and $f_0$ given by the RHS of Eq.\ (\ref{c-evo}). In particular, for the $f$-mode of Maclaurin spheroid we know \cite{Ho+Lai+99,Braviner+Ogilvie+14}
\begin{subequations}
\begin{align}
&\bm{\xi}_{2,2}=\frac{1}{\sqrt{2I^s_{xy}}} [(x^\prime+iy^\prime),i(x^\prime+iy^\prime),0], \label{xi22}  \\
&\bm{\xi}_{2,0}=iV\left[-x^\prime-2\frac{i\Omega_s}{\omega_0}y^\prime,-y^\prime+2\frac{i\Omega_s}{\omega_0}x^\prime,2z^\prime\right], \label{xi0} 
\end{align}
\label{f-xi}%
\end{subequations}
where the coefficients $V$ and $I^s_{xy}=I_{xx}+I_{yy}$ are determined by the normalization condition Eq.\ (\ref{norm-con}). Here $I_{xx}$ and $I_{yy}$ are the components of the moment of inertia $I_{ij}=\int \rho x^\prime_ix^\prime_jd\textit{V}^\prime$. We do not provide the expressions of $V$ and $I^s_{xy}$ since they are not needed in the future --- in the final equations of motion, these quantities will absorbed into tidal Love number and $f$-mode frequency of the NS, see Eq.\ (\ref{eq-lambda2}), (\ref{eq-lambda0}) and text around them. Then we get
\begin{subequations}
\begin{align}
&f_{2,\pm}=\frac{i\sqrt{I^s_{xy}}}{\omega_{2,\pm}+\Omega_s}\frac{3{M_2}}{4\sqrt{2}{r}^3}e^{2i({\phi}+\Omega_st)}, \\
&f_{0}=\frac{-i{M_2}}{{r}^3}\frac{\Omega_s}{4V\omega_0^2}.
\end{align}
\label{fs}%
\end{subequations}
In fact, Eqs.~(\ref{fs}) are not limited to Maclaurin spheroid. For a non-Maclaurin NS with low spin, we have [based on the definition of $(j=2,k=2)$ mode]
\begin{align}
\bm{\xi}_{22}=h_{22}(r_s)\bm{\nabla} Y_{22}(\theta^\prime,\iota^\prime), \label{xi22-generic}
\end{align}
where $h_{22}(r_s)$ depends on the EoS. This always leads to 
\begin{align}
f_{2,\pm}\sim \frac{1}{r^3}e^{2i(\phi+\Omega_st)}, \label{f-generic}
\end{align}
with the coefficient eventually absorbed into tidal Love numbers. For larger spins, the NS's $j=2$ modes will couple to $j \neq 2$ tidal gravity field (which are weaker), we ignore this coupling in this paper. 

\subsubsection{Orbital evolution}

By coupling the orbital motion to the NS modes, one can write the Hamiltonian of the whole system as \cite{Flanagan+Racine+07}
\begin{align}
&H=\frac{p_r^2}{2\mu}+\frac{p_\phi^2}{2\mu r^2}-\frac{\mu {M_t}}{r}+b_0(\omega_0|c_0|^2+if_0c_0^*-if_0^*c_0) \notag \\
&+\sum_{\nu=\pm}b_{2,\nu}(\omega_{2,\nu}|c_{2,\nu}|^2+if_{2,\nu}c_{2,\nu}^*-if_{2,\nu}^*c_{2,\nu}), \label{Hamiltonian}
\end{align}
where $\mu$ is the reduced mass and ${M_t}$ is the total mass.\ The generalized coordinates of the system consists of ($r$, $\phi$, $c_0$, $c_{2,\pm}$), and the conjugate momenta ($p_r$, $p_\phi$, $ib_0c_0^*$, $ib_{2,\pm}c_{2,\pm}^*$).\ From Hamilton's equations we obtain the equations of motion
\begin{subequations}
\begin{align}
\ddot{r}- r\dot{\phi}^2&=-\frac{{M_t}}{r^2}+\frac{3ib_0}{\mu r}(c_0^*f_0-c_0f_0^*) \notag \\
&+\sum_{\nu=\pm}\frac{3ib_{2,\nu}}{\mu r}(c_{2,\nu}^*f_{2,\nu}-c_{2,\nu}f_{2,\nu}^*), \label{r-eq-no-diss}\\
 r\ddot{\phi}+ 2\dot{r}\dot{\phi}&=\sum_{\nu=\pm}\frac{2b_{2,\nu}}{\mu}(c^*_{2,\nu}f_{2,\nu}+c_{2,\nu}f^*_{2,\nu}). \label{phi-eq-no-diss}
\end{align}
\label{orbit-r-phi-no-diss}%
\end{subequations}
Equations\ (\ref{c-eq-full}), together with Eqs. (\ref{orbit-r-phi-no-diss}), are a complete set of equations that describe the conservative evolution of the inspiraling BNS system. To include the effect of gravitational radiation, we add the Burke-Thorne dissipation term to the orbital evolution \cite{Flanagan+Hinderer+08}
\begin{align}
a_i=-\frac{2}{5}x_j\frac{d^5\textsl{Q}_{\textit{ij}}^\text{Total}}{dt^5}, \label{Burke-Thorne}
\end{align}
where $Q^\text{Total}_{ij}$ is the total quadrupole moment of the system in the inertial frame, which consists of the orbital part and the stellar part, i.e., $Q^\text{Total}_{ij}=Q_{ij}+\mu( x_ix_j-r^2\delta_{ij}/3)$. For simplicity, we neglect the effect of radiation reaction on the mode evolution. 

To express $Q_{ij}$ in terms of the mode amplitudes, we start from the definition of the stellar quadrupole moment in the co-rotating frame
\begin{align}
Q^{\prime ij}=\int d^3 \textit{x}^\prime\rho\left(\textit{x}^{\prime \textit{i}} \textit{x}^{\prime \textit{j}}-\frac{1}{3}\textit{r}^{\prime2}\delta^{\textit{ij}}\right).
\end{align}
The unperturbed quadrupole moment vanishes under the axisymmetric assumption. To linear order in perturbation, we get\footnote{The symbol $\delta$ on the RHS represents Eulerian perturbation, however, the symbol on the LHS only means the perturbation of the integral.} \cite{shapiro2008black}
\begin{align}
\delta Q^{\prime ij}=&\int d^3 \textit{x}^\prime \delta\rho\left(\textit{x}^{\prime \textit{i}}\textit{x}^{\prime \textit{j}}-\frac{1}{3}\textit{r}^{\prime 2}\delta^{\textit{i}\textit{j}}\right) \notag \\
&+\int d^3 \textit{x}^\prime \nabla\cdot\left[\rho\bm{\xi}\left(\textit{x}^{\prime \textit{i}}\textit{x}^{\prime \textit{j}}-\frac{1}{3}\textit{r}^{\prime 2}\delta^{\textit{ij}}\right)\right] \notag \\
&=\int d^3 \textit{x}^\prime\rho\left(\textit{x}^{\prime \textit{i}}\xi^{\prime \textit{j}}+\textit{x}^{\prime \textit{j}}\xi^{\prime \textit{i}}-\frac{2\textit{r}^\prime}{3}\xi_\textit{r}^\prime\delta^{\textit{ij}}\right), \label{Q-pert}
\end{align}
where we have used $\delta\rho=-\nabla\cdot(\rho\bm{\xi})$ to simplify the expression. The tensorial components of symmetric tracefree tensors are related to their harmonic components $q^\prime_{lm}$ through Clebsch-Gordan coefficients. The transformation can be expressed in a compact form \cite{Thorne+80}
\begin{subequations}
\begin{align}
&\delta Q^{\prime ij}={J}^{ij}_mq_m^\prime, \label{Q-qm}\\
&q_m^\prime=({J}_{m}^{ij})^*\delta Q^{\prime ij}, \label{qm-Q}
\end{align}
\end{subequations}
where we suppress the index $l$ of $q^\prime$ since we only consider $l=2$ components, and
\begin{align}
&{J}_{-2}^*={J}_2= \frac{1}{2}\left(\begin{matrix}
   1 & i & 0 \\
i & -1 & 0  \\
   0 & 0 & 0  \\
  \end{matrix}\right),  ~{J}_0= \frac{1}{\sqrt{6}}\left(\begin{matrix}
   -1 & 0 & 0 \\
0 & -1 & 0  \\
   0 & 0 & 2  \\
  \end{matrix}\right). \notag
\end{align}
Combining Eqs. (\ref{f-xi}), (\ref{Q-pert}) and (\ref{qm-Q}), we obtain
\begin{subequations}
\begin{align}
&q_{-2}^{\prime*}=q_2^\prime=\sqrt{2I^s_{xy}}(c_{2,+}+c_{2,-}), \label{q2-c2} \\
&q_0^\prime=\sqrt{\frac{2}{3}}\frac{V\omega_0}{\Omega_s}\left(c_{0,+}+c_{0,+}^*\right). \label{q0-c0}
\end{align}
\label{q-c-tot}%
\end{subequations}
Note that the harmonic component $q_2^\prime$ is a linear combination of retrograde and prograde modes, which oscillates at two different mode frequencies. So one can expect that $q_2^\prime$ satisfies a second order differential equation.

So far the expressions are in the co-rotating frame; to transform them to the inertial coordinate system, one can use the relationship between tensor components in the two frames
\begin{align}
Q^{ij}=R^i_mR^j_nQ^{\prime mn}, \notag 
\end{align}
where the operator $R$ first rotates $Q^{\prime mn}$ along the 
$z^\prime$-axis by $-\Omega_st$, and does the other rotation along the new $x$-axis by $\pi$, i.e.,
\begin{align}
&R=\left(\begin{matrix}
   \cos\Omega_st &  -\sin\Omega_st & 0 \\
-\sin\Omega_st & - \cos\Omega_st & 0  \\
   0 & 0 & -1  \\
  \end{matrix}\right). \notag
\end{align}
This results in
\begin{subequations}
\begin{align}
&q_2=e^{2i\Omega_st}q_{-2}^\prime,  \\
&q_0=q_0^\prime.
\end{align}
\label{q-trans}%
\end{subequations}
Plugging Eqs. (\ref{q-c-tot}) and (\ref{q-trans}) into Eqs.\ (\ref{orbit-r-phi-no-diss}), we finally get
\begin{subequations}
\begin{align}
&\ddot{r}- r\dot{\phi}^2=-\frac{{M_t}}{r^2}+\frac{3{M_2}}{2\mu r^4}\sqrt{\frac{3}{2}}q_0^\prime-\frac{9{M_2}}{2\mu r^4}A+\frac{1}{5}\sqrt{\frac{2}{3}}\frac{d^5\textit{q}_0^\prime}{d\textit{t}^5}r \notag \\
&-\frac{2r}{5}\text{Re}\left[e^{-2i\phi}\frac{d^5}{d\textit{t}^5}(\textit{q}_{2}^\prime e^{-2i\Omega_st})\right]-\frac{\mu}{15}\frac{d^5\textit{r}^2}{d\textit{t}^5}r \notag \\
&-\frac{\mu r}{5}\text{Re}\left[e^{-2i\phi}\frac{d^5}{d\textit{t}^5}(r^2e^{2i\phi})\right], \label{r-eq} \\
& r\ddot{\phi}+ 2\dot{r}\dot{\phi}=\frac{3{M_2}}{\mu r^4}B-\frac{2r}{5}\text{Im}\left[e^{-2i\phi}\frac{d^5}{d\textit{t}^5}(q_{2}^\prime e^{-2i\Omega_st})\right] \notag  \\
&-\frac{\mu r}{5}\text{Im}\left[e^{-2i\phi}\frac{d^5}{d\textit{t}^5}(r^2e^{2i\phi})\right],\label{phi-eq}\\
& \ddot{q}_2^\prime-2i\omega_3\dot{q}_2^\prime+\omega_2^2 q_2^\prime=\frac{3}{2}\frac{\omega_2^2\lambda_2 {M_2}}{r^{3}}e^{2i\phi+2i\Omega_st}-\frac{3{M_2}}{2r^3}e^{2i\phi+2i\Omega_st} \notag \\
&\times\frac{\omega_2^2\lambda_2(\Omega_s-\omega_3)}{\Omega_s^2-2\Omega_s\omega_3-\omega_2^2}\left(2\dot{\phi}+2\Omega_s-\omega_3+3i\frac{\dot{r}}{r}\right), \label{q2-eq} \\
&\ddot{q}_0^\prime+\omega_0^2q_0^\prime=-\sqrt{\frac{3}{2}}\frac{\omega_0^2\lambda_0{M_2}}{r^3}, \label{q0-eq}
\end{align}
\label{eq-of-motion}%
\end{subequations}
where we have defined two real variables $A$ and $B$ as
\begin{align}
q_2^\prime e^{-2i\phi-2i\Omega_st}=A+iB. \label{A-B-def}
\end{align}
In Eqs.\ (\ref{eq-of-motion}), $A$ is proportional to the radial force while $B$ to the azimuthal torque. We have also defined 
\begin{align}
&\lambda_2=I^s_{xy}/\omega_2^2, \label{eq-lambda2}\\
&\lambda_0=(I^s_{xy}+4I_{zz})/(3\omega_0^2). \label{eq-lambda0}
\end{align}
It is straightforward to identify these two quantities as the Love numbers of the $(2,2)$ and $(2,0)$ modes, respectively. 

When deriving Eqs.\ (\ref{eq-of-motion}), we have assumed the star is described as a Maclaurin spheroid.  Nonetheless, this affects only the values of the coupling constants, $\lambda_0$ and $\lambda_2$. The form of Eqs.\ (\ref{eq-of-motion}) holds generically [as we discussed in Eqs.~(\ref{xi22-generic}) and (\ref{f-generic})]. To generailize the result to a realistic EoS, one only needs to replace the values of $\lambda_0$ and $\lambda_2$ accordingly --- our equation of motion is an effective theory for the evolution of binary system (without relativistic corrections). Under the assumption of homogeneity and incompressibility, the Love numbers become $\lambda_0=\lambda_2=\ra^5/2$ for a non-spinning NS. This leads to $k_2=3/4$
[see Eq.\ (\ref{lambda-k}) and Ref.\ \citep{wahl2017concentric}]. However, this value differs significantly from those obtained from more realistic EoS (cf.~numbers provided in Sec.~\ref{sec:EoS}). Hence in this paper, we obtain values of $\lambda_0$ and $\lambda_2$ by inserting values of $\ra$ and $k_2$ from H4 and $\Gamma=2$ polytropic EoS into Eq.\ (\ref{lambda-k}); and we ignore the spin corrections to them. As a result, our calculations do not rely on the expressions of the auxiliary variables we introduced in Eq.~(\ref{f-xi}).

The two frequency parameters $\omega_2$ and $\omega_3$ in Eqs. (\ref{eq-of-motion}) are given by
\begin{align}
&\omega_2^2=-\omega_{2+}\omega_{2-},\label{omega2}\\
&\omega_3=-\frac{\omega_{2+}+\omega_{2-}}{2}.
\end{align}
The minus sign appears in Eq.\ (\ref{omega2}) because $\omega_{2\pm}$ have opposite signs. As discussed in the last subsection, we assume the mode frequencies dependence on $\Omega_s$, given in Eqs.\ (\ref{omega-total}), is still valid, which implies
\begin{align}
\omega_3=\Omega_s, \label{omega3-spin}
\end{align}
and the second term on the RHS of Eq. (\ref{q2-eq}) vanishes in our case.

We can see that Eqs.~(\ref{eq-of-motion}) reduce to the conventional mode-orbit equations when $\Omega_s\to0$ [cf.~Eq.~(6) of Ref.~\cite{Flanagan+Hinderer+08}].  As discussed by Ref. \cite{Flanagan+Hinderer+08}, high order time derivatives in the radiation reaction terms can be lowered by repeatedly replacing the second time derivatives by contributions from the conservative part alone. In this way, Eqs.\ (\ref{eq-of-motion}) become a set of second order ordinary differential equations.

\section{Model of DT: Stellar oscillations}
\label{sec:new-dyn-tides}
As we have discussed in the introduction, both L94 \cite{Lai+94} and FR07 \cite{Flanagan+Racine+07} focused on the total change in the orbital phase when the system evolves through a DT resonance. This is because for $g$- and/or $r$-modes that have weak tidal couplings, only the resonant regime plays a significant role in affecting the orbital evolution. On the other hand, H+16 \cite{Steinhoff+Hinderer+16,Hinderer+Taracchini+16} proposed an EOB formalism to study the strongly tidal-coupled $f$-mode by introducing an effective Love number, which works well when the driving frequency is comparable yet still less than the eigenfrequency of the $f$-mode. In this and the next sections, we will use semi-analytic methods to carry out a systematic study of DT, and provide an alternative way to describe the full dynamics of DT, including both stellar and orbital evolutions. This section mainly focuses on the stellar part, where we extend H+16 \cite{Steinhoff+Hinderer+16,Hinderer+Taracchini+16} and find analytic solutions of stellar evolution that are valid in all regimes (adiabatic, resonant and post-resonance) and for arbitrary spins. With the new analytic expressions, we can have a better understanding on DT. We first review the approximations presented in L94 \cite{Lai+94} and H+16 \cite{Steinhoff+Hinderer+16,Hinderer+Taracchini+16} in Sec.\ \ref{sec:DT-review}, and then in Sec.\ \ref{sec:new_DT} we propose our new approximations and compare them with numerical integrations. In the next section (Sec.\ \ref{sec:post-resonance orbit dynamics}), we will apply our approximation to describe tidal back-reaction.

\subsection{Previous studies on DT}
\label{sec:DT-review}
As studied in L94 \cite{Lai+94}, the $(2,2)$ mode $q_2^\prime$ in a non-spinning NS can be treated as a harmonic oscillator driven by tidal force
\begin{align}
\ddot{q}_2^\prime+\omega_2^2q_2^\prime=\frac{3}{2}\frac{\omega_2^2\lambda_2 M_2}{r^{3}}e^{2i\phi}. \label{q2-nospin}
\end{align}
When the orbital frequency $\Omega \ll \omega_2$, the NS adiabatically follows the tidal driving, with its main time dependence given by $e^{2i\phi}$. Therefore it is appropriate to define a variable $b=q_2^\prime e^{-2i\phi}$, which satisfies
\begin{align}
\ddot{b}+4i\Omega\dot{b}+(\omega_2^2-4\Omega^2)b=\frac{3}{2}\frac{\omega_2^2\lambda_2 M_2}{r^{3}}.
\end{align}
Here we have ignored time derivative of orbital frequency since its rate of change due to GW radiation is small compared with other variables. Note that the quantity $A+iB$ we defined in the last section reduces to $b$ when the spin vanishes. Since the major time dependence $e^{-2i\phi}$ has been factored out, we have $\ddot{b}\ll 4\Omega\dot{b}\ll (\omega_2^2 - 4\Omega^2)b$, it is safe to ignore $\ddot{b}$ and $\dot{b}$, leading to the well-known adiabatic approximation
\begin{align}
b=\frac{3\omega_2^2\lambda_2 M_2}{2r^{3}}\frac{1}{\omega_2^2-4\Omega^2}. \label{b-adia}
\end{align}
As $\Omega$ approaches $\omega_2/2$, the mode gets resonantly excited. L94 \cite{Lai+94} assumed that near resonance, the mode mainly oscillates at its natural frequency $\omega_2$, so they defined a slowly varying complex amplitude $c=q_2^\prime e^{-i\omega_2t}$, which satisfies\footnote{The other term proportional to $q_2^\prime e^{i\omega_2t}$ doesn't contribute to SPA in Eq. (\ref{c-no-spin-amp})}
\begin{align}
\ddot{c}+2i\omega_2\dot{c}=\frac{3}{2}\frac{\omega_2^2\lambda_2 M_2}{r^{3}}e^{2i\phi-i\omega_2t}.
\end{align}
Similarly, by neglecting $\ddot{c}$, this equation can be solved as
\begin{align}
c=\frac{3}{4i\omega_2}\int^t\frac{\omega_2^2\lambda_2 M_2}{r^{\prime3}}e^{2i\phi^\prime-i\omega_2t^\prime}d\textit{t}^\prime, \label{L94+c}
\end{align}
which can in turn be evaluated with SPA, giving the post-resonance amplitude:
\begin{align}
|c|=\frac{3}{4\omega_2}\frac{\omega_2^2\lambda_2 M_2}{r^{3}_r}\sqrt{\frac{\pi}{\omd}}. \label{c-no-spin-amp}
\end{align}
Hereafter we use the subscript $r$ to refer to the point of resonance. As we can see, the treatment in L94 \cite{Lai+94} is piecewise: they separated out distinct time dependence in different regimes. This is enough for evaluating the energy and angular momentum transfers from orbital motion to NS mode since they only depend on the post-resonance amplitude. However, neither the detailed time evolution of the mode, nor the orbital dynamics in the resonant regime were provided.

L94 \cite{Lai+94} was improved by H+16 \cite{Steinhoff+Hinderer+16,Hinderer+Taracchini+16}, who solved Eq. (\ref{q2-nospin}) with the Green function, obtaining
\begin{align}
q_2^\prime(t)=\frac{3}{2\omega_2}\int^t\frac{\omega_2^2\lambda_2 M_2}{r^{\prime3}}e^{2i\phi^\prime}\sin\omega_2(t-t^\prime)d\textit{t}^\prime. \label{q2p-green}
\end{align}
Near resonance, Eq.\ (\ref{q2p-green}) reduces to Eq.\ (\ref{L94+c}) if one writes $\sin\omega_2(t-t^\prime)=\frac{1}{2i}[e^{2i\omega_2(t-t^\prime)}-e^{-2i\omega_2(t-t^\prime)}]$ and neglects the term that does not contribute to SPA. However, Eq.\ (\ref{q2p-green}) is exact in all regimes. This lays the foundation to obtain a single continuous function to represent  the stellar motion during DT. Instead of using SPA to get the final amplitude of the mode, H+16 \cite{Steinhoff+Hinderer+16,Hinderer+Taracchini+16} expanded the integrand in Eq. (\ref{L94+c}) near resonance
\begin{align}
c=\frac{3}{4i\omega_2}\frac{\omega_2^2\lambda_2 M_2}{r^{3}_r}\int^te^{i\omd (t^\prime-t_r)^2}\textit{t}^\prime, \label{H+16+c+time}
\end{align}
which becomes a Fresnel function. This approximation is accurate within the duration of the resonance $T_\text{dur}$
\begin{align}
|t-t_r|\ll T_\text{dur}, \label{timewidth}
\end{align}
where $T_\text{dur}=\sqrt{\frac{\pi}{\omd}}$. They then asymptotically matched Eq.\ (\ref{H+16+c+time}) to Eq.\ (\ref{b-adia}). More specifically, they first observed that Eq.\ (\ref{b-adia}) diverges as $(t-t_r)^{-1}$ as $\Omega\to\omega_2/2$
\begin{align}
be^{2i\phi-i\omega_2t}\sim-M_2\lambda_2\omega_2^2\frac{3}{8\omega_2 r^{3}_r}\frac{e^{2i\phi_r-i\omega_2t}}{\omd (t-t_r)}. \label{conter-old}
\end{align}
H+16 \cite{Steinhoff+Hinderer+16,Hinderer+Taracchini+16} used the RHS of Eq. (\ref{conter-old}) as a counterterm: they added the adiabatic solution in Eq.\ (\ref{b-adia}) and the resonant one in Eq.\ (\ref{H+16+c+time}) up and then subtracted the counterterm. In this way, the divergence is cured, and the sum has the correct asymptotic behavior in both the adiabatic and resonant regimes. This new solution cannot describe the post-resonance evolution, as is expected because the asymptotic behavior in that regime was not yet considered. As pointed out in the introduction, this approximation is sufficient for non-spinning NS if the post-resonance regime is short. However, for highly spinning systems, we must extend this method to the post-resonance regime.

\subsection{New approximation and numerical comparisons}
\label{sec:new_DT}
\begin{figure}[htb]
        \includegraphics[width=\columnwidth,height=5.6cm,clip=true]{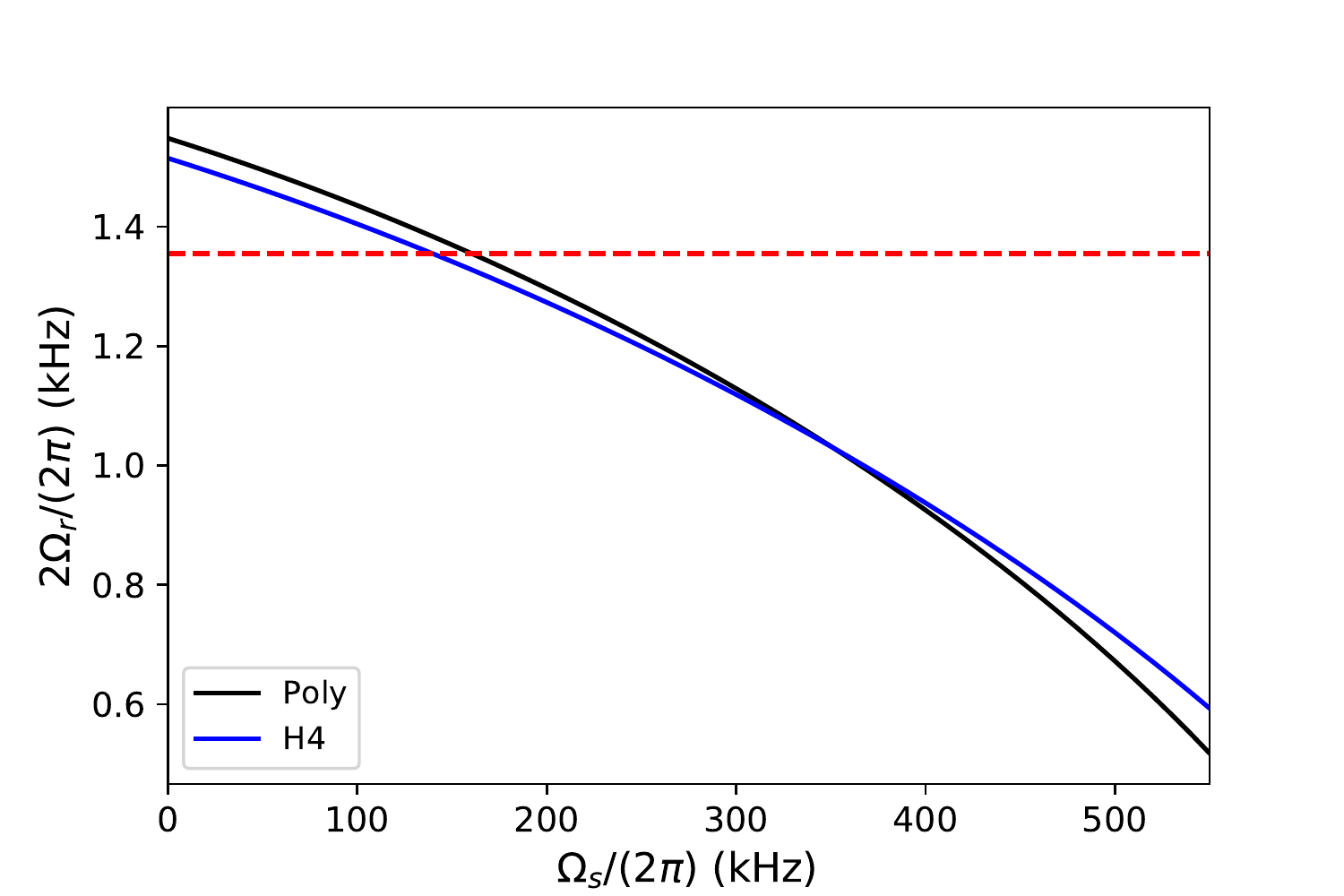}
  \caption{The resonant GW frequency ($2\Omega_r$) as functions of spin frequency for two EoS. We also plot the contact GW frequency as red dashed line for comparison. The retrograde mode frequency is shifted by spin to a smaller value, which makes DT possible during the inspiral.}
 \label{fig:reson-fre}
\end{figure}

Let us start from the equation that governs the $(2,2)$ mode [Eq. (\ref{q2-eq})]. By defining $x=q_2^\prime e^{-i\Omega_st}$, it becomes
\begin{align}
\ddot{x}+\zeta^2x=\frac{3}{2}\frac{\omega_2^2\lambda_2 M_2}{r^{3}}e^{2i\phi+i\Omega_st},  \label{eq-nogw-q2}
\end{align}
where
\begin{align}
&\zeta^2=\Omega_s^2+\omega_2^2.
\end{align}
Note that the second term on the RHS of Eq.\ (\ref{q2-eq}) vanishes because $\omega_3=\Omega_s$ [Eq.~(\ref{omega3-spin})]. The resonance is determined by the condition
\begin{align}
\dot{\phi}=\Omega_r=\frac{\zeta-\Omega_s}{2}.  \label{reson-condition}
\end{align} 
Under the assumed $\omega_2-\Omega_s$ relation, $\zeta$ can be simplified to $(\omega_+-\omega_-)/2$, then we have
\begin{align}
\dot{\phi}=\Omega_r=-\Omega_{s}-\frac{\omega_{2-}}{2}, \label{reson-ome-minus}
\end{align}
 but here we keep $\zeta$ for generality. Eq.~(\ref{reson-ome-minus}) shows that only the retrograde mode is excited.
 The dependence of $\Omega_r$ on $\Omega_s$ is shown in Fig.~\ref{fig:reson-fre}.

By incorporating spin into procedures discussed in the previous subsection, H+16's result \cite{Steinhoff+Hinderer+16,Hinderer+Taracchini+16} can be written as
\begin{subequations}
\begin{align}
&A(t)=\frac{3M_2\lambda_2\omega_2^2}{2r^3}\frac{1}{\zeta^2-(\Omega_s+2\Omega)^2}+\frac{3M_2\lambda_2\omega_2^2}{8\sqrt{\omd}\zeta r^{3}_r}\frac{1}{\tha} \notag \\
&+\frac{3M_2\lambda_2\omega_2^2}{4r^{3}_r\zeta}\sqrt{\frac{\pi}{2\omd}} \left[-\frac{1}{\sqrt{2}}\sin\left(\tha^2-\frac{\pi}{4}\right)\right.\notag \\
&\left.-\FC\left(\sqrt{\frac{2}{\pi}}\tha\right)\sin\tha^2+\FS\left(\sqrt{\frac{2}{\pi}}\tha\right)\cos\tha^2\right] , \label{steinhoff-A} \\
&B(t)=\frac{3M_2\lambda_2\omega_2^2}{4r^{3}_r\zeta}\sqrt{\frac{\pi}{2\omd}} \left[-\frac{1}{\sqrt{2}}\sin\left(\tha^2+\frac{\pi}{4}\right)\right.\notag \\
&\left.-\FC\left(\sqrt{\frac{2}{\pi}}\tha\right)\cos\tha^2-\FS\left(\sqrt{\frac{2}{\pi}}\tha\right)\sin\tha^2\right], \label{steinhoff-B} 
\end{align}
\label{steinhoff-AB}%
\end{subequations}
where variables $A$ and $B$ are defined in Eq. (\ref{A-B-def}). We can see that the phase of $A$ and $B$'s oscillations is governed by:
\begin{align}
\tha=\sqrt{\omd}(t-t_r).
\end{align}
$\FC$ and $\FS$ in Eqs.\ (\ref{steinhoff-AB}) are Fresnel functions defined as $\int^{\tha}_{-\infty} \sin s^2d\textit{s}=\sqrt{\pi/8}[1+2\FS(\tha\sqrt{2/\pi})]$ and $\int^{\tha}_{-\infty} \cos s^2d\textit{s}=\sqrt{\pi/8}[1+2\FC(\tha\sqrt{2/\pi})]$.

\begin{figure}[hbt!]
        \includegraphics[width=\columnwidth,clip=true]{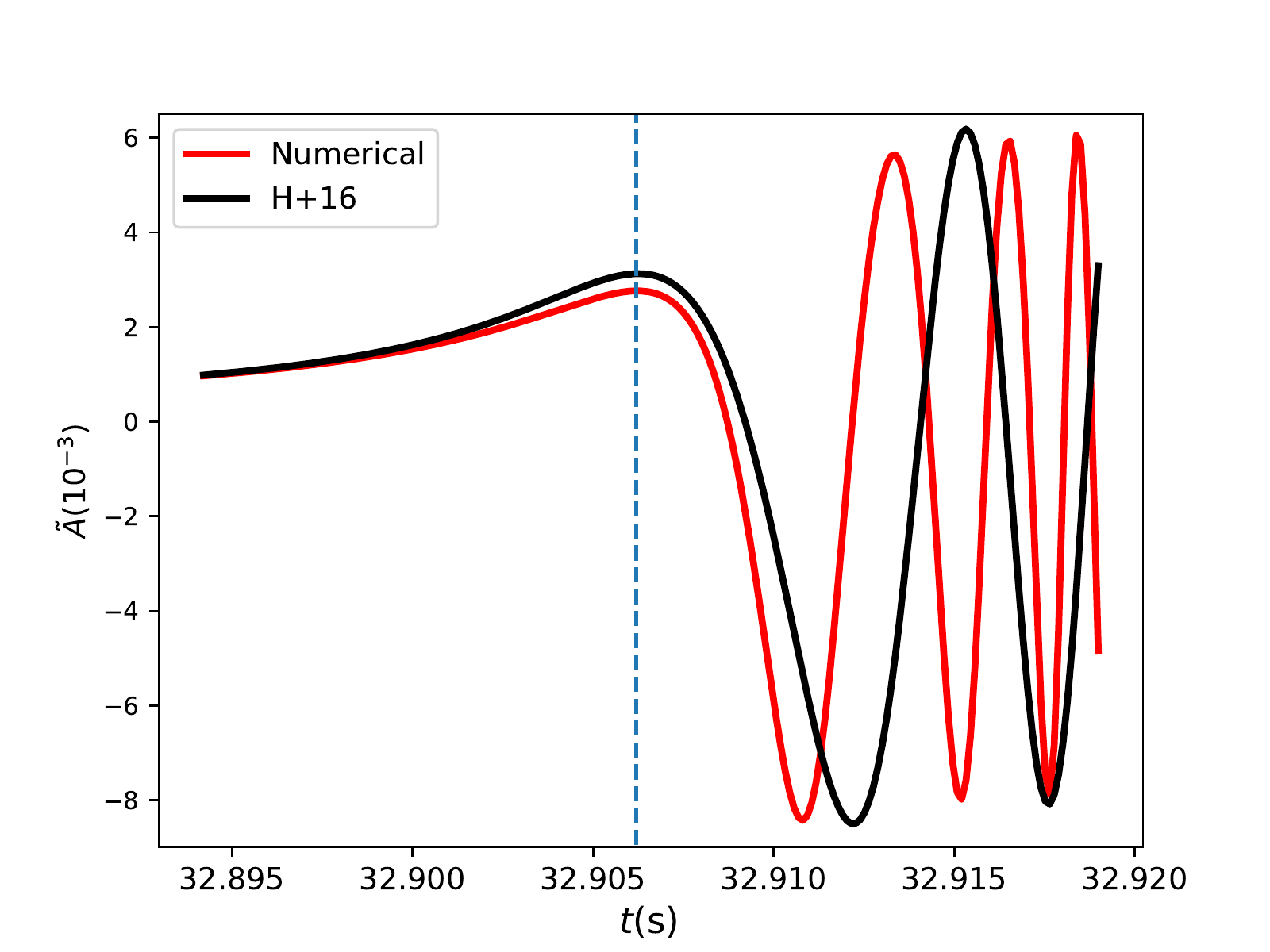}
        \includegraphics[width=\columnwidth,clip=true]{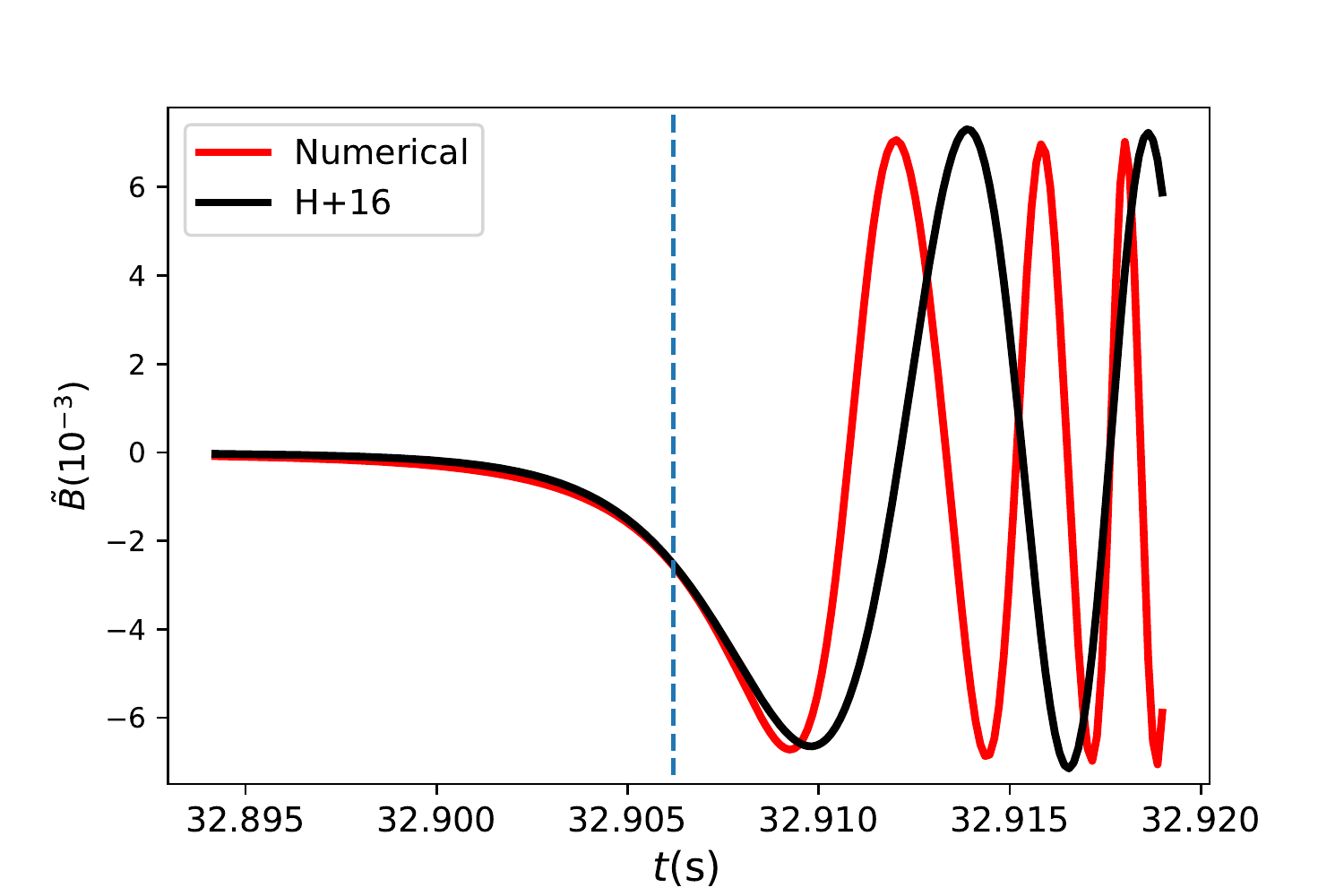}
  \caption{Dimensionless quadrupole moments [normalized by ${\ra^3}$ in Eq. (\ref{normalization-AB})] induced by DT as functions of time. Red curves are results from fully numerical evolution and black curves are from Eqs.\ (\ref{steinhoff-AB}). The vertical dashed blue line denotes the time of resonance. Eqs.\ (\ref{steinhoff-AB}) are accurate in pre-resonance regime, but fail to describe the phasing of post-resonance oscillation.}
 \label{fig:q-formula-H16}
\end{figure}

To check the accuracies of these formulae, we compare them with numerical integrations of Eqs.\ (\ref{eq-of-motion}). We choose the H4 EoS and spin frequency of 550Hz. This gives $e_0=0.63$, $\omega_0=2\pi\times1.71$kHz, $\omega_+=2\pi\times0.59$kHz and $\omega_-=-2\pi\times1.69$kHz. Eq.\ (\ref{reson-condition}) indicates that  resonance happens at the orbital angular frequency $2\pi\times0.30$kHz. Using these numbers, we solve Eqs.\ (\ref{eq-of-motion}) numerically with the following initial conditions:
 \begin{align}
 &\dot{\phi}^{(0)}=2\pi F_0=2\pi\times18\text{Hz}, ~ r^{(0)}=\left(\frac{M_t}{\dot{\phi}^{(0)2}}\right)^{1/3}, \notag \\
 &\dot{r}^{(0)}=-\frac{64}{5}\eta\left(\frac{M_t}{r^{(0)}}\right)^{3}, ~q_0^{(0)}=-M_2\lambda_0\sqrt{\frac{3}{2}}\frac{1}{r_0^3}, \notag \\
 &\dot{q}_0^{(0)}=-3\frac{\dot{r}_r^{(0)}}{r_r^{(0)}}q_0^{(0)}, ~A^{(0)}=\frac{3M_2\lambda_2\omega_2^2}{2r^{(0)3}}\frac{1}{\zeta^2-(2\dot{\phi}^{(0)}+\Omega_s)^2}, \notag \\
 &\dot{A}^{(0)}=0, B^{(0)}=0, ~\dot{B}^{(0)}=0. \label{full-initial}
 \end{align}
The evaluation of Eq.\ (\ref{steinhoff-AB}) requires the information of orbital evolution, like $r(t)$, $\Omega(t)$, and $\omd$. Here we take them from the numerical integrations (with tidal back-reaction). In Fig. \ref{fig:q-formula-H16}, we plot the numerical solutions (red) versus predictions of Eqs.\ (\ref{steinhoff-AB}) (black). Dimensionless variables $\tilde{A}$ and $\tilde{B}$ are defined by
\begin{align}
\tilde{A}=\frac{3}{2}\frac{A}{\ra^3}, ~~~~\tilde{B}=\frac{3}{2}\frac{B}{\ra^3}. \label{normalization-AB}
\end{align}
The vertical dashed line labels the time of resonance. We can see that Eqs.\ (\ref{steinhoff-AB}) can describe pre-resonance evolutions of $A$ and $B$ to a high accuracy, despite a small discrepancy in $\tilde{A}$ at $t_r$. They smoothly connect the adiabatic and resonant regimes. In the post-resonance regime, the formulae give the correct amplitude of mode oscillation, same as L94 \cite{Lai+94}, but do not predict the correct phasing of post-resonance oscillation. Let us attempt to improve the treatment in H+16 \cite{Steinhoff+Hinderer+16,Hinderer+Taracchini+16}, in several steps.

\begin{figure}[hbt!]
        \includegraphics[width=\columnwidth,clip=true]{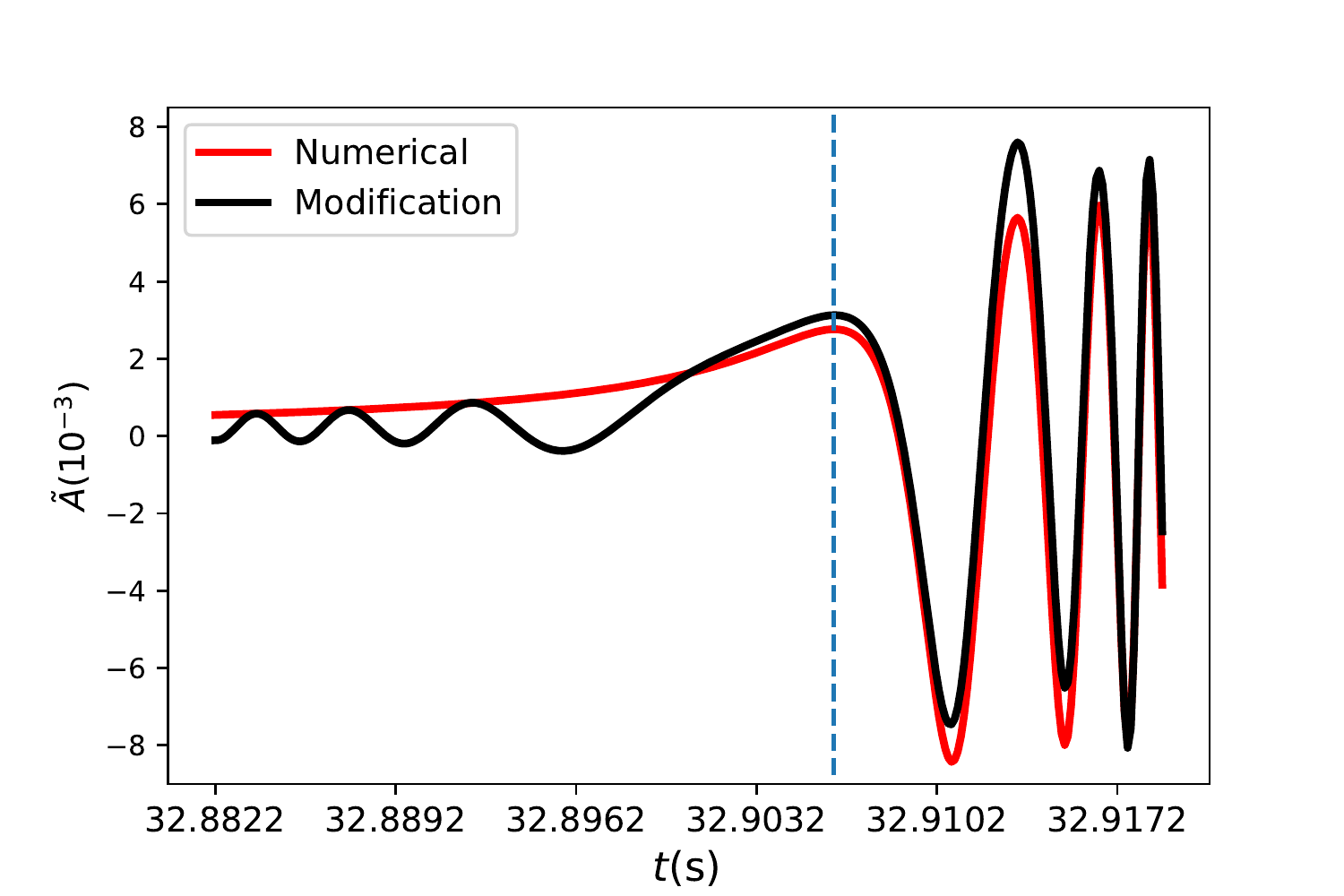}
  \caption{Time evolution of dimensionless quadrupole moment $\tilde{A}$. The black line represents the formula in Eq.\ (\ref{steinhoff-A}) with $\tha^2$ that appears in trigonometric functions replaced by $\Theta$ [Eq.\ (\ref{Theta})], while the red line is from numerical integrations. The vertical dashed line is the time of resonance. This modification gives the correct post-resonance phasing, but does not give accurate post-resonance amplitude nor adiabatic evolution. }
 \label{fig:Astein-modi}
\end{figure}

The post-resonance oscillation can be viewed as trigonometric functions modulated by Fresnel functions $\FC$ and $\FS$. In this regime, $\FC$ and $\FS$ both approach $1/2$ when $\tha\to\infty$, Eqs.\ (\ref{steinhoff-AB}) then predict
\begin{subequations}
\begin{align}
&A\sim\frac{3{M_2}\lambda_2\omega_2^2}{4r^3_r\zeta}\sqrt{\frac{\pi}{\omd}}\cos(\tha^2+\frac{\pi}{4}), \\
&B\sim-\frac{3{M_2}\lambda_2\omega_2^2}{4r^3_r\zeta}\sqrt{\frac{\pi}{\omd}}\sin(\tha^2+\frac{\pi}{4}), \label{B-asym-H16}
\end{align}
\end{subequations}
which lead to
\begin{align}
x\sim\frac{3M_2\lambda_2\omega_2^2}{4r^3_r\zeta}\sqrt{\frac{\pi}{\omd}}e^{-i\tha^2-i\pi/4+2i\phi+i\Omega_st}. \label{x-sym}
\end{align}
However, as pointed out by L94 \cite{Lai+94}, $x$ should oscillate at its eigenfrequency $\zeta$ in the post-resonance regime. Re-writing the phase of $x$ in Eq.\ (\ref{x-sym}) as $(2\phi-\zeta t+\Omega_st-\tha^2)+\zeta t-\pi/4$, it is straightforward to see that the term in the bracket is supposed to vanish in order to meet this requirement. Therefore we can attempt to replace all $\tha^2$ in trigonometric functions in Eq. (\ref{steinhoff-AB}) by 
\begin{align}
\Theta=-\chi_r-\zeta t+2\phi+\Omega_st, \label{Theta}
\end{align}
where $\chi_r=2\phi_r-\zeta t_r+\Omega_st_r$. The constant $\chi_r$ is chosen so that $\Theta$ is 0 at $t_r$ to match $\tha$. Note that $\tha^2$ is the leading order of Taylor expansion of $\Theta$ around $t_r$. Figure \ref{fig:Astein-modi} shows the result of our new approximation, which gives the correct phasing in the post-resonance regime, but still fails to explain the amplitude of the first cycle as well as the evolution in the adiabatic regime.

\begin{figure}[hbt!]
        \includegraphics[width=\columnwidth,clip=true]{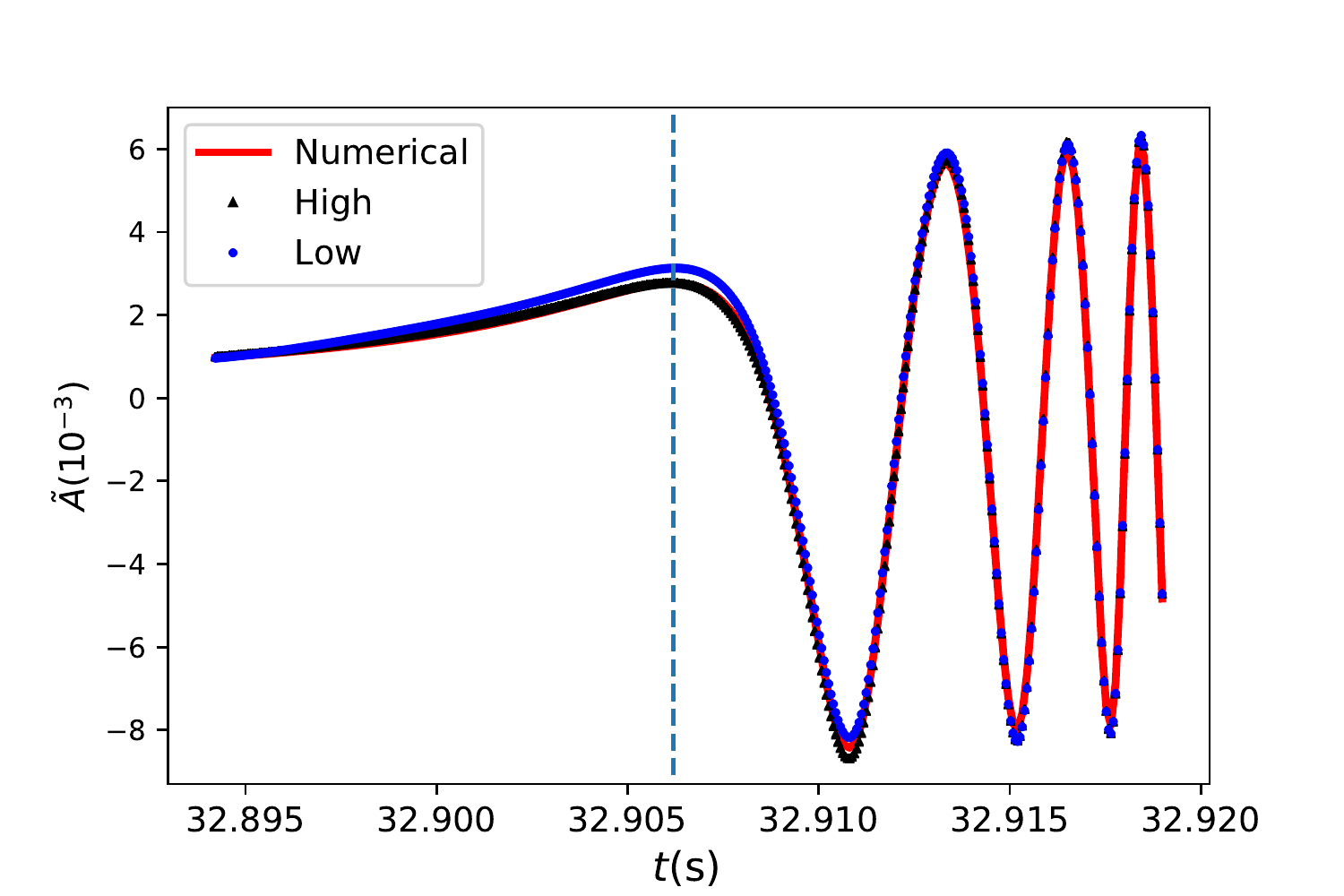}
        \includegraphics[width=\columnwidth,clip=true]{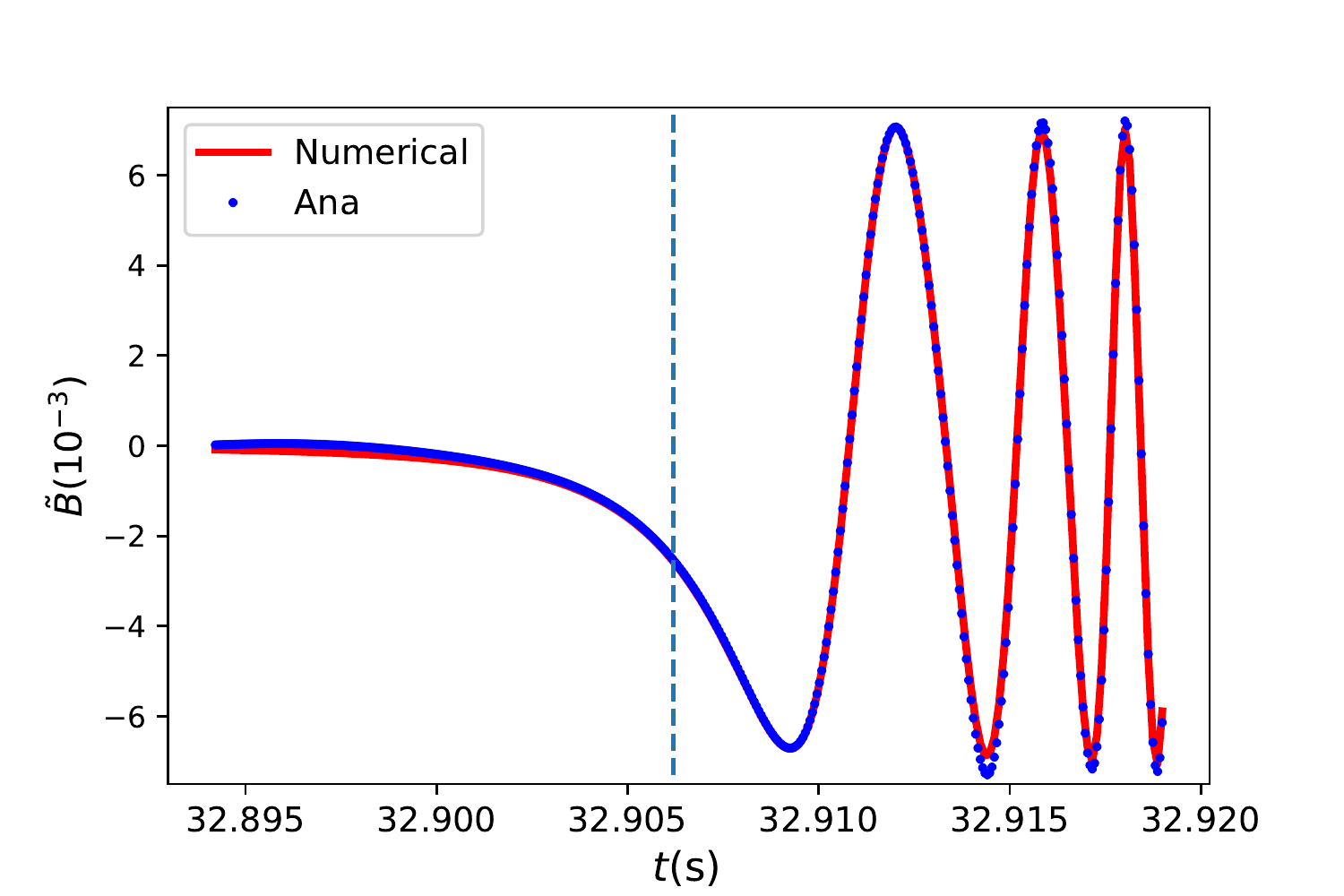} 
  \caption{Same as Fig.\ \ref{fig:q-formula-H16}, but the numerical solutions are compared with Eqs. (\ref{new-formulae-tide}) and (\ref{mine-A-ex}). Formula of $\tilde{B}$ is already accurate enough to fit the numerical results. While the formula of $\tilde{A}$ without higher order correction (blue dots) predicts larger value near $t_r$. The problem is fixed after the inclusion of Eq. (\ref{mine-A-ex}), which we plot as black triangles.}
 \label{fig:q-formula}
\end{figure}
These undesired features can be cured by making a further change to the counterterm Eq.\ (\ref{conter-old}) and adding a new term to $B$, resulting in:
\begin{subequations}
\begin{align}
&A=\frac{3M_2\lambda_2\omega_2^2}{2r^3}\frac{1}{\zeta^2-(2\Omega+\Omega_s)^2}+\frac{3M_2\lambda_2\omega_2^2}{8\sqrt{\omd}\zeta r^3_r}\frac{\cos(\tha^2-\Theta)}{\tha} \notag \\
&+\frac{3M_2\lambda_2\omega_2^2}{4r^3_r\zeta\dot{\Omega}_r}\sqrt{\frac{\pi}{2\omd}}\left[-\frac{1}{\sqrt{2}}\sin(\Theta-\frac{\pi}{4})\right. \notag \\
&\left.-\FC\left(\sqrt{\frac{2}{\pi}}\tha\right)\sin\Theta+\FS\left(\sqrt{\frac{2}{\pi}}\tha\right)\cos\Theta \right],\label{mine-A} \\
&B=\frac{3M_2\lambda_2\omega_2^2}{8r^3_r\zeta\sqrt{\omd}}\frac{\sin(\tha^2-\Theta)}{\tha} \notag \\
&+\frac{3M_2\lambda_2\omega_2^2}{4r^3_r\zeta}\sqrt{\frac{\pi}{2\omd}}\left[-\frac{1}{\sqrt{2}}\sin(\Theta+\frac{\pi}{4})\right. \notag \\
&\left.-\FC\left(\sqrt{\frac{2}{\pi}}\tha\right)\cos\Theta-\FS\left(\sqrt{\frac{2}{\pi}}\tha\right)\sin\Theta\right].  \label{mine-B}
\end{align}
\label{new-formulae-tide}%
\end{subequations}
We refer the interested readers to Appendix \ref{app:mode-oscil} for detailed derivations. The new expressions still need orbital information as input. For example, one cannot obtain $A(t)$ and $B(t)$ without the knowledge of $\omd$, $t_r$ and so on. In the next section, we will combine our new formulae with orbital evolutions to give analytic estimations on these parameters.

Results from Eq.\ (\ref{new-formulae-tide}) are plotted as blue dots in Fig.\ \ref{fig:q-formula}, and compared with numerical solutions (red lines). We can see that our new results are more accurate. In comparison  with H+16 \cite{Steinhoff+Hinderer+16,Hinderer+Taracchini+16}, the second term in the first line of Eq.\ (\ref{mine-A}) is multiplied by $\cos(\tha^2-\Theta)$. The modification can be understood as follows. The adiabatic term, i.e., the first term in Eq.\ (\ref{mine-A}), diverges as the system reaches the resonance point. H+16 \cite{Steinhoff+Hinderer+16,Hinderer+Taracchini+16} chose Eq.\ (\ref{conter-old}) as the counterterm to cancel the undesired infinity. Our better counterterm, $\cos(\tha^2-\Theta)/\tha$, not only diverges as $1/\tha$, but also has the correct oscillatory behavior. This cures the problems shown in Fig.\ \ref{fig:Astein-modi}. In $B$, we have a new term $\sim\sin(\tha^2-\Theta)/\tha$ [the first line in Eq.\ (\ref{mine-B})], which vanishes both as $|t_r|\to\infty$ and at $t_r$ (recall that $\displaystyle\lim_{x\to0}\sin x^3/x=0$, hence no infinity issue at $t_r$), therefore does not modify the asymptotic behaviors of $B$ in the adiabatic or in the post-resonance regimes. 

In comparison with Fig.\ \ref{fig:Astein-modi}, changes in Fig.\ \ref{fig:q-formula} not only cancel the undesired features in adiabatic regime, but also move the first cycle of post-resonance evolution downward to match the amplitude. Prior to resonance, $A$ gradually grows while $B$ remains 0. Approximately, the resonance time is the local maximum of $A$, but the value of $A$ on resonance is less than its final amplitude, only reaching it after one cycle. The evolution of $B$ is similar but lags behind $A$. Although Eq.\ (\ref{mine-A}) predicts slightly larger $A$ in the resonant regime, they are accurate enough for the purpose of studying the tidal back-reaction onto the orbital motion, as we shall see in the next section. 

If one wants to obtain more accurate expressions, especially to remove the discrepancy near resonance, a higher order correction can be made by adding 
\begin{align}
&\Delta A(t)=\frac{3M_2\lambda_2\omega_2^2}{16\zeta r^3_r\sqrt{\omd}}\frac{\sin(\tha^2-\Theta)}{\tha^3}, \label{mine-A-ex}
\end{align}
into Eq. (\ref{mine-A}). Readers can find derivations in Appendix \ref{app:mode-oscil}.
The result is shown in Fig.\ \ref{fig:q-formula} with black triangles, where we can see the formula with higher order correction gives more accurate description on $A$ near $t_r$. 

To quantify the accuracies of the analytic results, we calculate the values of $A$ and $B$ at $t_r$
\begin{subequations}
\begin{align}
&A_r=\frac{3M_2\lambda_2\omega_2^2}{8\zeta r^3_r}\left(\sqrt{\frac{\pi}{2\omd}}+\frac{3}{\omd}\frac{\dot{r}_r}{r_r}+\frac{1}{\zeta}\right)+\frac{{M_2}\lambda_2\omega_2^2}{8\zeta r^3_r}\frac{\ddot{\Omega}_r}{\omd^2},   \\
&B_r=-\frac{3{M_2}\lambda_2\omega_2^2}{8\zeta r^3_r}\sqrt{\frac{\pi}{2\omd}},  
\end{align}
\label{ab-proper}%
\end{subequations}
where the last term in $A_r$ comes from the higher order correction Eq.\ (\ref{mine-A-ex}). It is interesting to see that $B_r$ is equal to half of the final amplitude [cf. Eq.\ (\ref{B-asym-H16})]. For completeness, we also list $q_0$ below
\begin{align}
q_0=-{M_2}\lambda_0\sqrt{\frac{3}{2}}\frac{1}{r^3}, \label{q0-adiba}
\end{align}
which comes from the adiabatic approximation. These values are compared with numerical results in Table \ref{table:AB}, which shows that our analytic results of $A$ with higher order correction and $B$ only differ from numerical results by several percents. We can see the error decreases as spin rises. We also compare the formula of $A$ without the higher correction Eq. (\ref{mine-A-ex}), errors are around tens of percents. Hence the correction is important if we require high accuracy around the resonance.

Finally, we want to note that discussions in this subsection may not be useful in practice, because one can get tidal evolution by directly integrating Eqs.\ (\ref{eq-of-motion}). However, the structure of Eqs.\ (\ref{new-formulae-tide}) helps us gain more physical insights, especially after combining with orbital dynamics in the next section.


\begin{table}
    \centering
    \caption{Relative errors of Eqs.\ (\ref{ab-proper}) and (\ref{q0-adiba}) for different spins, where 'High' and 'Low' means including and not including the higher order correction Eq.\ (\ref{mine-A-ex}), respectively. }
    \begin{tabular}{c c c c c} \hline\hline
        $\Omega_s/(2\pi)$ &   $|\Delta q_0|/q_0$ &  \multicolumn{2}{c}{$|\Delta A|/A$ $(\times 1\%)$ } &  $|\Delta B|/B$   \\
(Hz)&$(\times0.1\%)$ &High & Low&$(\times 1\%)$ \\ \hline
550 & 0.2 & 0.2 &13.1 & 1.4 \\ \hline
450 & 1.3 & 1.1  &14.0 &0.6 \\ \hline
350 & 4.0 & 2.2 &14.4 &0.1 \\ \hline
250 & 8.7 & 3.2 &14.4 &0.8  \\ \hline
150 & 15.1 & 4.0 &14.4 &1.4 \\ \hline\hline
     \end{tabular}
     \label{table:AB}
\end{table}

\section{Model of DT: Orbital dynamics near resonance}
\label{sec:post-resonance orbit dynamics}
In this section we will discuss the post-resonance orbital dynamics. As we will review in Sec.\ \ref{sec:pre-works}, currently there are mainly two analytic approximations to DTs:\ the method of averaged PP orbit in FR07 \cite{Flanagan+Racine+07} and the method of effective Love number in H+16 \cite{Steinhoff+Hinderer+16,Hinderer+Taracchini+16}. Here we provide an alternative way to describe the post-resonance dynamics. In Sec.\ \ref{sec:osculation}, we derive a set of first order differential equations for osculating variables: the Runge-Lenz vector (whose magnitude is proportional to the eccentricity of the orbit), angular momentum and the orbital phase. These equations, with our new formulae for $A$ and $B$ [Eqs. (\ref{new-formulae-tide})], are self-contained except that they need $\omd$ as input. But as we will discuss in Sec. \ref{sec:init}, osculating equations lead to an analytic expression (or more accurately, a quintic equation) for $\omd$, which is accurate for the systems we study. Therefore we do not need to use non-tidal orbit as a prior knowledge to feed into the formulae of $A$ and $B$. Then in Sec.\ \ref{sec:orbit-num-comp}, we compare our analyses and the method of effective Love number with fully numerical results. Finally in Sec.\ \ref{sec: averaged-pp}, we propose an alternative way to obtain the post-resonance PP orbit, which turns out to agree with FR07 \cite{Flanagan+Racine+07} to the leading order in tidal interaction. By combining our approach and FR07 \cite{Flanagan+Racine+07}, we derive an analytic expression for $t_r$, i.e., the time of resonance. 

\subsection{Review of previous works}
\label{sec:pre-works}
The model in FR07 \cite{Flanagan+Racine+07} is based on the fact that the DT only causes significant energy and angular momentum transfers to the star near resonance, within the time
\begin{align}
\Delta t=\frac{\Delta L}{\dot{L}_\text{GW}}, \label{delta-t-shift}
\end{align}
where $\Delta L$ is the angular momentum transfer from the orbit to the star due to resonance and $\dot{L}_\text{GW}$ is the rate at which angular momentum radiated in GWs \cite{Poisson+Will+14}
\begin{align}
\dot{L}_\text{GW}=\frac{32}{5}\mu^2\frac{{M_t}^{5/2}}{r_r^{7/2}}. \label{Ldotpp}
\end{align}
We note that $r_r$ in Eq.\ (\ref{Ldotpp}) should be the actual separation of the star at $t_r$, instead of the one predicted by pre-resonance PP orbit. After resonance, the NS is treated as freely oscillating, with the interaction between the star and the orbit neglected, and the post-resonance orbit is another PP trajectory. The pre- and post-resonance orbital separations are related by the time shift $\Delta t$
\begin{align}
r(t)=
\begin{cases} 
      r^{\text{PP}}(t) & t-t_r\ll T_\text{dur}, \\
      r^{\text{PP}}(t+\Delta t) & t-t_r\gg T_\text{dur},
   \end{cases}
   \label{rpp-postres}
\end{align}
where $T_\text{dur}$ comes from the same reasoning that leads to Eq.\ (\ref{timewidth}). We can see that this method is based on the estimation of time shift $\Delta t$ due to resonance, where the non-tide $\dot{L}_\text{GW}$ is used. We will discuss these in details in Sec.\ \ref{sec: averaged-pp}.

A more detailed model was developed in H+16 \cite{Steinhoff+Hinderer+16,Hinderer+Taracchini+16}, where the authors incorporated DT to the EOB formalism by introducing an effective Love number $\lambda_\eff$, as defined in Eq. (\ref{eff-love-h}). This quantity is based on the non-tidal orbit as a prior knowledge, and does not incorporate the imaginary part of $q_2^\prime e^{-2i\phi-2i\Omega_s t}$. In fact, with the help of Eqs.\ (\ref{Q-qm}), the effective Love number can be written in our notation as
\begin{align}
\lambda_\eff=-\frac{r^3}{2{M_2}}\Re(q_2^\prime e^{-2i\phi-2i\Omega_st})=-\frac{r^3}{2{M_2}}A. \label{eff-love-def}
\end{align}
This term does not contain the full information of the NS oscillation, since $B$ is missing. By comparing this term with the RHS of Eq.\ (\ref{r-eq}), one can find that the effective Love number only describes the radial force due to the star's deformation. The ignored part, which characterizes the torque between the star and the orbit, actually plays an important role, as we shall see in Sec.\ \ref{sec:orbit-num-comp}. Furthermore, their calculations of effective Love number were obtained from non-tidal orbital evolution. This will cause inaccuracy when the spin is large.

\subsection{Osculating equations}
\label{sec:osculation}
Since the traditional method of osculating orbits(cf. Ref. \cite{Poisson+Will+14}) is singular for vanishing orbital eccentricity, we need to adopt a special perturbation method here \cite{Roy+73}. This method uses specific angular momentum $\bm{h}$, the Runge-Lenz vector $\bm{\epsilon}$ and the orbital phase $\phi$ as osculating variables. Assume that the perturbation force $\bm{F}$ is described by
\begin{align}
\frac{\bm{F}}{\mu}={\mathcal{W}}\bm{n}+S\bm{\lambda},
\end{align}
where $\bm{n}$ is the unit vector along the radial direction and $\bm{\lambda}$ the unit vector along the azimuthal direction. ${\mathcal{W}}$ and $S$ are the components of the acceleration. Equations of motion in terms of the osculating variables are given by
\begin{equation}
\begin{aligned}
&\frac{d\textbf{\textit{h}}}{d\textit{t}}=\bm{r}\times\bm{F}, \\
&\frac{d\bm{\epsilon}}{d\textit{t}}=\bm{F}\times\bm{h}+\bm{\dot{r}}\times\bm{\dot{h}}, \\
&\frac{d\phi}{d\textit{t}}=\frac{h}{r^2}.
\end{aligned}
\end{equation}
Note that the magnitude of $\bm{\epsilon}$ is proportional to the orbital eccentricity. In our case, only the $z$ component of $\bm{h}$, denoted by $h$, and in-plane components of $\bm{\epsilon}$=($\epsilon_r$, $\epsilon_\phi$) matter. The orbital separation $r$, and its rate of change $\dot{r}$, can be expressed as
\begin{subequations}
\begin{align}
&r=\frac{h^2}{{M_t}+\epsilon_r}, \label{osc-r}\\
&\dot{r}=-\frac{\epsilon_\phi}{h}. \label{osc-rdot}
\end{align}
\label{rrdot-osc}
\end{subequations}
Equations of motion of the osculating variables can then be re-written as
\begin{subequations}
\begin{align}
&\frac{d\phi}{d\textit{t}}=\frac{h}{r^2}, \label{h-initial} \\
&\frac{d\textit{h}}{d\textit{t}}=rS, \label{hdot-per}\\
&\frac{d\epsilon_r}{d\textit{t}}=\frac{h}{r^2}\epsilon_\phi+2Sh, \\
&\frac{d\epsilon_\phi}{d\textit{t}}=-\frac{h}{r^2}\epsilon_r-{\mathcal{W}}h-\dot{r}rS.
\end{align}%
\end{subequations}
The perturbation forces $S$ and ${\mathcal{W}}$ can be separated into radiation and tidal parts. The former comes from the Burke-Thorne radiation reaction potential. By neglecting tidal corrections, they are given by
\begin{subequations}
\begin{align}
&{\mathcal{W}}_\text{orb}=\frac{2}{5}\mu\left(\frac{32 {M_t}^2 \dot{r}}{3 r^4}+\frac{48 {M_t} \dot{r} \dot{\phi}^2}{r}+\frac{8 {M_t}\dot{r}^3}{r^3}\right), \\
&S_\text{orb}=\frac{2}{5}{M_t}\mu\left(\frac{8{M_t} \dot{\phi}}{r^3}+\frac{36 \dot{r}^2 \dot{\phi}}{r^2}-24\dot{\phi}^3\right).
\end{align}
\label{RS-orb}%
\end{subequations}
The tidal perturbation forces ${\mathcal{W}}_{\text{tid}}$ and $S_{\text{tid}}$ are given by
\begin{subequations}
\begin{align}
&{\mathcal{W}}_{\text{tid}}=\frac{3M_2}{2\mu}\frac{({M_t}+\epsilon_r)^4}{h^8}\left(\sqrt{\frac{3}{2}}q_0-3A\right), \\
&S_{\text{tid}}=\frac{3M_2}{\mu}B\frac{({M_t}+\epsilon_r)^4}{h^8}. \label{Stid}
\end{align}
\end{subequations}
For the time evolution of $q_0$, $A$ and $B$ we use our analytic formulae, as shown in Eqs.\ (\ref{new-formulae-tide}) and (\ref{q0-adiba}). Here we do not include the higher order correction to $A$ in Eq. (\ref{mine-A-ex}) since the leading order already turns out to be accurate enough.
By plugging Eqs.\ (\ref{rrdot-osc}) into equations above we get
\begin{widetext}
\begin{subequations}
\begin{align}
&\frac{d\phi}{d\textit{t}}=\frac{({M_t}+\epsilon_r)^2}{h^3},\label{phi-per}\\
&\frac{d\textit{h}}{d\textit{t}}=\frac{2}{5}{M_t}\mu\frac{({M_t}+\epsilon_r)^3}{h^7}\left[8 {M_t}({M_t}+\epsilon_r)+36\epsilon_\phi^2-24({M_t}+\epsilon_r)^2\right]+\frac{h^2}{{M_t}+\epsilon_r}S_\text{tid}, \label{h-per}\\
&\frac{d\epsilon_r}{d\textit{t}}=2hS_\text{tid}+\frac{({M_t}+\epsilon_r)^2}{h^3}\epsilon_\phi+\frac{4}{5}{M_t}\mu\frac{({M_t}+\epsilon_r)^4}{h^8}[8{M_t}({M_t}+\epsilon_r)+36\epsilon_\phi^2-24({M_t}+\epsilon_r)^2], \\
&\frac{d\epsilon_\phi}{d\textit{t}}=-h{\mathcal{W}}_\text{tid}+\frac{h\epsilon_\phi}{{M_t}+\epsilon_r}S_\text{tid}-\epsilon_r\frac{({M_t}+\epsilon_r)^2}{h^3}+\frac{2}{5}\mu {M_t}\epsilon_\phi\frac{({M_t}+\epsilon_r)^3}{h^8}\left[\frac{56{M_t}}{3}({M_t}+\epsilon_r)+24({M_t}+\epsilon_r)^2+44\epsilon_\phi^2\right], \label{epsilonphi-per}
\end{align}
\label{osc-all}%
\end{subequations}
\end{widetext}
Eqs.\ (\ref{RS-orb})---(\ref{osc-all}) are a complete set of equations of $\phi, h, \epsilon_r$ and $\epsilon_\phi$, except that we are missing the value of $\omd$ that appears in the formulae of $A$ and $B$, this will be determined in Sec.\ \ref{sec:init}. With these at hand, one can obtain the post-resonance orbital dynamics without solving tidal variables (e.g. $q_0$, $A$ and $B$) simultaneously. 

In practice, we numerically evolve the system slightly after the resonance point, i.e., $t_r+\delta$, to get rid of the numerical infinity due to the term $\sin(\tha^2-\Theta)/\tha$ in $B$. In our code, $\delta=10^{-8}$s. Two infinities in $A$ (adiabatic term and the counterterm) needs more care. The cancellation of these two infinities requires they have the exact the same behavior near the resonance point, this is difficult to achieve in practice, especially when there are osculating variables in $A$. In our simulations, we approximate the first divergence term by the following formula
\begin{align}
&\frac{3{M_2}\lambda_2\omega_2^2}{2r^3}\frac{1}{\zeta^2-(2\Omega+\Omega_s)^2} =-\frac{3{M_2}\lambda_2\omega_2^2}{8\sqrt{\omd}\zeta r^3\tha},  \label{infy-appro}
\end{align}
where the denominator is expanded around $t_r$. In this manner, both divergence terms go to infinity as $1/\tha$, so they cancel each other nicely. In order to improve the accuracy, one can include more terms of the Taylor expansion. However, this only works well for low spin, since the time for post-resonance evolution should be short enough such that the series converges. For high spin we only keep the leading term\footnote{As we shall see in Sec. \ref{sec:orbit-num-comp}, the orbital frequency is oscillatory for high spin in the post-resonance regime. Under this situation, the leading term alone is more accurate than including higher order corrections.}.

We should note that one can evolve the post-resonance system without knowing the value of $t_r$, of which our analytic estimations are not very accurate in some regimes of spin (we will discuss the estimation on it in Sec. \ref{sec:orbit-num-comp}), since the formulae of $A(t)$ and $B(t)$ only depend on $\tha$. One can shift the time of resonance to $t=0$ and simultaneously set $t_r=0$. Similarly, the orbital phase of the resonance $\phi_r$ in Eq.\ (\ref{Theta}) can be eliminated by an appropriate initial condition for $\phi$, here we choose $\phi_r=0$ and $\phi^{(0)}=0$, where $\phi^{(0)}$ is the initial value of $\phi$. Correspondingly, the constant $\chi_r$ becomes 0. What remains unknown in the osculating equations are $\omd$ and the initial conditions for $(\epsilon_{r,\phi},h,\phi)$. We will address them in the next subsection.

\subsection{The applications of osculating equations}
\label{sec:init}
In this subsection, we will discuss the applications of osculating equations introduced in the previous subsection. 
\subsubsection{Orbit at resonance}
Let us first derive algebraic equations for $\omd$, $\dot{r}_r$ and the initial conditions of Eqs.\ (\ref{osc-all}). The basic idea is that variables like $\omd$ and $\dot{r}_r$ at resonance are determined by the tidal variables $A$ and $B$ through the osculating equations. Conversely, $A$ and $B$ are governed by $\omd$ in Eqs.\ (\ref{ab-proper}). The relationship allows us to write down equations of $\omd$ and $\dot{r}_r$.

To calculate $\dot{r}$, we start with Eq.\ (\ref{osc-r}). In our cases, the value of $\epsilon_r$ rises as the spin of the NS decrease, but it remains a small number. So we can approximate $r$ by $h^2/{M_t}$. Using the equation of $\dot{h}$ [Eq. (\ref{h-per})], we get
\begin{align}
\frac{d\textit{r}}{dt}=2\sqrt{\frac{r^3}{{M_t}}}S. \label{rdot-reson}
\end{align}
For a quasicircular orbit, the radius and orbtial frequency approximately satisfy:
\begin{align}
r_r=\left(\frac{{M_t}}{\Omega_r^2}\right)^{1/3}. \label{r-init}
\end{align}
In Table \ref{table:orbit-ini} we verify that the error of Eq.\ (\ref{r-init}) is less than 0.4\% within the regime we concern. With this observation, together with $B_r$ in Eqs.\ (\ref{ab-proper}), one can simplify the expression of $S$ into
\begin{align}
&S=S_\text{tid}+S_\text{orb}=-\frac{3{M_2}^2\lambda_2}{\mu\zeta}\frac{3\omega_2^2}{8r^7_r}\sqrt{\frac{\pi}{2\omd}} \notag \\
&+\frac{4}{5}{M_t}\mu\Omega_{r}\left(\frac{18 \dot{r}_r^2}{r_r^2}-8\Omega_{r}^2\right), \label{S-reson}
\end{align}
which is completely determined by $\dot{r}_r$ and $\omd$. Substituting this into Eq.\ (\ref{rdot-reson}) leads to a equation for $\dot{r}_r$ and $\omd$
\begin{align}
&\dot{r}_r=-\frac{3{M_2}^2\lambda_2}{\Omega_r\mu\zeta}\frac{3\omega_2^2}{4r^7_r}\sqrt{\frac{\pi}{2\omd}}+\frac{8}{5}{M_t}\mu\left(\frac{18 \dot{r}_r^2}{r_r^2}-8\Omega_{r}^2\right).\label{ini-1}
\end{align}
In order to solve for these two variables, one can use Eqs. (\ref{h-initial}) and (\ref{hdot-per}) to establish another equation
\begin{align}
&\omd=\frac{\dot{h}}{r_r^2}-2\frac{\dot{r}_r}{r_r^3}h=\frac{S}{r_r}-2\Omega_{r}\frac{\dot{r}_r}{r_r},
\end{align}
which gives
\begin{align}
2r_r\omd=-3\Omega_{r}\dot{r}_r. \label{ini-2}
\end{align}
This relation can also be directly obtained by differentiating Eq.\ (\ref{r-init}). Plugging Eq.\ (\ref{ini-2}) back into Eq.\ (\ref{ini-1}) gives a quintic function for $\omd$. The calculation can be simplified by the approximation $\dot{r}_r/r_r\ll\Omega_r$, so that the first term in the bracket of Eq. (\ref{ini-1}) can be neglected. In this manner, we obtain an explicit expression for $\omd$:
\begin{align}
\omd=\left(\frac{u}{2}+\sqrt{\frac{u^2}{4}-\frac{v^3}{27}}\right)^{1/3}+\left(\frac{u}{2}-\sqrt{\frac{u^2}{4}-\frac{v^3}{27}}\right)^{1/3}, \label{omd-appro}
\end{align}
where
\begin{align}
&u=\frac{27{M_2}^2\lambda\omega_2^2}{8r_r^8\zeta\mu}\sqrt{\frac{\pi}{2}}, 
&v=\frac{96{M_t}\mu}{5r_r}\Omega_r^3=\frac{3}{\mu r_r^2}\dot{L}_\text{GW}^{(r)}. \label{uv}
\end{align}
Eq.\ (\ref{omd-appro}) can be further simplified by Taylor expanding in $w$, defined by
\begin{align}
w=\frac{2^{1/3}v}{3u^{2/3}}, \label{uvw}
\end{align}
leading to
\begin{align}
\omd=u^{2/3}\left[1+2^{2/3}w+w^2-\frac{w^3}{3}+\mathcal{O}(w^4)\right]. \label{tdur-deltat}
\end{align}
Recall that the duration of the resonance is $T_\text{dur}=\sqrt{\pi/\omd}$ [Eqs.\ (\ref{timewidth}) and (\ref{rpp-postres})], Eq.\ (\ref{tdur-deltat}) is in fact an analytic relation between $T_\text{dur}$ and the orbital time shift $\Delta t$ due to resonance. The variable $\dot{r}_r$ is determined once $\omd$ is known. Finally, the initial value of $\epsilon_\phi$ is related to $\dot{r}_r$ through its definition in Eq.\ (\ref{osc-rdot}). With the values of $\omd$ and $\dot{r}_r$, Eq.\ (\ref{new-formulae-tide}) for $A(t)$ and $B(t)$ does not require input from numerical integrations.

In Table \ref{table:orbit-ini}, we compare predictions of our formulae with numerical results.\ The parameters of NSs are the H4 EoS with component masses $(1.4,1.4)M_\odot$. Results show that the accuracies of our analyses are higher than $93\%$. We can also see that accuracy is lower for low spins. Since H+16 \cite{Steinhoff+Hinderer+16,Hinderer+Taracchini+16} used non-tidal $\omd$ in the effective Love number, we compare $\omd$ of non-tide orbits with realistic ones. The ratios of two quantities are shown in the last column of Table \ref{table:orbit-ini}, we can see that $\omd^\text{non}$ is only half of $\omd^\text{tide}$, hence the use of $\omd^\text{non}$ will cause inaccuracies.

\begin{table}
    \centering
    \caption{Comparisons between results from our formulae for $\omd$, $\dot{r}$, $\epsilon_\phi$, $r$ and numerical integrations, where ``Num.'' of  $\Delta \omd/\omd$ are the results by numerical solving Eqs.\ (\ref{ini-1}) and (\ref{ini-2}); ``Appr.'' are the results of Eq. (\ref{tdur-deltat}). The parameters of NSs are still the H4 EoS with component masses $(1.4,1.4)M_\odot$. The relative error becomes large when the spin decreases. The last column is the ratio of the non-tidal $\omd$ to the realistic $\omd$ when the orbital frequencies satisfy the resonance condition in Eq. (\ref{reson-condition}).}
    \begin{tabular}{c c c c c c c} \hline\hline
$\Omega_s/(2\pi)$ & \multicolumn{2}{c}{$\Delta \omd/\omd (\times10^{-2})$} & $\Delta \dot{r}/\dot{r}$ & $\Delta \epsilon_\phi/\epsilon_\phi$& $\Delta r/r$ &\multirow{2}{*}{$\frac{\omd^\text{non}}{\omd^\text{tide}}$}  \\
 (Hz) & Num. & Appr.& $(\times10^{-2})$&$(\times10^{-2})$ &$(\times10^{-2})$ \\ \hline
550 &0.9   & 0.8& 0.4 &0.1 &0.1 &0.56 \\ \hline
450 &2.7 &2.7 & 1.6 &1.8&0.2 &0.53  \\ \hline
350 & 4.4 & 4.5& 2.7 & 2.2&0.3 &0.52 \\ \hline
250 & 5.9 & 6.1 & 3.7 & 3.0 & 0.4 &0.52\\ \hline 
150 & 7.1 &7.3 & 4.5 & 3.6 & 0.4 &0.52\\ \hline \hline 
     \end{tabular}
     \label{table:orbit-ini}
\end{table}
\subsubsection{Angular momentum and energy transfers}
Another application of the osculating equations is to calculate the angular momentum and energy exchange between the star and the orbit. The transfer in $L$ can be directly  calculated from Eq.\ (\ref{hdot-per}). Following the procedure in Ref. \citep{Lai+94}, we get
\begin{align}
&\Delta L=-\mu h_t=-\int \mu rS_\text{tid}d\textit{t}=-\int3{M_2}\frac{\textit{B}}{\textit{r}^3}d\textit{t} \notag \\
&=-\text{Im}\int3{M_2}\frac{q_{2}^\prime e^{-2i\phi-2i\Omega_st}}{r^3}d\textit{t} \notag \\
&=-\frac{2}{\mu\omega_2^2\lambda_2}\text{Im}\int q_2^\prime(\ddot{q}_{-2}^\prime+2i\Omega_s\dot{q}_{-2}^\prime+\omega_2^2q_{-2}^\prime)d\textit{t} \notag \\
&=-\frac{2}{\mu\omega_2^2\lambda_2}\left[\dot{A}B-\dot{B}A-(A^2+B^2)(\Omega_s+2\Omega)\right], \label{h-tran}
\end{align}
where we have used Eq.\ (\ref{q2-eq}). Assuming the deformation of the star is small initially, this exact formula gives the angular momentum deposited in the star. In fact, the quantity is the generalization of the ``tidal spin'', defined by (up to a constant) $\epsilon_{ijs}Q^{mi}\dot{Q}^{jm}$ for a non-spinning star \citep{Steinhoff+Hinderer+16}.

By combining our formulae for $A$ and $B$ with the $\Delta L$ shown above, one can obtain a lengthy expression of angular momentum transfer as a function of time, but little can be learned from it. To give a more useful description, we follow the idea of FR07 \cite{Flanagan+Racine+07}, who assumed the net transfer only takes place near resonance.  Within the post-resonance regime, $\Delta L$ is periodic and the net transfer is zero. In fact, we can see this clearly with the asymptotic behavior of $A$ and $B$. From Eqs.\ (\ref{new-formulae-tide}) we know
\begin{subequations}
\begin{align}
&A\sim\frac{3{M_2}\lambda_2\omega_2^2}{4r^3_r\zeta}\sqrt{\frac{\pi}{\omd}}\cos(\chi_r+\zeta t-2\phi-\Omega_st-\frac{\pi}{4}), \\
&B\sim\frac{3{M_2}\lambda_2\omega_2^2}{4r^3_r\zeta}\sqrt{\frac{\pi}{\omd}}\sin(\chi_r+\zeta t-2\phi-\Omega_st-\frac{\pi}{4}), 
\end{align}
\label{AB-larget}%
\end{subequations}
where we have used the fact that the Fresnel functions go to $1/2$ as $\tha\to\infty$.
Plugging the above equations into Eq. (\ref{h-tran}) and averaging over orbital phase, we get the net angular momentum transfer as
\begin{align}
\Delta L=\frac{9{M_2}^2\pi\lambda_2\omega_2^2}{8\omd r_r^6\zeta}. \label{ang-mom-tran}
\end{align}
This formula reduces to the result in L94 \cite{Lai+94} when spin vanishes. The energy transfer is related to the angular momentum transfer by
\begin{align}
\Delta E=\Omega_r\Delta L.
\end{align}
By the expression of $\Delta L$ in Eq.\ (\ref{ang-mom-tran}), variables $u$ and $w$ defined in Eqs.\ (\ref{uv}) and (\ref{uvw}) can be expressed as:
\begin{align}
&u=\frac{9\Omega_r^2}{\sqrt{2\pi}}\frac{\Delta L}{L_r}\frac{\dot{L}^{(r)}_\text{GW}}{L_r}, 
&w=\left(\frac{2}{81}\frac{T_\text{orb}}{\Delta t}\frac{L_r}{\Delta L}\right)^{1/3},
\end{align}
with $T_\text{orb}=2\pi\Omega_r$ and $L_r$ the orbital angular momentum at resonance.

%

\begin{figure*}[!bt]
        \includegraphics[width=0.32\textwidth]{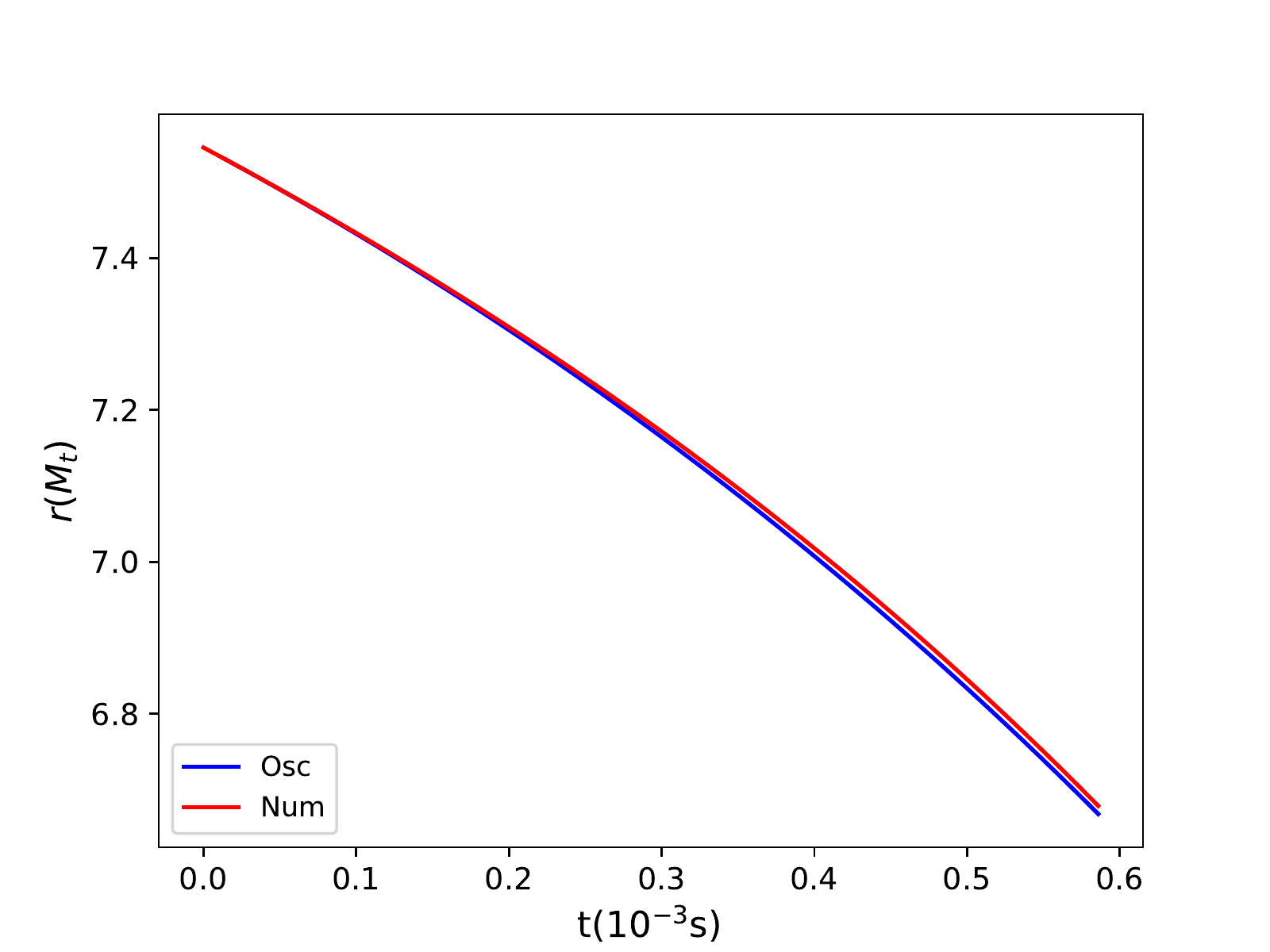} 
        \includegraphics[width=0.32\textwidth]{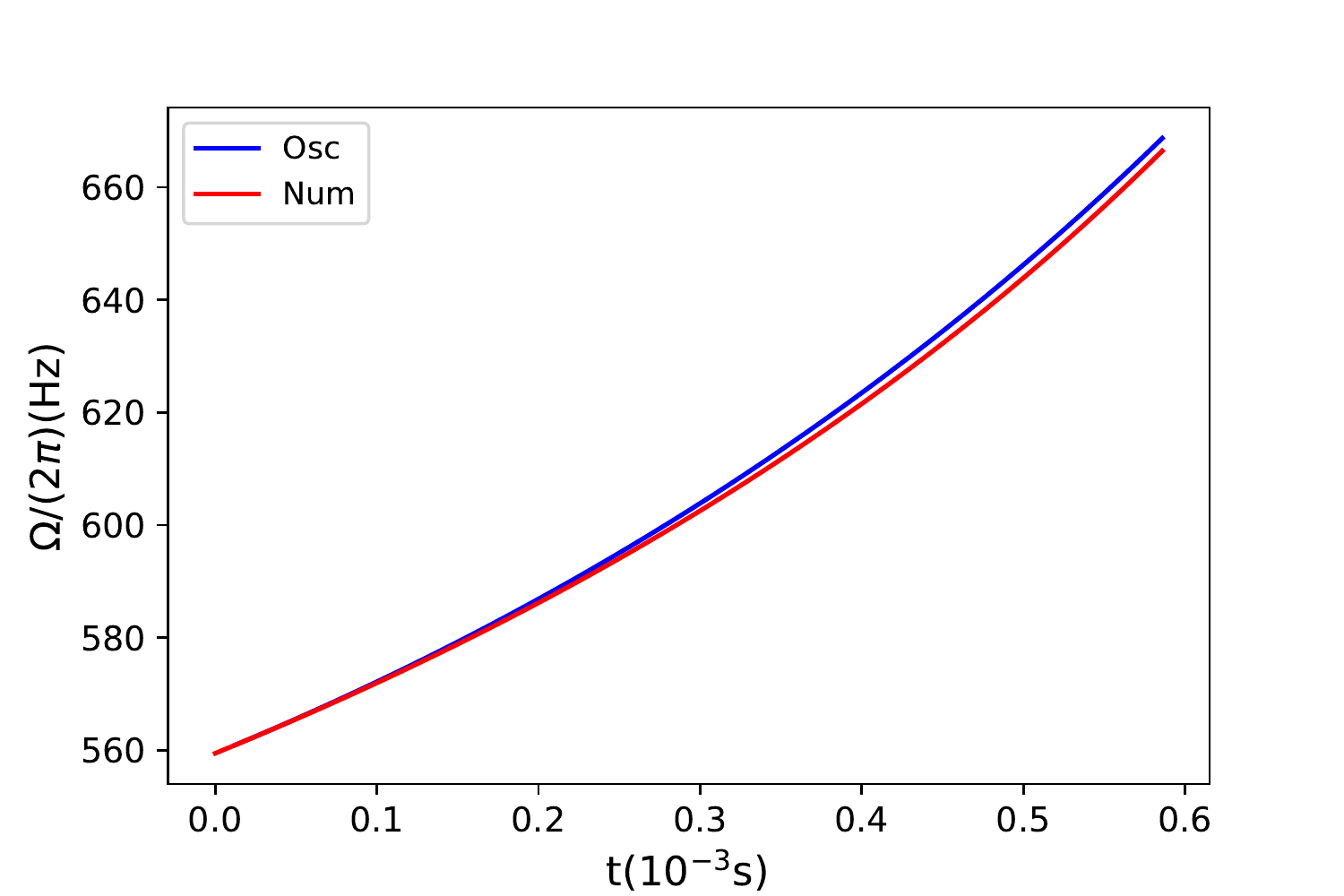}
        \includegraphics[width=0.32\textwidth]{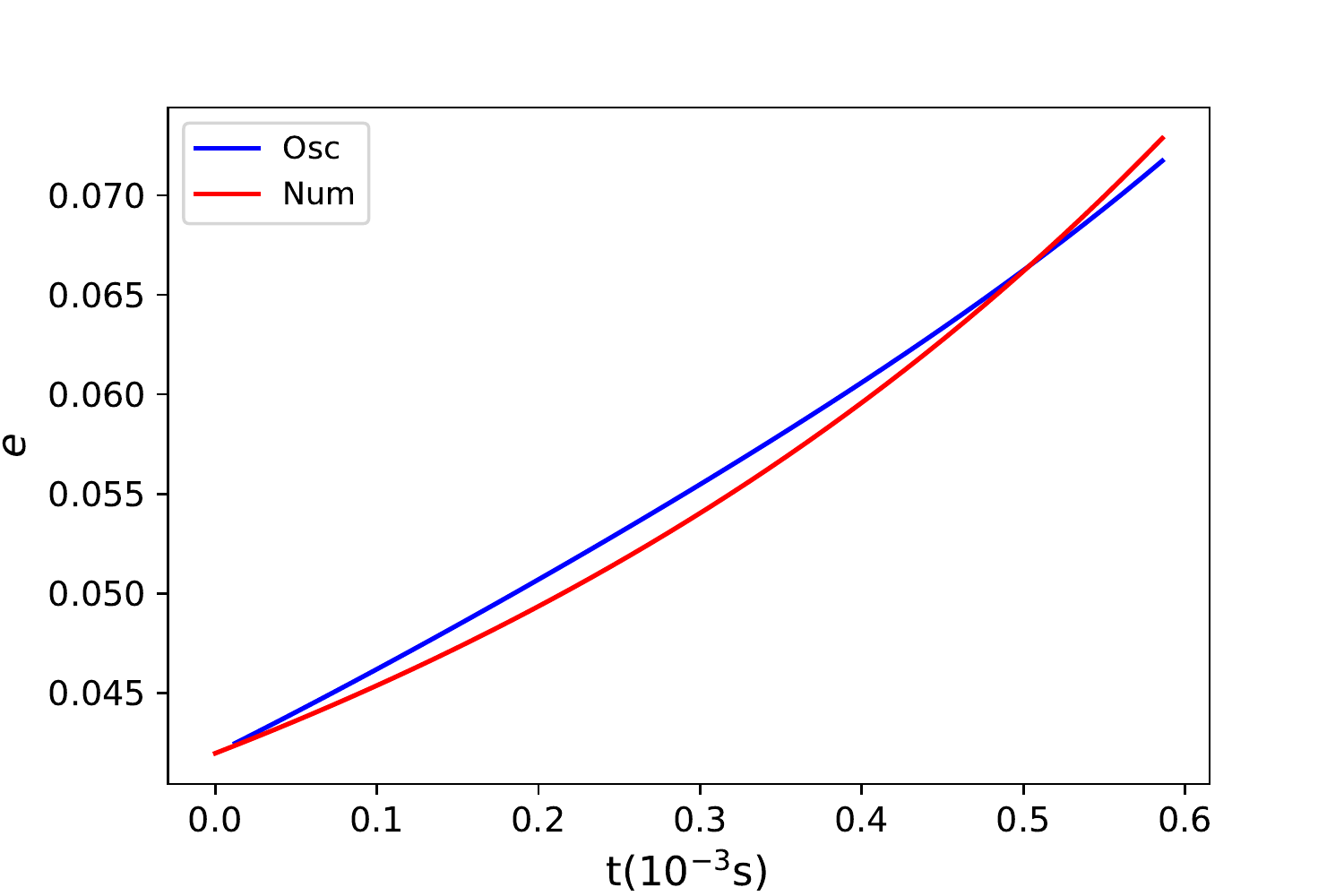} \\
        \includegraphics[width=0.32\textwidth]{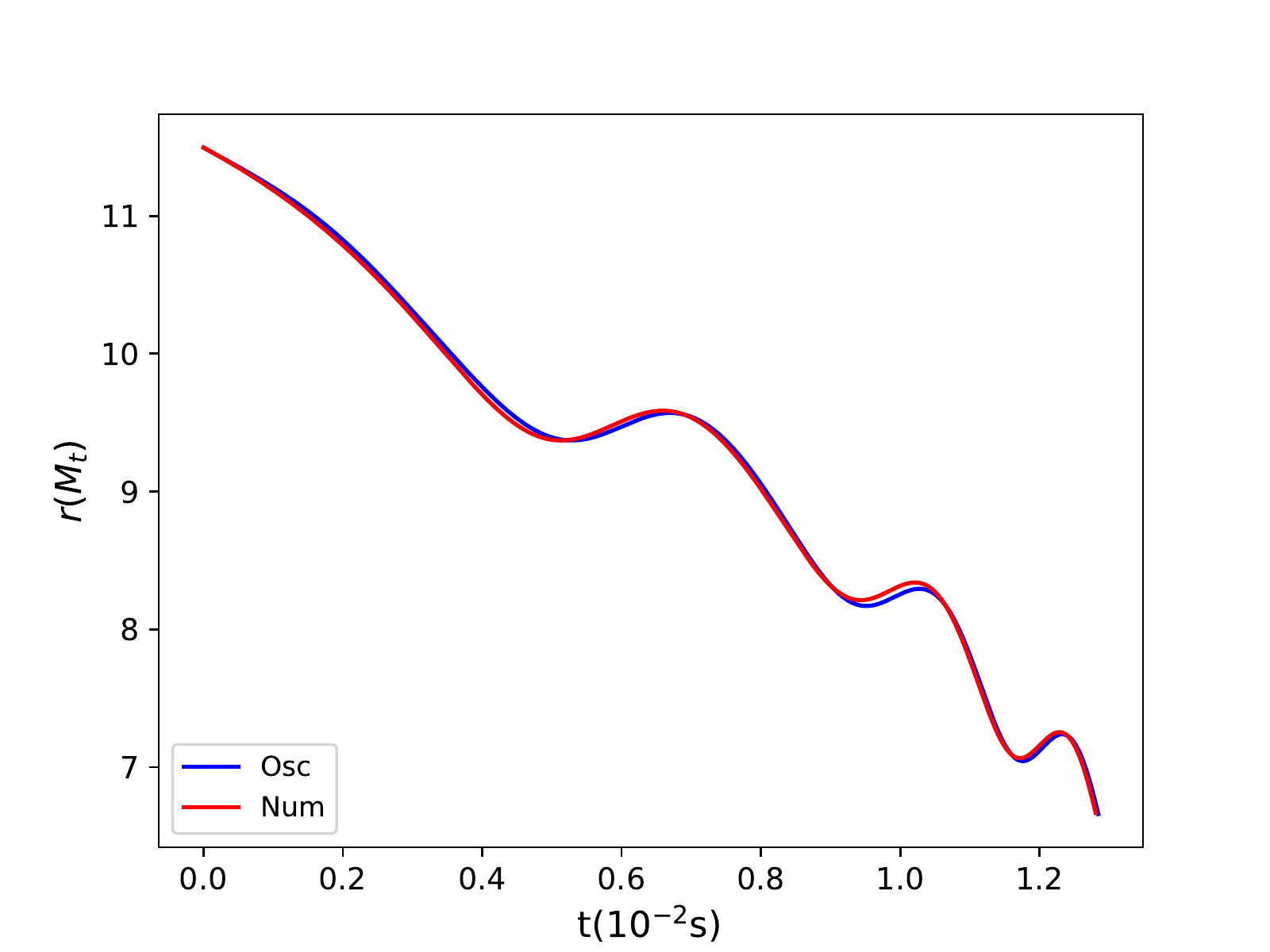}
        \includegraphics[width=0.32\textwidth]{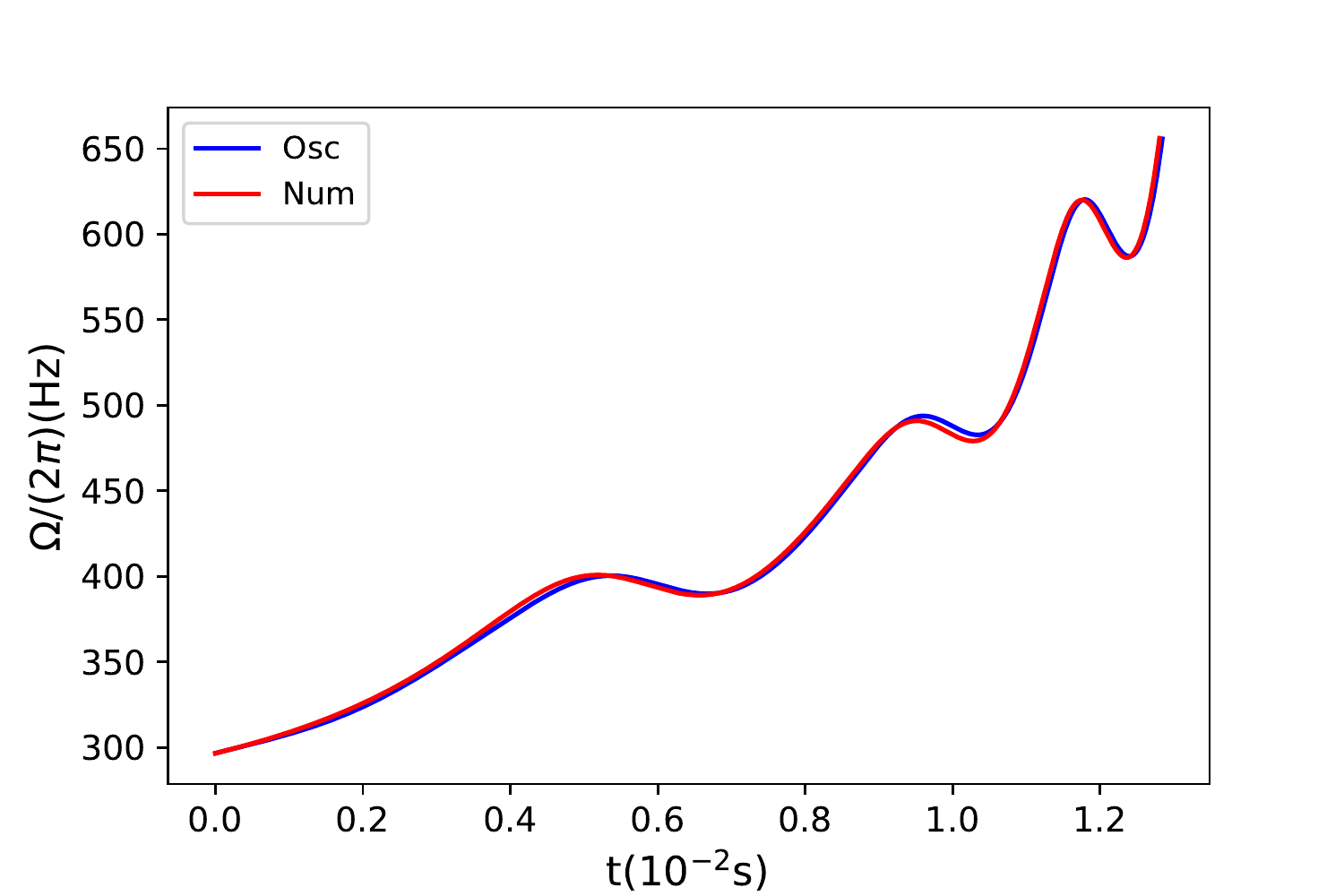}
         \includegraphics[width=0.32\textwidth]{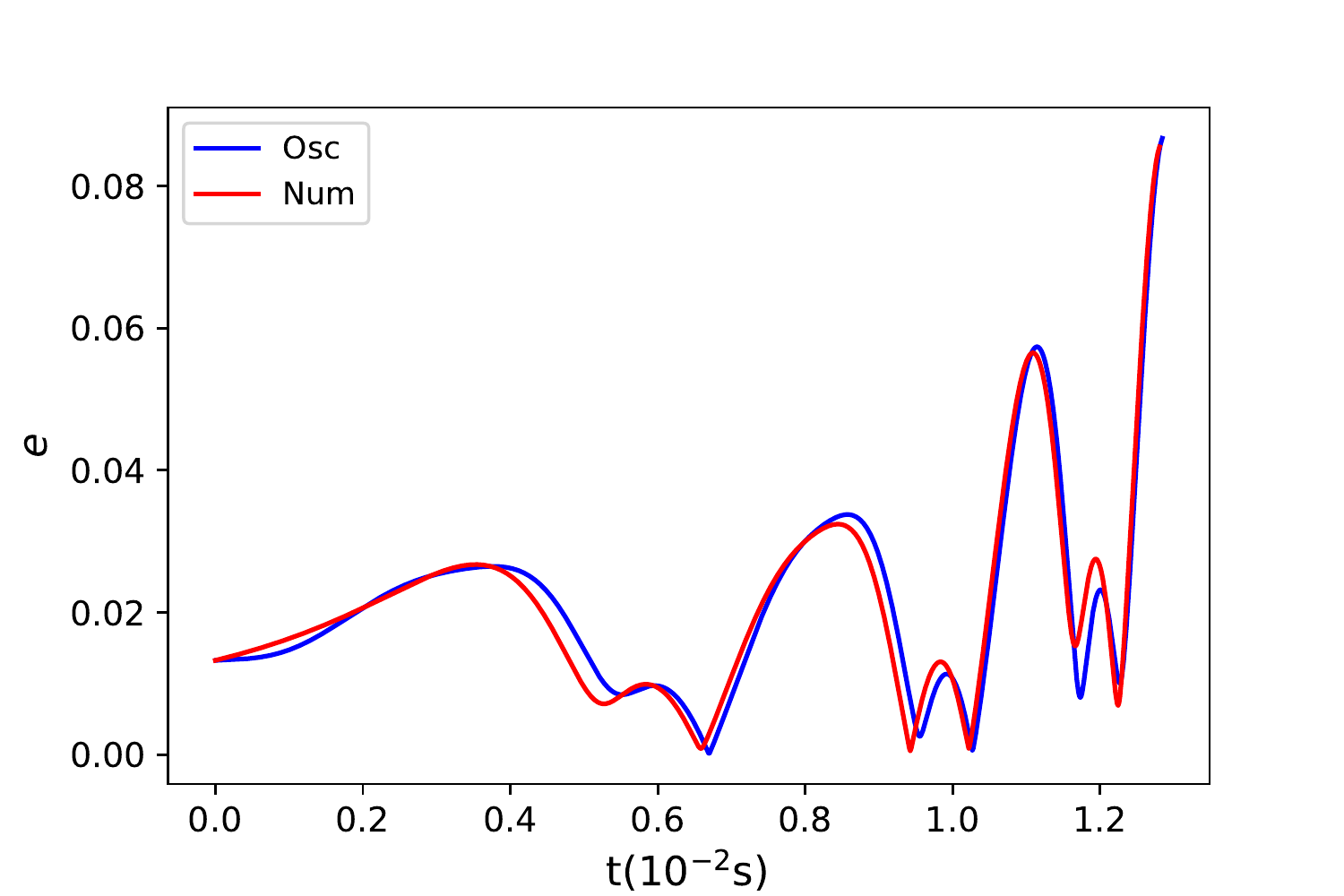}
  \caption{The seperation $r$, orbital frequency $\Omega/(2\pi)$ and the eccentricity $e$ as functions of time. The initial time $t=0$ represents the location of resonance and the end point corresponds to the contact separation. Red lines are from fully numerical solutions and blue lines are the results of osculating equations Eqs.\ (\ref{osc-all}). The spin of the upper panel is 300Hz, and the bottom one is 550Hz. We keep both the leading and the sub-leading terms in Eq.\ (\ref{infy-appro}) in low spin case while only the leading term in high spin case.}
 \label{fig:per-r}
\end{figure*}

\subsection{Comparisons with numerical results}
\label{sec:orbit-num-comp}
In this subsection we will compare our approximations, as well as the method of effective Love number in H+16 \cite{Steinhoff+Hinderer+16,Hinderer+Taracchini+16}, with fully numerical results, in the post-resonance regime. We still choose the H4 EoS with spin frequencies $300$Hz and $550$Hz.
\subsubsection{Validating osculating equations}
We numerically solve Eqs.\ (\ref{osc-all}) starting from $t=\delta=10^{-8}$s, where we have shifted the resonance time to 0 and set $t_r=0$. The initial values of $h, \epsilon_r$ and $\epsilon_\phi$ are from Eqs. (\ref{osc-rdot}), (\ref{h-initial}), (\ref{r-init}), (\ref{ini-1}) and the resonance condition in Eq. (\ref{reson-condition}). In the absence of analytic estimations for $\epsilon_r$, we assume $\epsilon_r$ is 0 in Eq. (\ref{r-init}), since it remains small within the domain we are interested in.


In Fig.\ \ref{fig:per-r}, we plot orbital separation $r$ (left panels), orbital frequency $\Omega/(2\pi)$ (middle panels), and eccentricity $e$ (right panels) as functions of time, for NS spins 300Hz (upper panels) and 550Hz (lower panels). For the low spin case, we approximate the adiabatic term in Eq.\ (\ref{infy-appro}) by both the leading and sub-leading terms, while for the high spin case we only keep the leading term. Predictions of our osculating equations agree well with the real post-resonance orbital dynamics. This again verifies that our formulae for $A(t)$ and $B(t)$ are accurate enough to describe the star's oscillation and its back reaction on the orbit. Furthermore, in our osculating equations we have only included the orbital part of the radiation-reaction force. The comparison confirms that 
the other part, i.e. the stellar radiation-reaction force, can be safely ignored. One interesting feature of the post-resonance dynamics is the eccentricity of the orbit. Once the oscillations of NSs are excited, the tidal torque and the radial tidal force lead to energy and angular momentum exchanges between the orbit and the star periodically. As a result, the eccentricity of the orbit increases and oscillates. Results show that the final eccentricities are nearly 0.08 for both cases.

\subsubsection{Deficiency of the method of effective Love number}
\label{sec:defi-eff-love}
\begin{figure}[htb]
        \includegraphics[width=\columnwidth,height=5.6cm,clip=true]{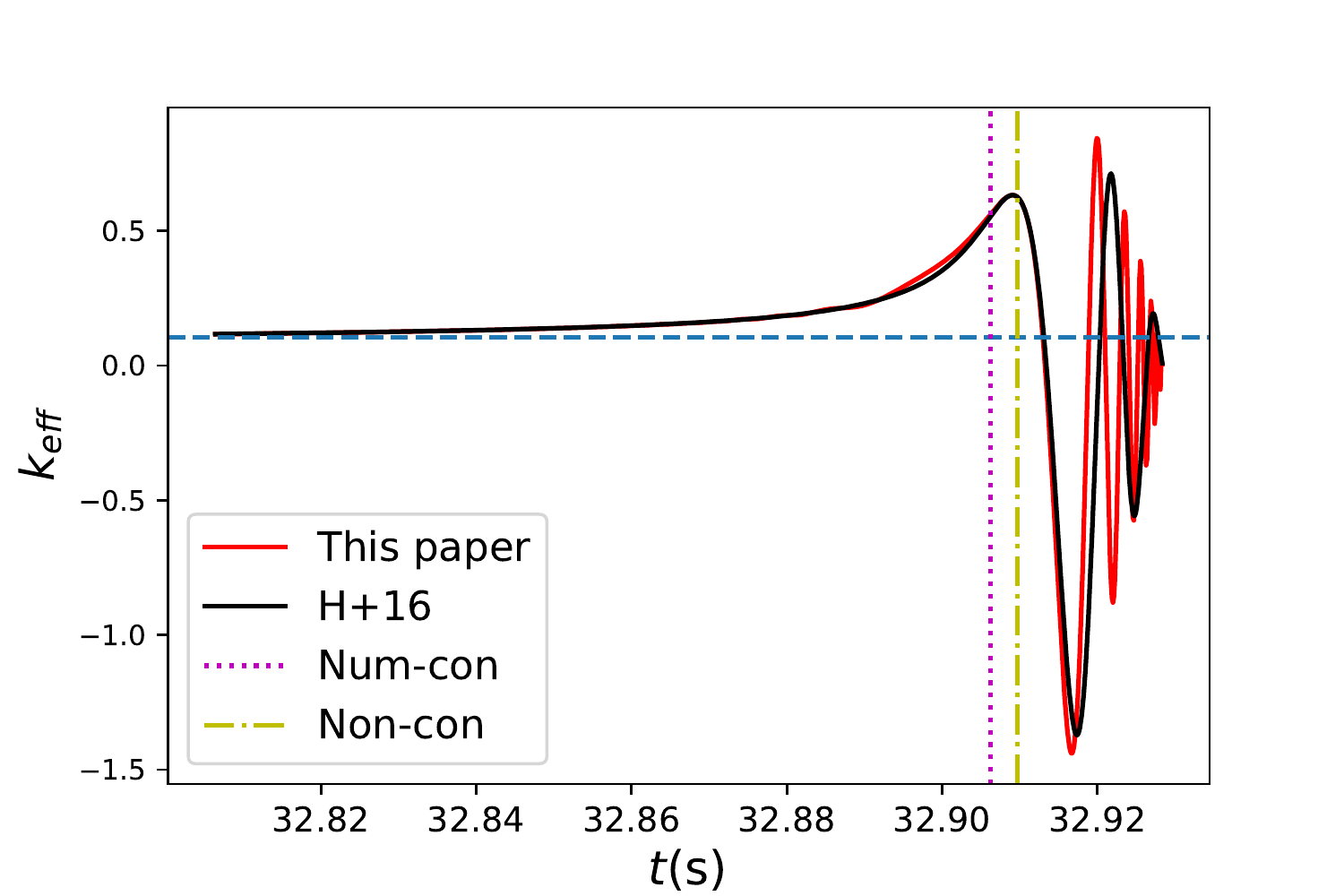}
  \caption{The time evolution of effective Love number based on the PP orbit. The red line is from our new formulae of $A$ and $B$ while the black one is from H+16 \cite{Steinhoff+Hinderer+16,Hinderer+Taracchini+16}. As represented by the horizontal dash line, the effective $k$ asymptotically approaches to $k_2=0.104$ in the adiabatic regime. The dotted vertical line represents the real resonant time and the dash-dotted vertical line is from the pre-resonance PP orbit.}
 \label{fig:eff-love-t}
\end{figure}

According to the definition of effective Love number in Eq.\ (\ref{eff-love-def}), we first construct the non-tidal binary orbit with the same initial conditions in Eq.\ (\ref{full-initial})
\begin{subequations}
\begin{align}
&\phi(t)=\frac{1}{32\eta}\frac{1}{(2\pi {M_t}F_0)^{5/3}}\notag \\
&\times\left\{1-\left[1-\frac{256}{5}t{M_t}^{2/3}\mu(2\pi F_0)^{8/3}\right]^{5/8}\right\},\\
&r(t)=\left(r^{(0)4}-\frac{256\eta {M_t}^3}{5}t\right)^{1/4},  \label{r-t-notide}
\end{align}
\end{subequations}
with initial value $r^{(0)}$ obtained from Eq.\ (\ref{full-initial}). Following the procedure in H+16 \cite{Steinhoff+Hinderer+16,Hinderer+Taracchini+16}, we use the PP orbit's time of resonance $t_r^\text{(PP)}$ and the time derivative of angular frequency as the true $t_r$ and $\omd$. Substituting them and the formulae of $A$ and $B$ into the equation of effective Love number in Eq.\ (\ref{eff-love-def}) gives the time evolution of the effective Love number. In Fig.\ \ref{fig:eff-love-t}, we plot the results by using both H+16 \cite{Steinhoff+Hinderer+16,Hinderer+Taracchini+16} and our new formulae of $A$ and $B$. The dotted one represents the resonance time from the full numerical integrations,  and the dash-dotted line is from the PP orbit. We can see that the true resonance time is earlier than that of the PP orbit. This is expected because the mode excitation extracts energy and angular momentum from the orbit, and accelerates the inspiraling process. The amplitude of the two models decay at the same rate but have different phases. Our formulae predict more oscillation cycles. 

\begin{figure}[htb]
        \includegraphics[width=\columnwidth,height=5.6cm,clip=true]{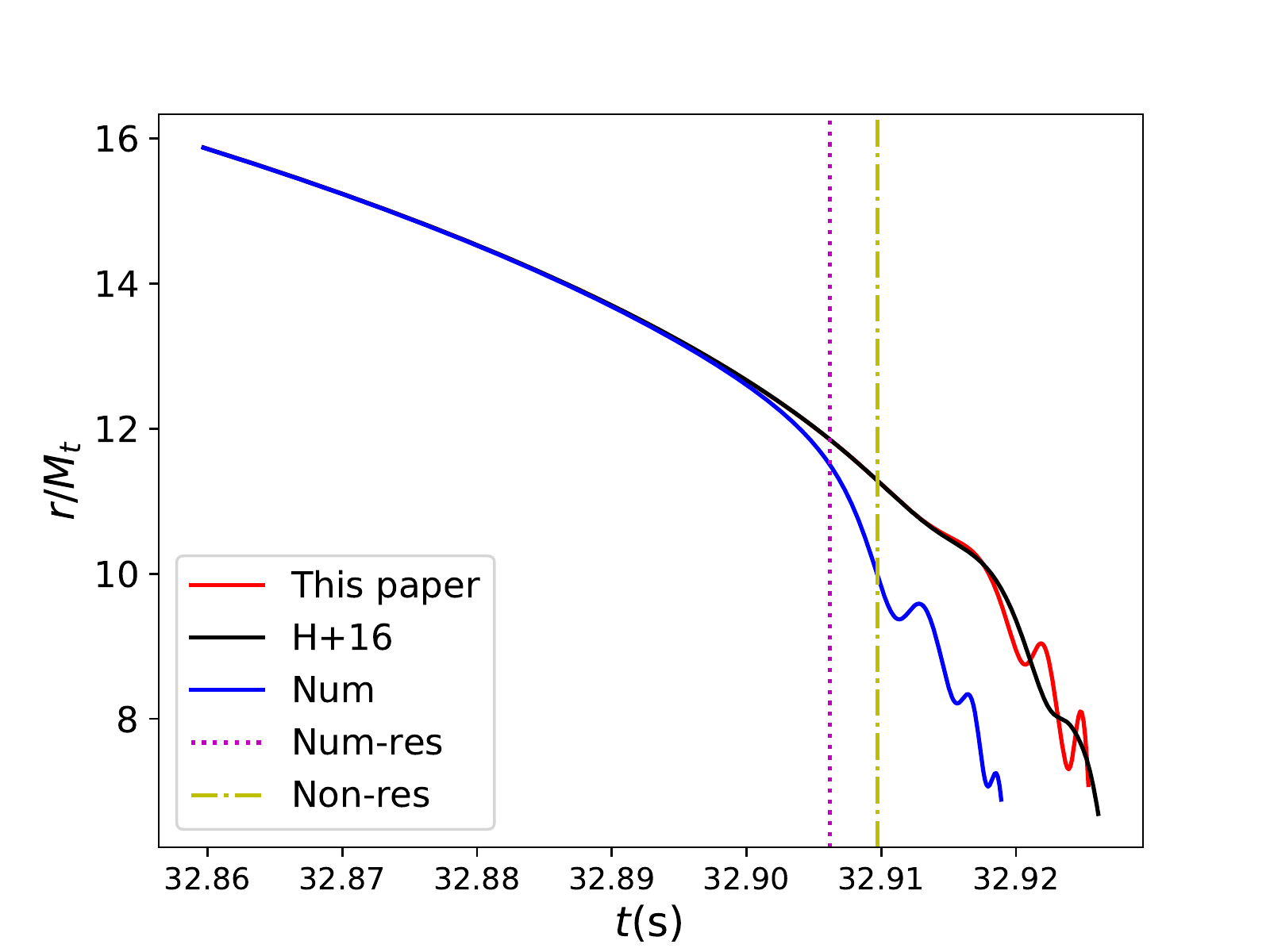}
  \caption{The orbital dynamics near the resonance, by means of effective Love number. The blue line is the result of fully numerical integration. The red line is from our new formulae of DT, while the black one is from H+16 \cite{Steinhoff+Hinderer+16,Hinderer+Taracchini+16}.  Same as Fig. \ref{fig:eff-love-t}, the dotted line and dash-dotted line represent the resonance condition of numerical and PP evolution, respectively.}
 \label{fig:eff-love-r}
\end{figure}

By feeding $k_\eff(t)$ into the orbital dynamics, we obtain the evolution of orbital separation $r(t)$ in Fig.\ \ref{fig:eff-love-r}. We can see that neither formulae could capture the feature of post-resonance dynamics. The similarity between two results show that it is the formalism of effective love number itself that is inaccurate. Such inaccuracy mainly comes from the fact that the torque is missing, and the orbit does not shrink as fast as it should be, as we have discussed around Eq.\ (\ref{eff-love-def}).

\subsection{The averaged orbit in the post-resonance regime}
\label{sec: averaged-pp}
\begin{figure}[htb]
        \includegraphics[width=\columnwidth,height=5.6cm,clip=true]{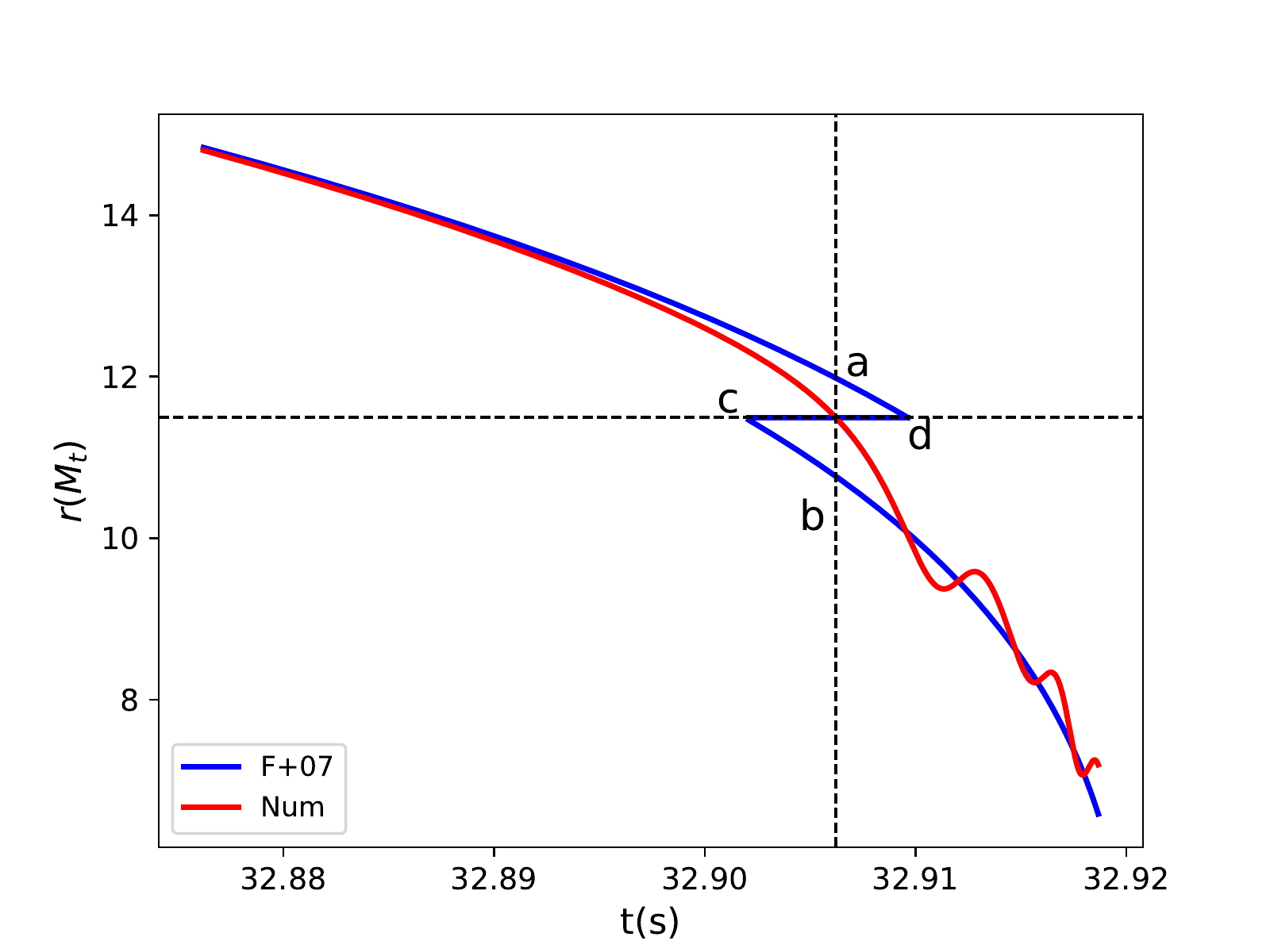}
        \includegraphics[width=\columnwidth,height=5.6cm,clip=true]{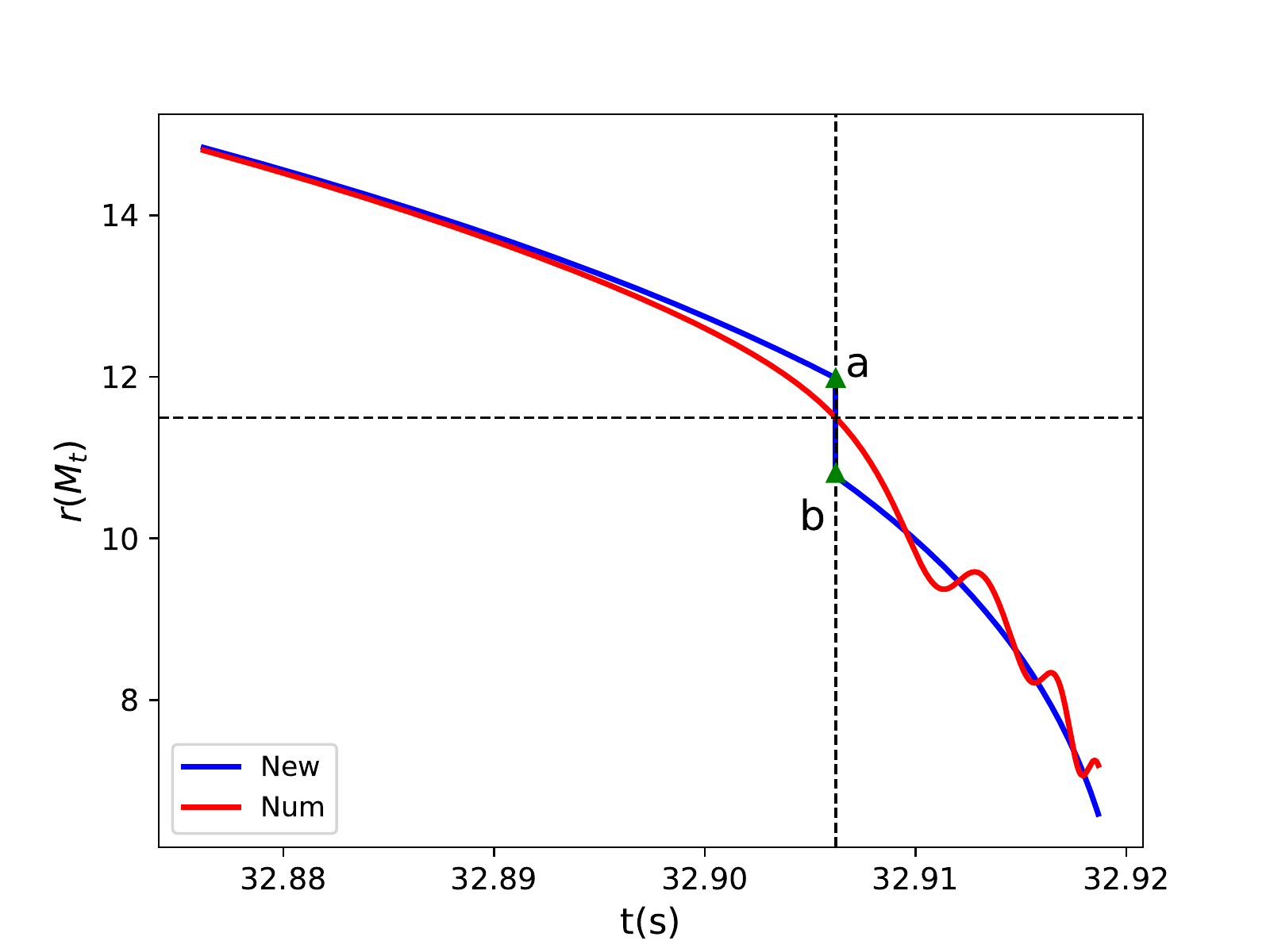}
  \caption{The orbital separation as a function of time, with NS spinning at 550Hz. The vertical dashed lines indicate the time of resonance, and the horizontal dashed lines represent the actual separation of the system at resonance. The red curves are from numerical integrations, while the blue curves are predictions of PP orbits. The upper blue curves have the same initial conditions as the system we study. They intersect with the vertical and horizontal dashed lines at ``a'' and ``d''. The lower blue curves are predictions of FR07 \cite{Flanagan+Racine+07} (upper panel) and our new method (lower panel), which intersect with the vertical and horizontal dash lines at ``b'' and ``c''. To connect the pre- and post-resonance PP orbits, FR07 \cite{Flanagan+Racine+07} proposed the time jump $\Delta t$ from ``d'' to ``c'' at the fixed separation, while we use the angular momentum jump (or equivalently, the separation jump) from ``a" to ``b" at the fixed time $t_r$.}
 \label{fig:orbit-average}
\end{figure}

As discussed in FR07 \cite{Flanagan+Racine+07}, there are three timescales in the system's dynamics, although their values in our case may not be well-separated. The shortest one is orbital timescale, characterized by the orbital angular frequency $\Omega$; the middle one is the tidal timescale, characterized by the angular frequency $\sim\dot{\Theta}=2\Omega+\Omega_s-\zeta$ [Eq.\ (\ref{Theta})]; and the final one is the gravitational radiation reaction timescale, characterized by the frequency $\dot{L}_\text{GW}/L$. The separation between tidal and radiation reaction timescales is shown more clearly in Fig.\ \ref{fig:orbit-average}, where we plot $r(t)$ near resonance with $\Omega_s=2\pi\times550$Hz. Let us first focus on the upper panel, which is from FR07 \cite{Flanagan+Racine+07}. The vertical dashed line indicates the time of resonance, and the horizontal dashed line represents the actual separation of the system at resonance. Both quantities are obtained from the numerical integration. In the radiation-reaction timescale, the system evolves as PP. The upper blue curve corresponds to the non-tidal quasi-circular orbit with the same initial conditions as our system. It intersects with the vertical and horizontal dashed lines at ``a'' and ``d''. We can see that there is little difference between full orbit and the PP orbit in the adiabatic regime. After resonance, the actual separation oscillates around another PP orbit in the tidal timescale, which is determined by Eq.\ (\ref{rpp-postres}) and shown as the lower blue curve; this curve intersects with the vertical and horizontal dash lines at ``b'' and ``c''. The pre- and post-resonance PP orbits are related by an instantaneous time shift $\Delta t$ [cf. Eq.\ (\ref{delta-t-shift})] when the pre-resonance PP orbit satisfies the resonance condition Eq.\ (\ref{reson-condition}), i.e., the horizontal line between ``c'' and ``d''. We should note that the regimes between ``ad'' and ``cb'' are not real evolution stages that the system undergoes. This is only an effective way to describe the resonance between two PP orbits. The time of ``d'', $t_d$, is actually $t_r^\text{(PP)}$ that we used to construct the effective Love number in Sec.\ \ref{sec:defi-eff-love}, it is larger than the actual resonance time $t_r$ because the tide effect accelerates the inspiral process and makes resonance earlier. We can see that FR07 \cite{Flanagan+Racine+07} can track the post-resonance PP orbit to a high accuracy.

Here we provide an additional description on the averaged orbit. As shown in the lower panel of Fig.\ \ref{fig:orbit-average}, instead of evolving the pre-resonance PP orbit to ``d'' and making a jump in time at a fixed separation, we propose that the orbit has an immediate jump in angular momentum (or equivalently, separation) at the fixed time $t_r$, i.e., the vertical line between ``ab''. The jump can be determined as follows. The orbital angular momentum at ``a'' is given by
\begin{align}
L_\text{PP}^{(a)}=\mu {M_t}^{1/2}r_\text{PP}^{(a)1/2}, \label{L-omega}
\end{align}
while at ``b'' the angular momentum is determined by the angular momentum transfer in Eq. (\ref{ang-mom-tran}),
\begin{align}
L_\text{PP}^{(b)}=L_\text{PP}^{(a)}-\Delta L,
\end{align}
which leads to the orbital separation $r_{\text{PP}}^{(b)}$
\begin{align}
r_{\text{PP}}^{(b)}=r_{\text{PP}}^{(a)}\left(1-\frac{\Delta L}{L_\text{PP}^{(a)}}\right)^{2}.\label{rnew-jump}
\end{align}
Evolving the PP orbit with the above initial condition gives the lower panel of Fig.\ \ref{fig:orbit-average}. This method is very similar to FR07 \cite{Flanagan+Racine+07}. However, it also has a disadvantage: since so far we do not have an independent analytic estimation on the time of resonance, we cannot know the value of $r_{\text{PP}}^{(a)}$ without solving the full equations. Nevertheless, this method provide us an alternative understanding on the post-resonance PP orbit, i.e., it is related to the pre-resonance PP orbit by an instantaneous jump in a angular momentum, by contrast to a time shift $\Delta t$ at a fixed separation. In fact, one can prove that two methods agree with each other to the leading order in $\Delta t$. By expanding Eq.\ (\ref{rnew-jump}), we find the jump between ``a'' and ``b'' to be
\begin{align}
r_{\text{PP}}^{(a)}-r_{\text{PP}}^{(b)}=\frac{2\Delta L}{L_\text{PP}^{(a)}}r_\text{PP}^{(a)}=\dot{r}_\text{PP}^{(a)}\Delta t,
\end{align}
where the last equality comes from the fact that $L\propto r^{1/2}$ and the relation between $\Delta L$ and $\Delta t$ in Eq.\ (\ref{delta-t-shift}). The result is exactly the jump predicted by Eq.\ (\ref{rpp-postres}) if one expands $r(t_r+\Delta t)-r(t_r)$ to the leading order in $\Delta t$. In fact, we can work conversely. By imposing that the two methods predict the same orbital separation for the post-resonance PP orbit at resonance, we get an analytic equation for $t_r$
\begin{align}
r_{\text{PP}}^{(b)}=r(t_r+\Delta t)=r(t_r)\left(1-\frac{\Delta L}{L_r}\right)^{2}, \label{tr}
\end{align}
where
\begin{align}
&L_r=\mu {M_t}^{1/2}r(t_r)^{1/2},
\end{align}
and $r(t)$ is shown in  Eq. (\ref{r-t-notide}). Eq.\ (\ref{tr}) is an algebraic equation for $t_r$. In Table \ref{table:delta-t}, we show the accuracies of results by calculating the ratio between $\Delta t$ and $|t_d-t_r|$, where $\Delta t$ is the difference between $t_r$ obtained from Eq.\ (\ref{tr}) and the true $t_r$; and $|t_d-t_r|$ is the time difference between ``a'' and ``d'' in Fig.\ \ref{fig:orbit-average}. The ratios are between 5\%---20\%.

From the above discussion, we can see the method of averaged orbit is qualitatively accurate. By connecting two PP orbits with a jump, one can already extract some information of the system (e.g. $t_r$) without solving fully coupled differential equations. However, this method has two disadvantages. The first one is that it ignores the oscillation on the top of the averaged orbit in the post-resonance regime, which carries the information of $f$-mode. Secondly, averaging is only valid when the spin is large.  As shown in Fig. \ref{fig:per-r}, since the system does not undergo a full tidal oscillation cycle when spin is 300Hz or below, it is not appropriate to discuss the averaged orbit in this case.

\begin{table}
    \centering
    \caption{The comparisons between our analytic estimates for $t_r$ in Eq.\ (\ref{tr}) and full numerical integrations. For reference, the errors of results are compared with $|t_d-t_r|$, i.e., the time difference between ``d'' and ``a'' in Fig. \ref{fig:orbit-average}.}
    \begin{tabular}{c c c c c c} \hline\hline
$\Omega_s/(2\pi)$(Hz) & 550 & 450 & 350 & 250 &150  \\ \hline
$\frac{|\Delta t_r|}{|t_r-t_a|}$ ($\times10^{-2}$) &20.3 &5.3 &5.4 & 13.6 &20  \\ \hline\hline
     \end{tabular}
     \label{table:delta-t}
\end{table}

\section{Gravitational waveforms and extraction of parameters}
\label{sec:GW}
In the last two sections, we mainly discussed near zone dynamics. We obtained new formulae Eqs.\ (\ref{new-formulae-tide}) for the tidal deformation amplitudes $A$ and $B$; obtained osculating equations Eqs. (\ref{osc-all}) for the orbit; and developed analytic treatments that coupled stellar and orbital motions and carried out comparisons between analytic and numerical results.

In this section, we will go to the far zone to study GWs. We first quantify the accuracy of the method of effective Love number and the method of averaged PP orbit in the framework of the match filtering. We then compute the SNR of GWs emitted during and after resonance. Results show that post-resonance GWs may be strong enough to be observed by future GW detectors. We finally show that DT can provide more precise estimations on the parameters of NSs. We want to emphasize again that the major goal of this section is to provide a qualitative feature of impact of DT on GW observations. As we have discussed above, the EoS we used, as well as high spin rate, might be unlikely in realistic scenarios.


\subsection{Accuracies of DT models}
To the lowest order, GW emitted by a system is related to the near-zone dynamics through \cite{Poisson+Will+14}
\begin{align}
h_{ij}^\text{TT}=\frac{2}{{D_L}}\ddot{Q}^\text{TT}_{ij}, \label{h-q}
\end{align} 
where ${D_L}$ is the distance between the detector and the source, which we choose as $100$Mpc. $Q_{ij}$ is the quarupole moment of the system. The superscript ``TT'' stands for the transverse-traceless components of the tensor. Amplitudes of the two polarizations of the GW are given by \cite{Poisson+Will+14}
\begin{subequations}
\begin{align}
&h_+=-\frac{1}{4}s_i^2(Q_{xx}+Q_{yy})+\frac{1}{4}(1+c_i^2)c_{2\beta}(Q_{xx}-Q_{yy}) \notag \\
&+\frac{1}{2}(1+c_i^2)s_{2\beta} Q_{xy}-s_ic_ic_\beta Q_{xz}-s_ic_is_\beta Q_{yz} \notag \\
&+\frac{1}{2}s_i^2Q_{zz}, \\
&h_\times=-\frac{1}{2}c_is_{2\beta}(Q_{xx}-Q_{yy})+c_ic_{2\beta} Q_{xy} \notag \\
&+s_is_{\beta} Q_{xz}-s_ic_\beta Q_{yz},
\end{align}
\end{subequations}
where $c_i=\cos\iota$, $s_i=\sin\iota$, $c_{2\beta}=\cos 2\beta$, and $s_{2\beta}=\sin 2\beta$. The angle $\iota$ is the inclination of the orbital plane with respect to the line of sight toward the detector, and $\beta$ is azimuthal angle of the line of nodes. The detector measures the linear combination of the two polarizations
\begin{align}
h(t)=F_+h_++F_\times h_\times,
\end{align}
where the detector antenna pattern functions $F_+$ and $F_\times$ are given by
\begin{subequations}
\begin{align}
&F_+=\frac{1}{2}(1+\cos^2\theta)\cos2\phi\cos2\psi-\cos\theta\sin2\phi\sin2\psi,\\
&F_\times=\frac{1}{2}(1+\cos^2\theta)\cos2\phi\sin2\psi+\cos\theta\sin2\phi\cos2\psi,
\end{align}
\end{subequations}
with $\theta$ and $\phi$ the angular location of the source relative to the detector, $\psi$ the polarization angle \cite{Poisson+Will+14}. 

\begin{figure}[htb]
        \includegraphics[width=\columnwidth,height=5.6cm,clip=true]{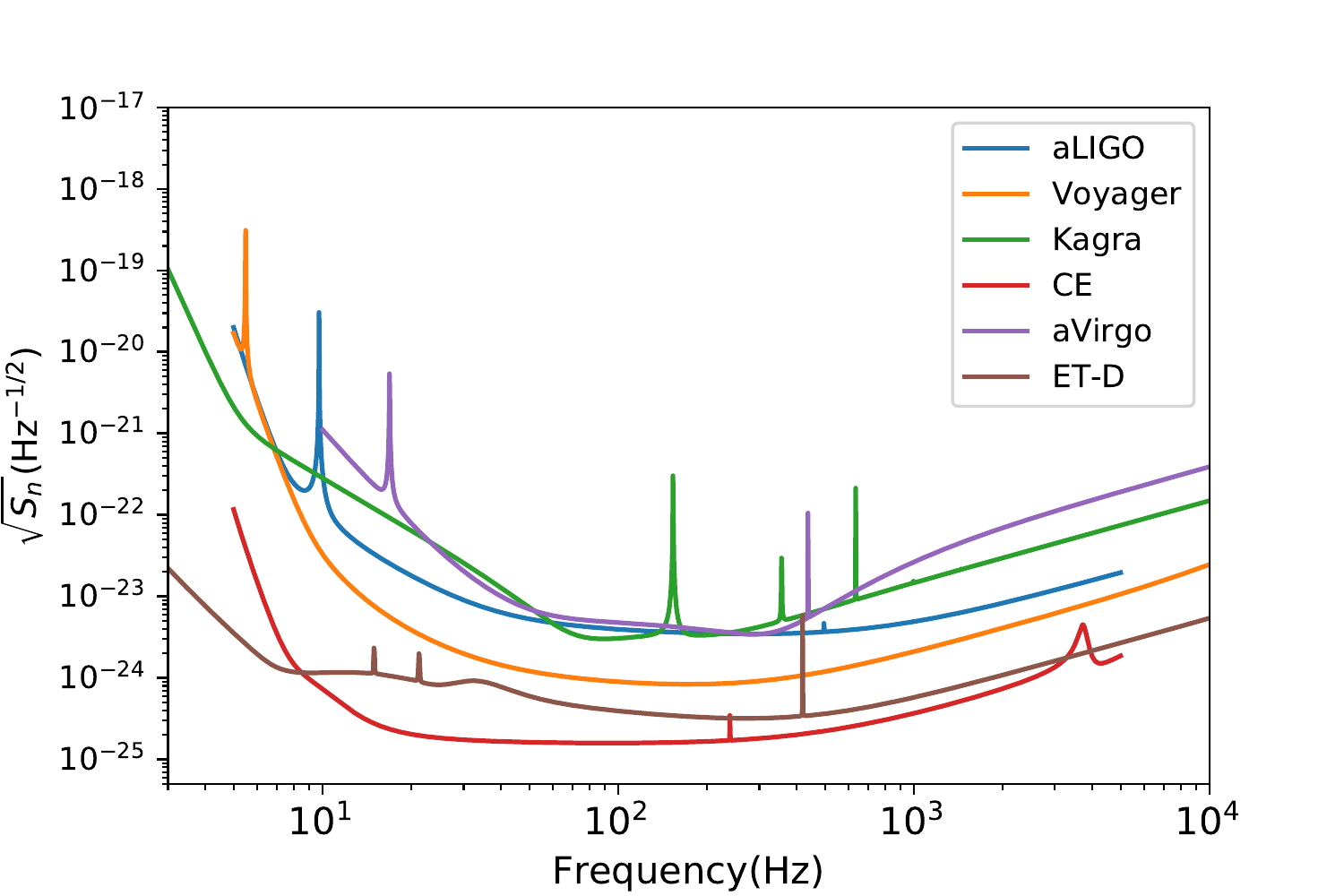}
  \caption{The noise spectral densities of several ground-based detectors.}
 \label{fig:psd}
\end{figure}

In order to measure the similarity between two waveforms $h$ and $g$, we define their match \cite{Cutler+Flanagan+94}
\begin{align}
\mathcal{O}[h,g]=\max_{t_c,\phi_c}\frac{(h|g)}{\sqrt{(h|h)(g|g)}},
\end{align}
and mismatch $1-\mathcal{O}$. The inner product $(h|g)$ between two waveforms is defined as
\begin{align}
(h|g)=4\Re\int\frac{\tilde{h}^*(f)\tilde{g}(f)}{S_n(f)}d\textit{f},
\end{align}
with the superscript $*$ standing for complex conjugation, and $S_n(f)$ the noise spectral density of the detector. In Fig.\ \ref{fig:psd}, we plot the noise spectral densities of aLIGO \cite{aligo-noise,Aasi:2013wya}, aVirgo \cite{virgo-noise,Aasi:2013wya}, KAGRA \cite{kagra-noise,Aasi:2013wya}, Voyager \cite{voyager}, CE \cite{CE}, and ET \cite{ET}.

\begin{figure}[htb]
        \includegraphics[width=\columnwidth,height=5.6cm,clip=true]{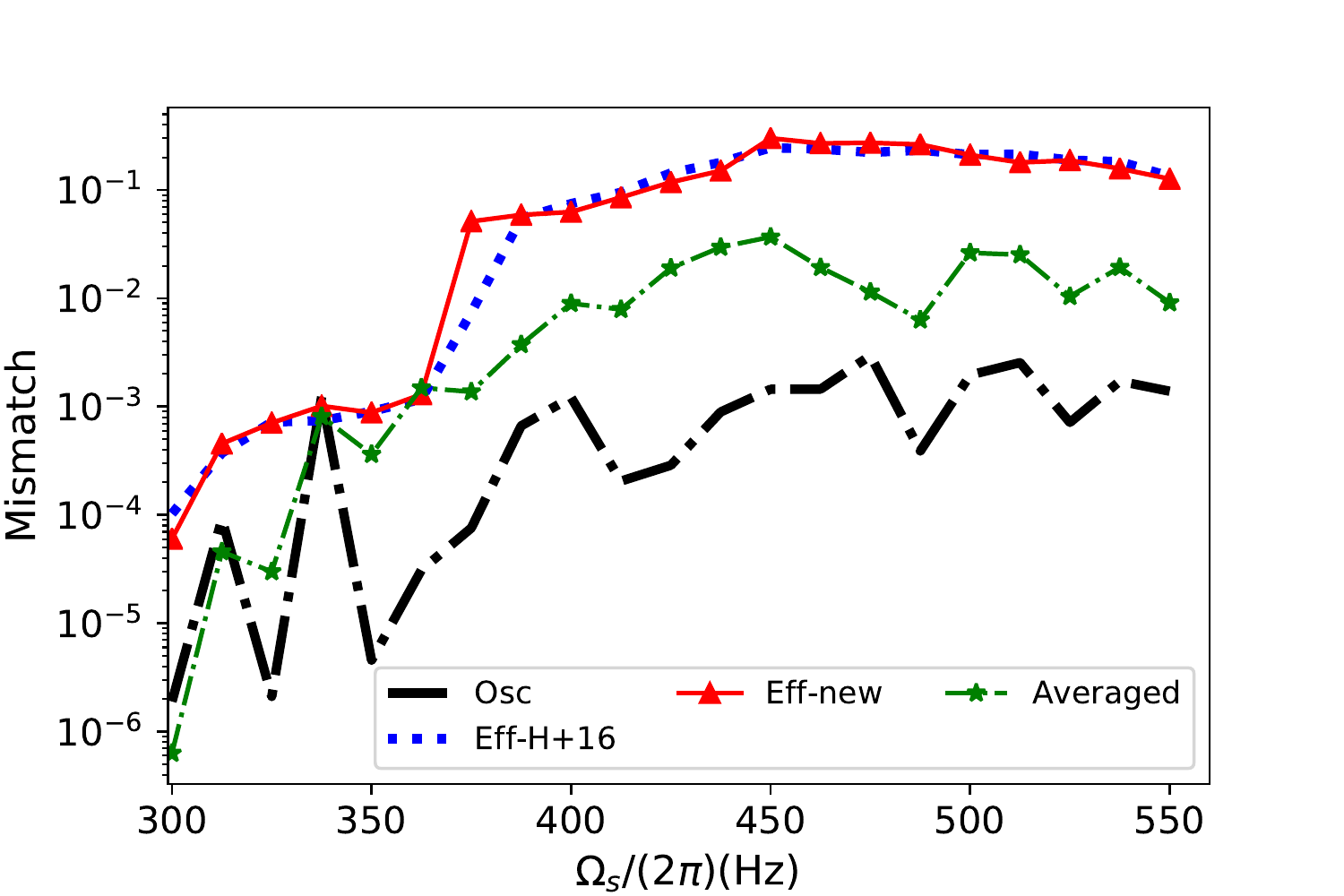}
  \caption{The mismatches as functions of spin frequency. We only use the signals with frequency higher than $2\Omega_r/(2\pi)$ because we only focus on the post-resonance dynamics. The fully numerical integrations are compared with four models, effective Love number with H+16 \cite{Steinhoff+Hinderer+16,Hinderer+Taracchini+16} (blue dashed line), effective Love number with our new DT formulae (red line), our new averaged PP orbit (green line), and osculating equations (black line). The mismatches of osculating equations are lower than  $10^{-3}$, while the method of the effective Love number gives $\sim 0.1-0.2$ for spin higher than 370Hz. This approach is insensitive to which DT model we use. Our new averaged PP orbit, on the other hand, is in the middle of two other approaches. The worst mismatch is around $3\times10^{-2}$. }
 \label{fig:mismatch}
\end{figure}


The fully numerical simulated waveforms can be computed in the following way. We first numerically solve the equations of motion Eqs.\ (\ref{eq-of-motion}), which gives the total quarupole moment of the system $Q^\text{Total}_{ij}=Q_{ij}+\mu x_ix_j-\mu r^2\delta_{ij}/3$ by Eq.\ (\ref{Q-qm}). We then obtain the waveform $h(t)$ from Eq.\ (\ref{h-q}). In this paper, we choose $\iota=\beta=\theta=\phi=\psi=0$ for simplicity. We then sample the solutions in the time domain with the rate $1/8192$s, and use the fast Fourier Transform (FFT) algorithm to perform the discrete Fourier transform on the sampled data. Following the procedure of Ref. \cite{Droz+Knapp+99}, we zero-pad the strain data on both sides to satisfy periodic boundary condition before FFT. Our choice of sample rate already ensures that the Nyquist frequency is larger than the contact frequency. We define the frequency-domain waveform within the frequency band $[2F_0,2F_\text{contact}]$ as the full signal and $[2\Omega_r/(2\pi),2F_\text{contact}]$ as post-resonance signal. Here $F_\text{contact}$ is the orbital contact frequency, the factor of 2 comes from the correspondence between the orbital frequency and GW frequency at quadrupole order.

In Fig.\ \ref{fig:mismatch}, we plot the mismatch between post-resonance waveforms obtained from different DT models, as functions of spin frequency. One waveform is calculated from the fully numerical integration; against this target waveform, we compare waveforms  obtained from 4 different models: effective Love number with H+16 \cite{Steinhoff+Hinderer+16,Hinderer+Taracchini+16} (blue dashed line), effective Love number with our new formulae Eqs.\ (\ref{new-formulae-tide}) (red line), our new post-resonance averaged PP orbit defined in Eq.\ (\ref{rnew-jump}) (green line), and osculating equations (black line).  Here we do not include the averaged orbit model in FR07 \cite{Flanagan+Racine+07} because it is very close to our model. Since the match depends weakly on detector noise curve, we shall use that of aLIGO. One can see that the mismatches of all models are smaller that $10^{-3}$ for spins below 370Hz, since in this case the post-resonance signals are very short, such that the phase mismatches does not accumulate with frequency. The mean mismatches of our osculating equations are around $10^{-4}$, with the worst one still below $10^{-3}$. Accordingly, this approach describes the post-resonance dynamics accurately. This confirms that our new formulae of $A$ and $B$ are precise enough for describing the tidal back-reaction on the orbit. Methods that use the effective Love number, on the other hand, give the large mismatch of around 0.2 when the spin frequency reaches $\sim 450$Hz. The fact that both versions lead to similar mismatches, even with our accurate formulae for $A$ and $B$, shows that the formalism itself is imprecise. The mismatch of our averaged PP-orbit treatment is less than 0.03 within the entire regime we study. Therefore this approach gives a fairly accurate description of post-resonance GW signals.

\subsection{Detectability and Fisher analyses}
\label{sec:fim}
\begin{figure}[htb]
        \includegraphics[width=\columnwidth,height=5.6cm,clip=true]{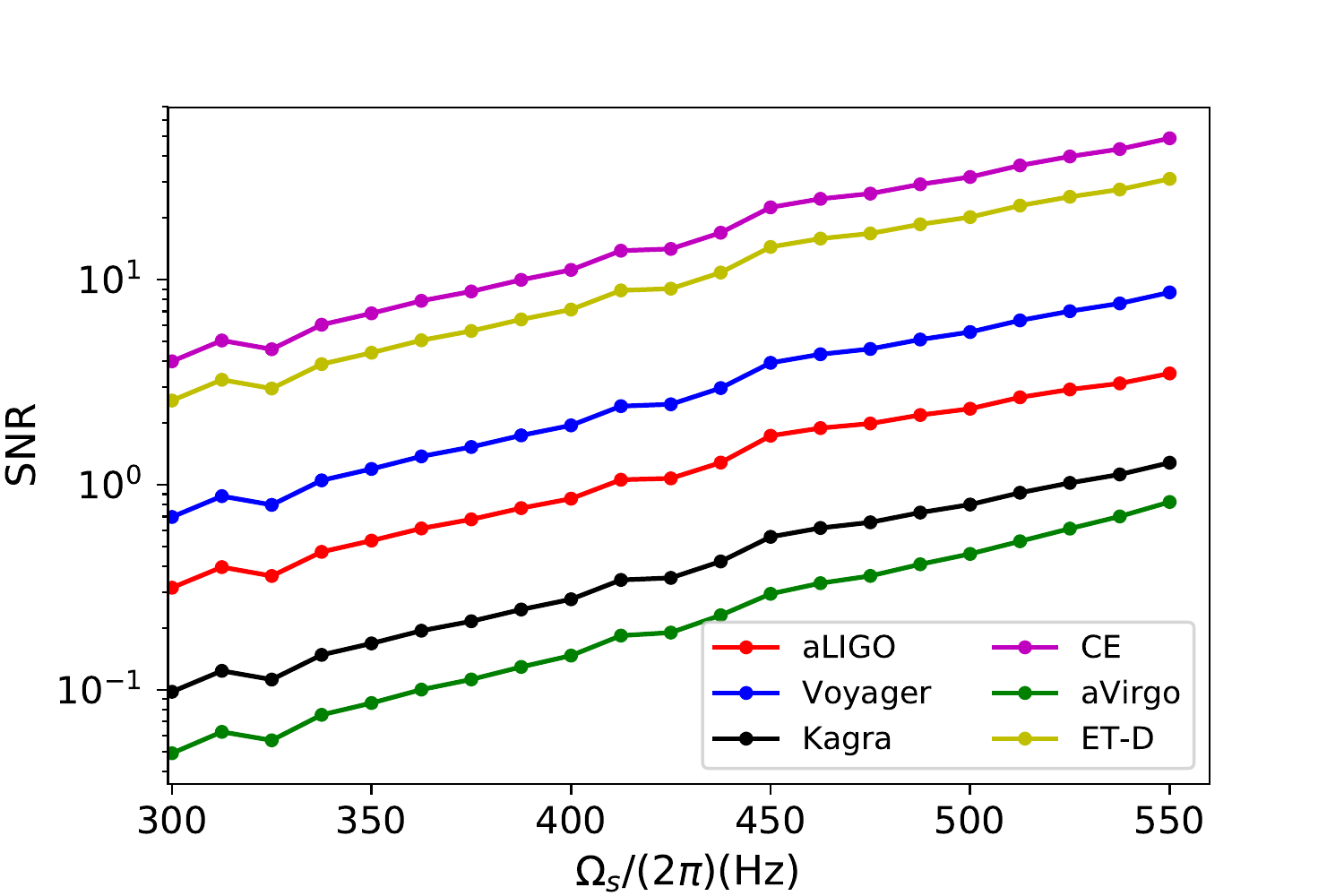}
  \caption{The SNRs from the resonant part of GW signals, with frequency higher than $2\Omega_r/(2\pi)$. The faster the NS spins, the higher the SNR. The SNR is around 0.3-3 for current detectors, but $\sim10-50$ for 3G detectors.}
 \label{fig:snr_reson}
\end{figure}

In Fig.\ \ref{fig:snr_reson}, we plot the signal-to-noise ratios (SNRs) of post-resonance GW (within the band $[2\Omega_r/(2\pi),2F_\text{contact}]$) as functions of spin frequency $\Omega_s$. As expected, it grows with spin frequency. For aLIGO, $\Omega_s$ needs to be above $\sim 425$Hz to lead to SNR $>1$. For 3G detectors, SNRs are around 4 for spin $\sim300$Hz; It can reach 50 if the spin is around $500$Hz. For comparison, we also calculate the SNRs of full signals within the band $[2F_0,2F_\text{contact}]$ in Table \ref{table:snr_full}. Since the full SNRs depend weakly on the spin frequency, here we choose $\Omega_s=2\pi\times300$Hz.

\begin{table}
    \centering
    \caption{The SNRs of full GW signals within the band $[2F_0,2F_\text{contact}]$ for different detectors. The spin frequency of NS is 300Hz.}
    \begin{tabular}{c c c c c c} \hline\hline
aLIGO & aVIRGO & KAGRA &Voyager & ET-D & CE \\ \hline
31.6 & 25.4 & 31.4 & 135.1 & 305.7 & 884.0 \\ \hline\hline
     \end{tabular}
     \label{table:snr_full}
\end{table}

\begin{figure*}[!tb]
        \includegraphics[width=\columnwidth,height=5.6cm,clip=true]{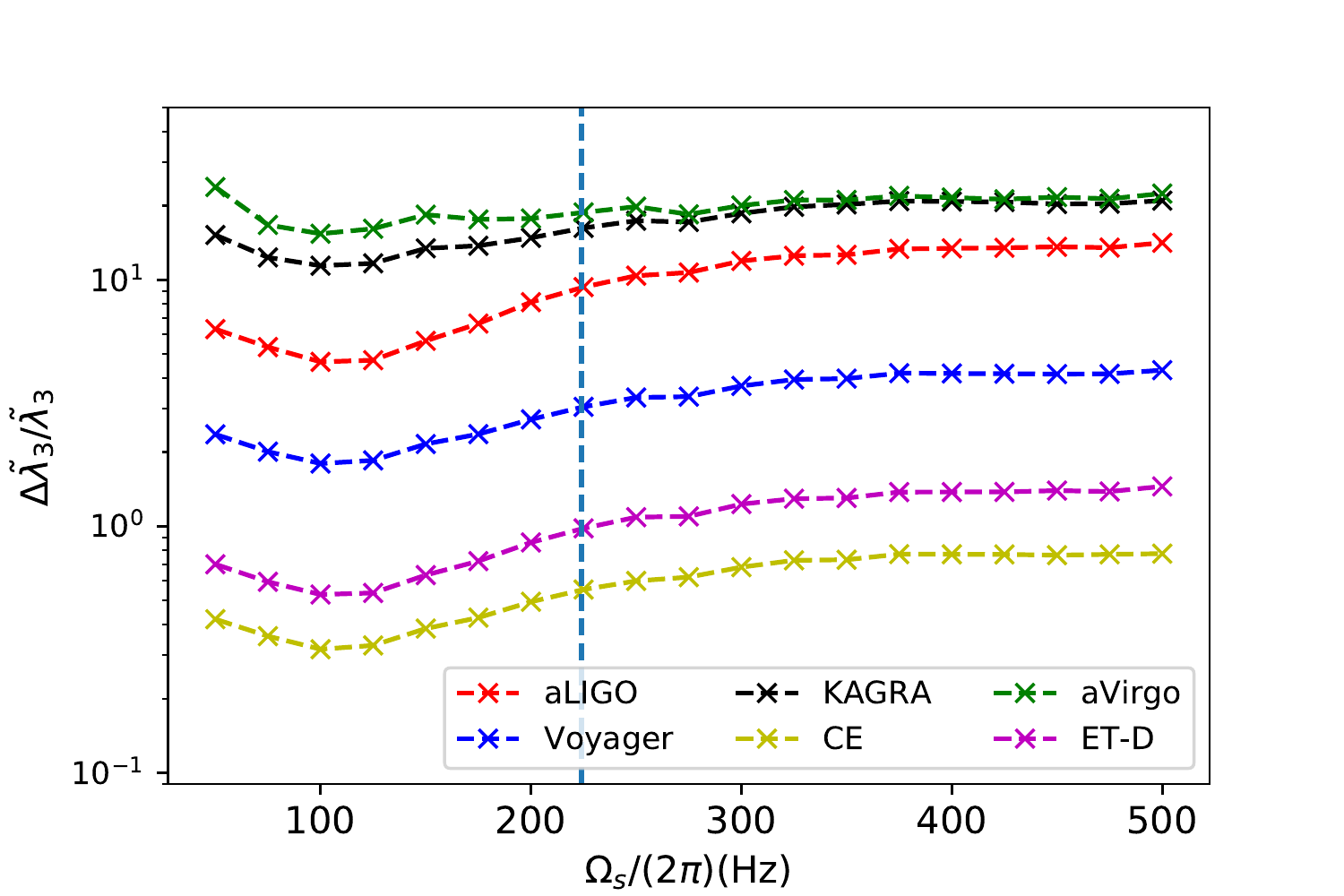}
        \includegraphics[width=\columnwidth,height=5.6cm,clip=true]{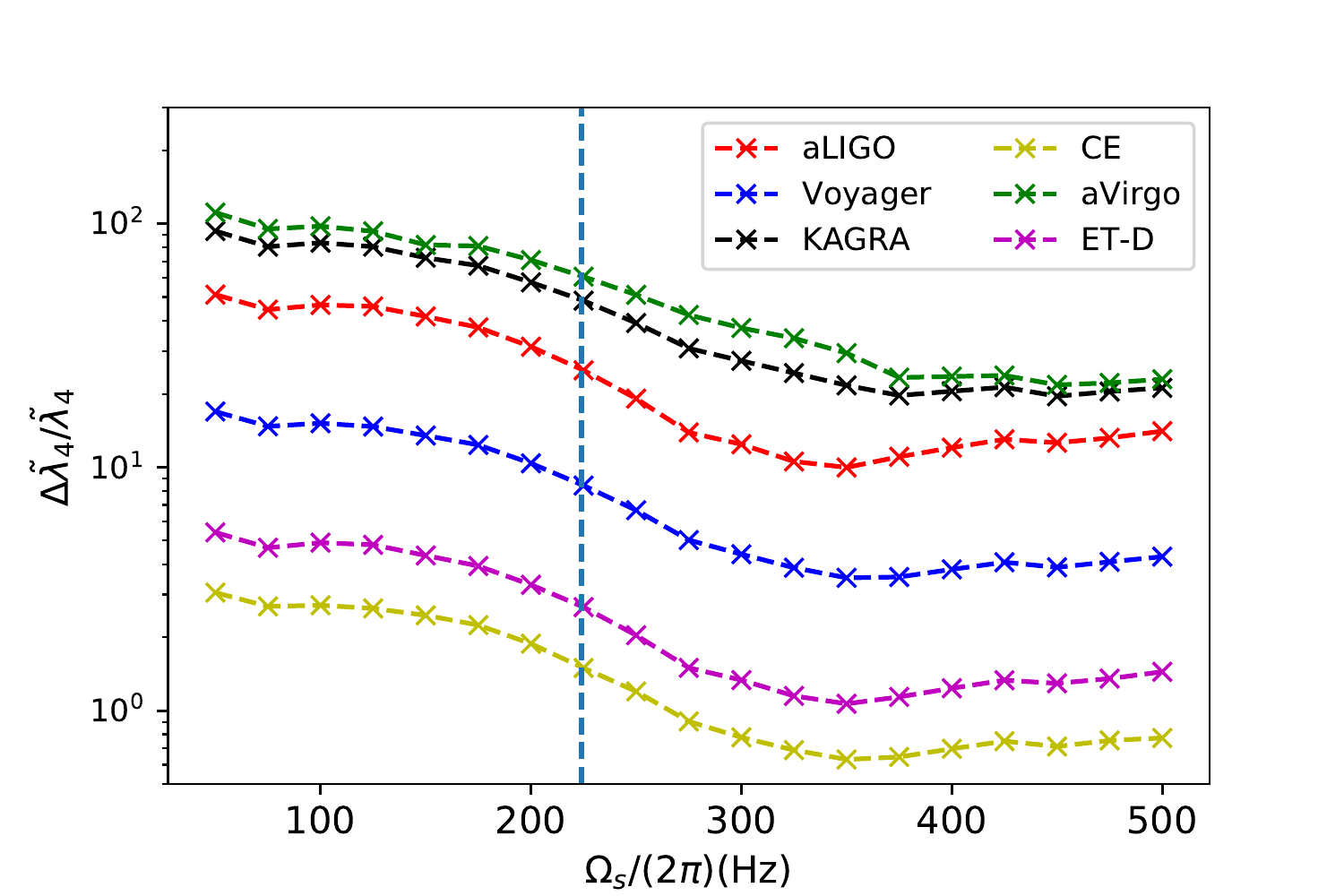}\\
         \includegraphics[width=\columnwidth,height=5.6cm,clip=true]{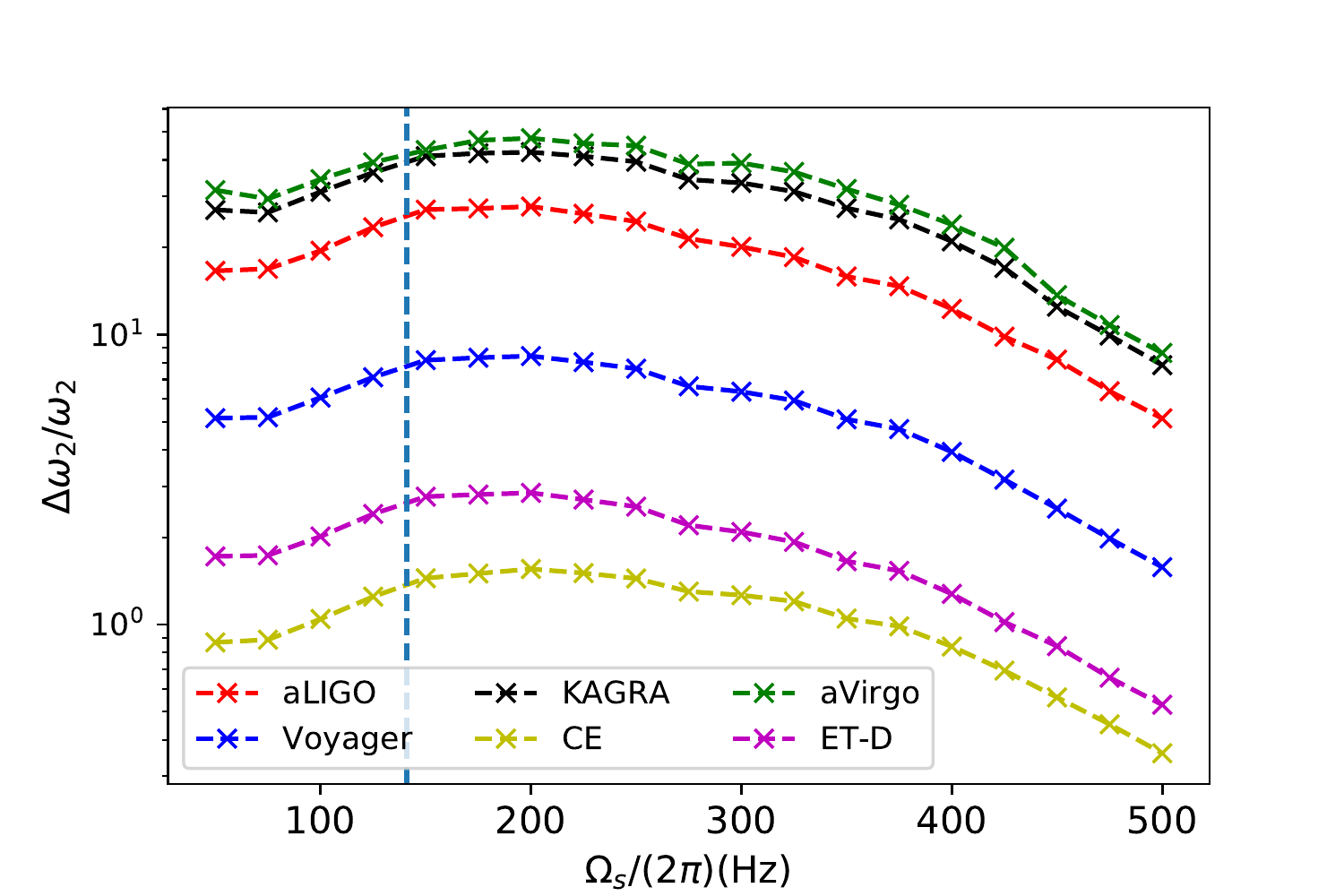}
          \includegraphics[width=\columnwidth,height=5.6cm,clip=true]{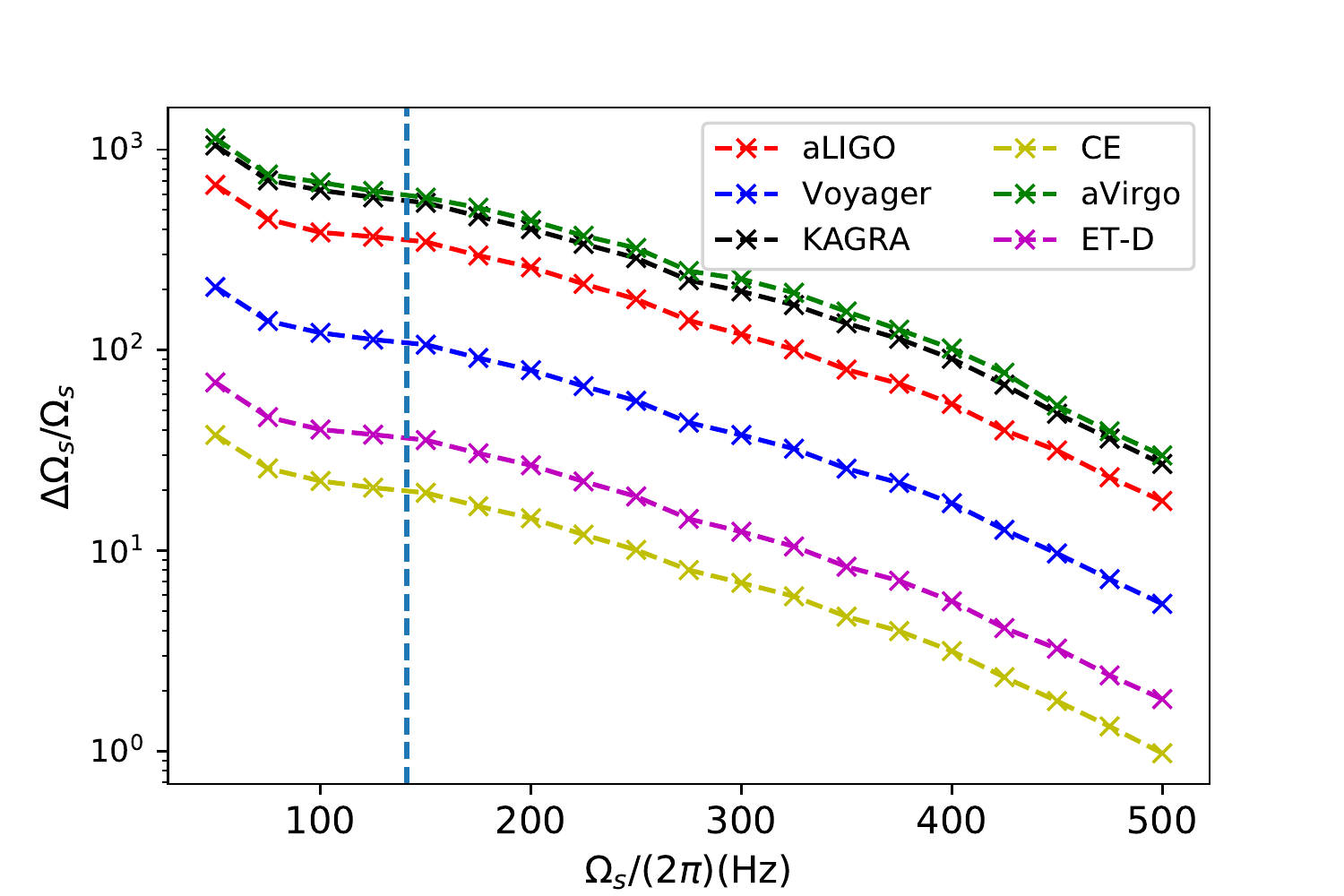}
  \caption{Relative errors of $\lol_3$, $\lol_4$, $\omega_2$ and $\Omega_s$ as functions of spin from Fisher analyses. The GW waveform is at the Newtonian order. The vertical dotted line stands for the location where resonance happens. The system is optimally oriented at 100Mpc, with component masses $(1.4M_\odot,1.4M_\odot)$. The H4 EoS is used.}
 \label{fig:h4-fim}
\end{figure*}

\begin{figure*}[!tb]
        \includegraphics[width=\columnwidth,height=5.6cm,clip=true]{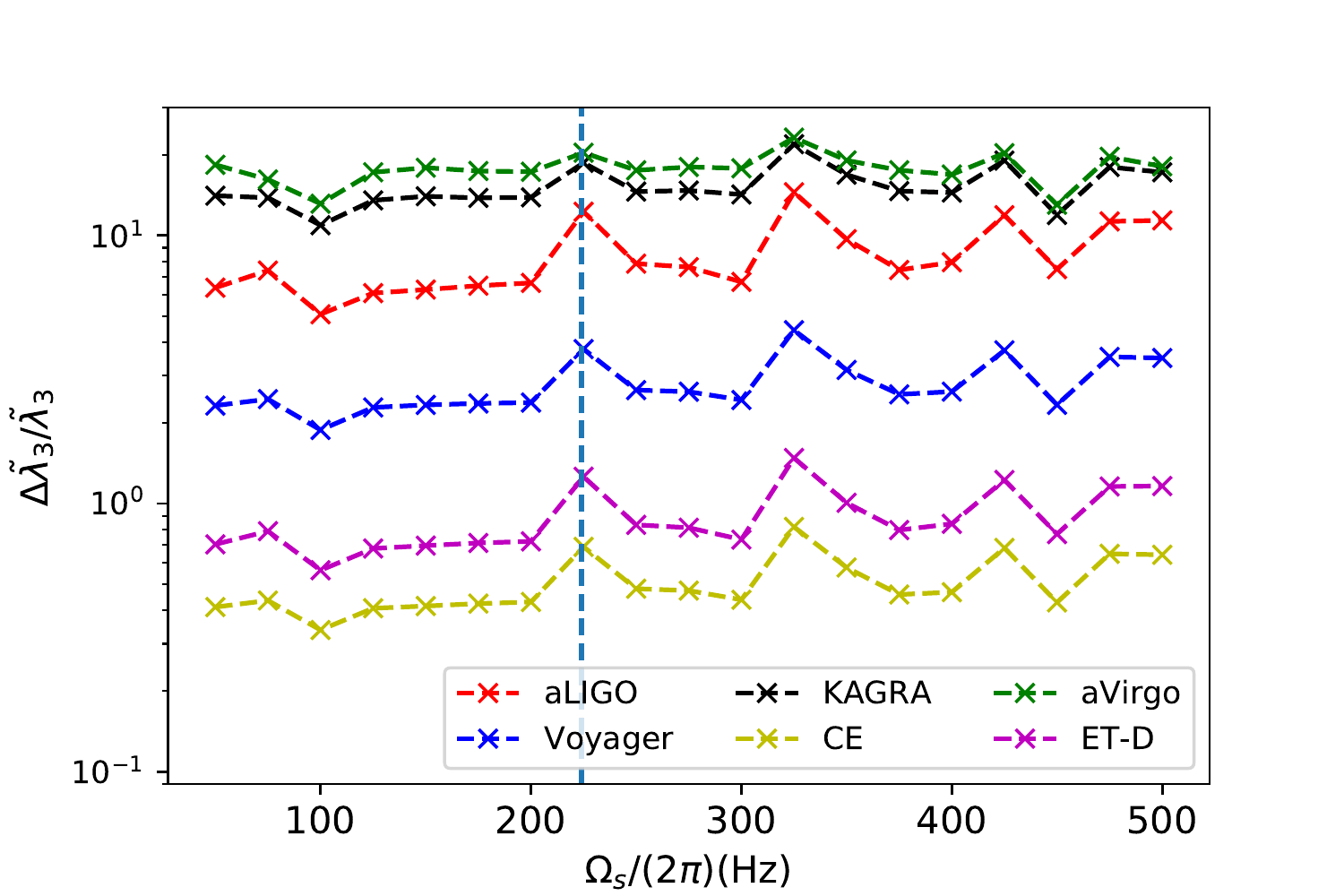}
        \includegraphics[width=\columnwidth,height=5.6cm,clip=true]{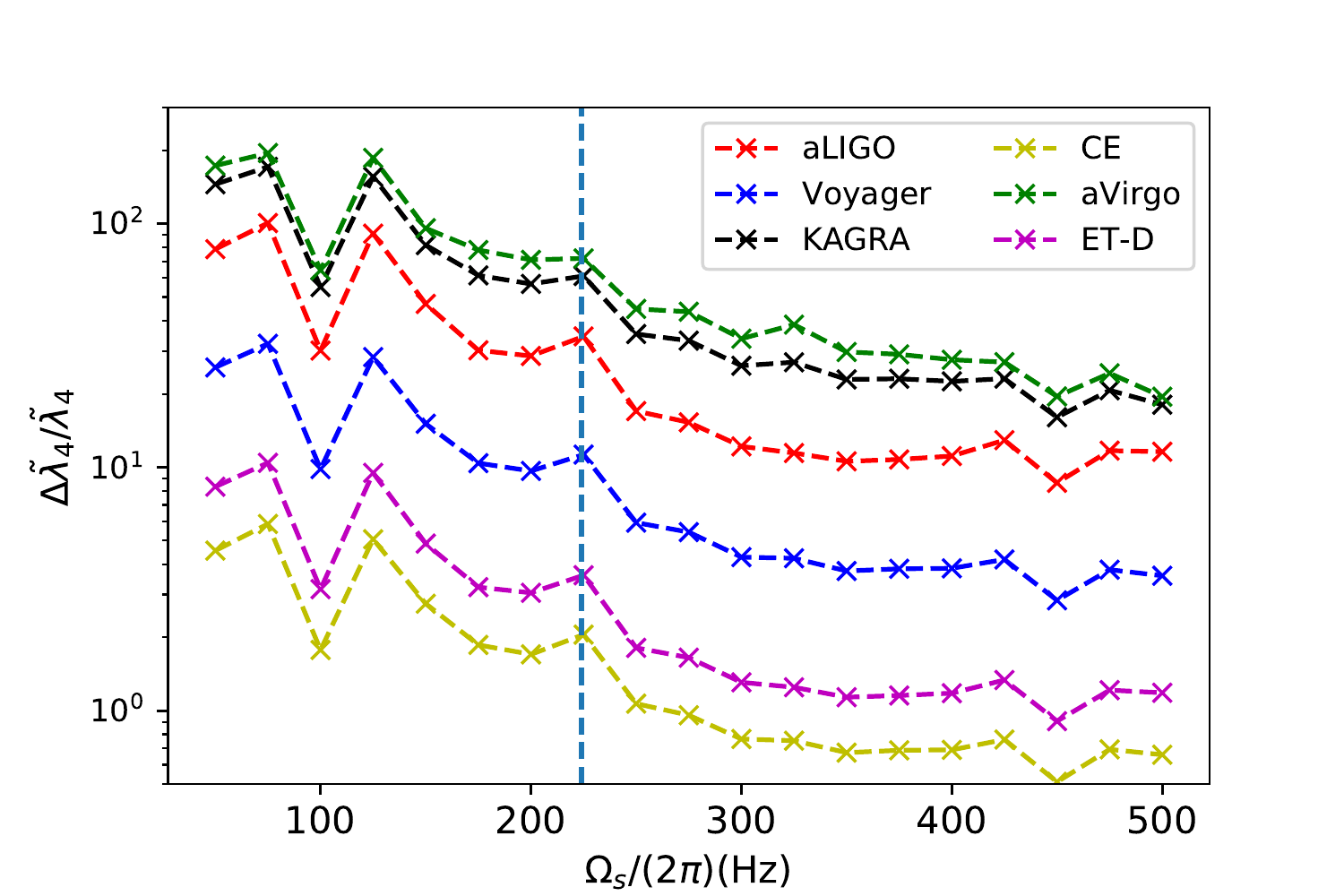}\\
         \includegraphics[width=\columnwidth,height=5.6cm,clip=true]{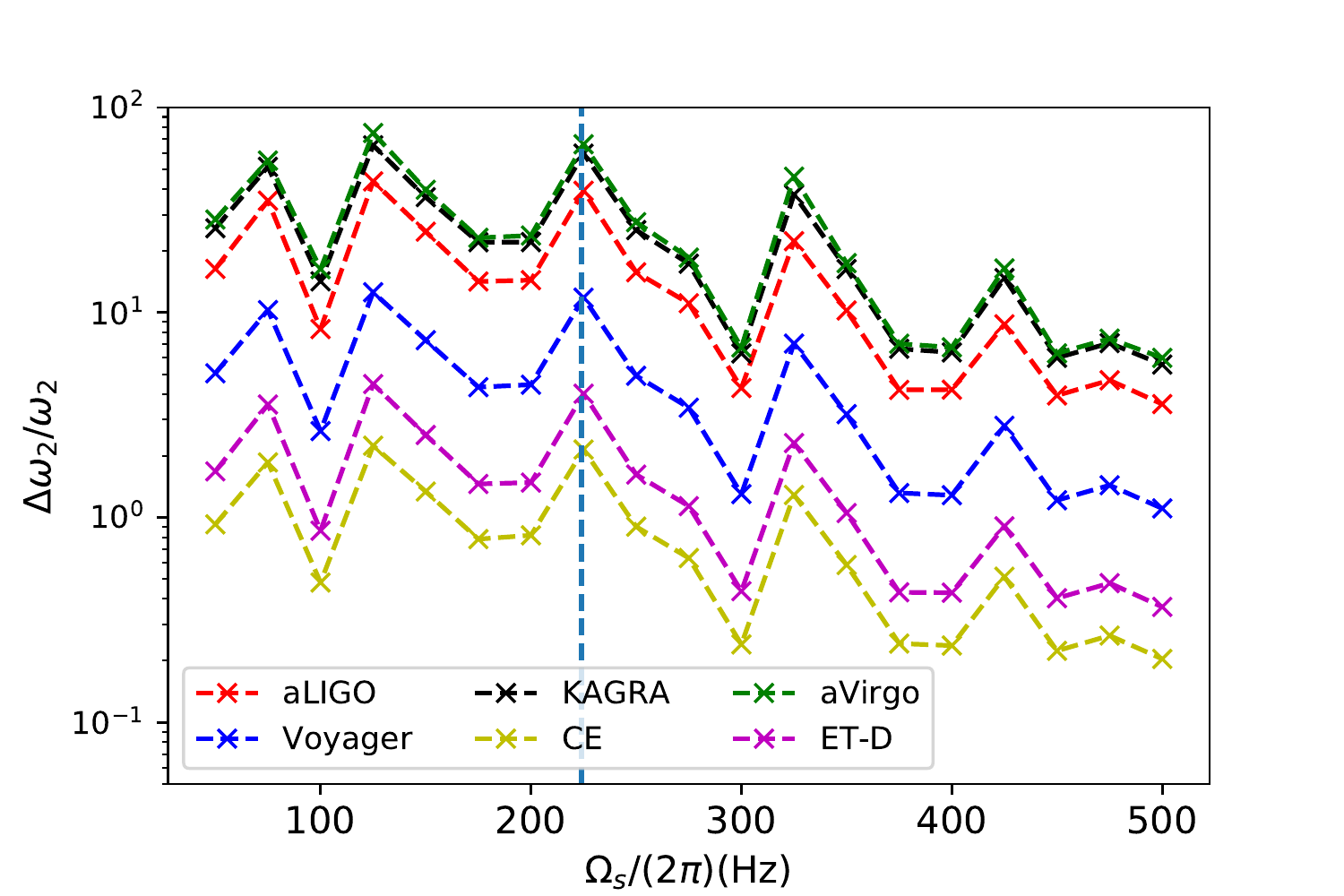}
          \includegraphics[width=\columnwidth,height=5.6cm,clip=true]{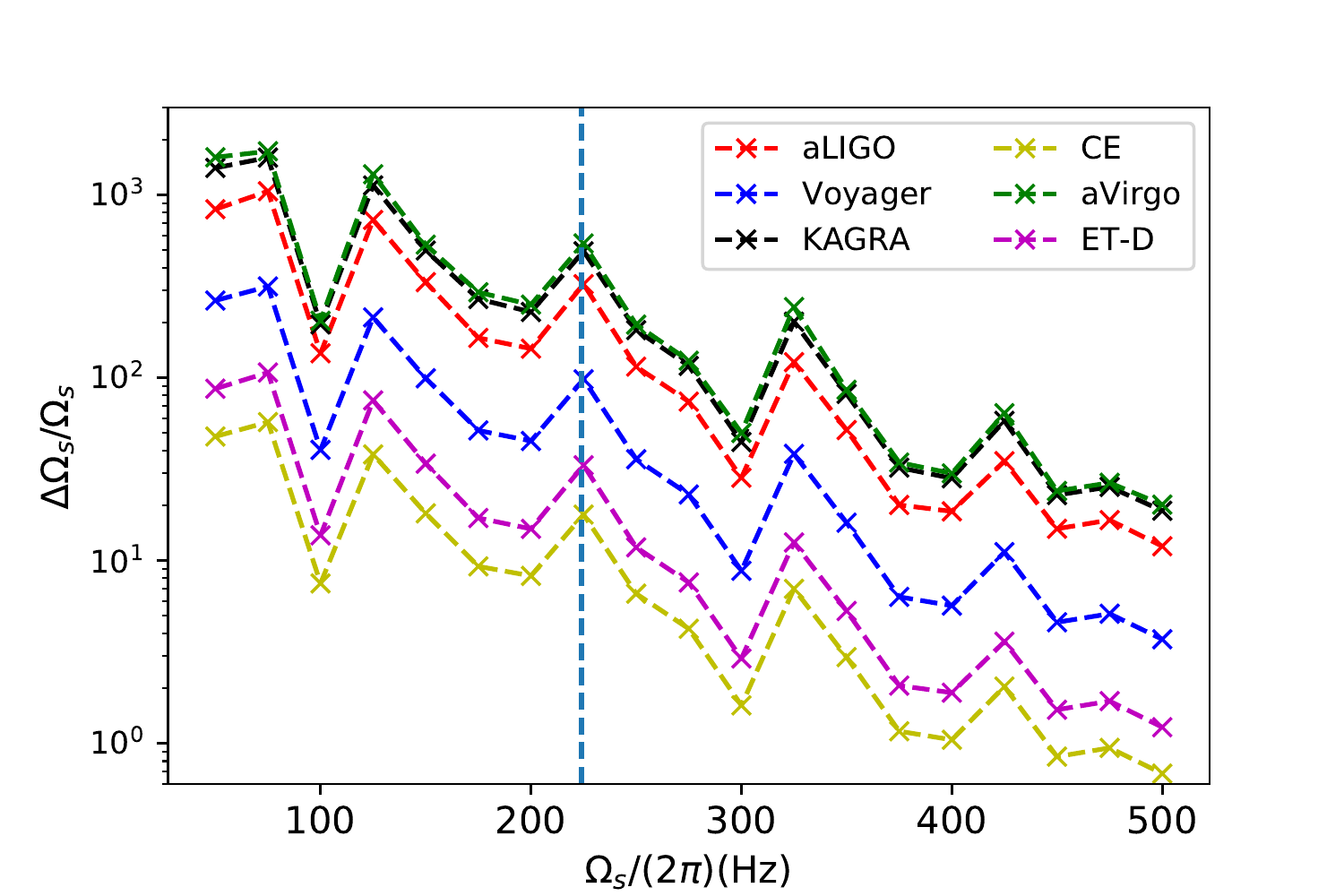}
  \caption{Same as Fig.\ \ref{fig:h4-fim}, except the polytropic EoS is used.}
 \label{fig:polytropic-fim}
\end{figure*}

These results of SNRs show the potential to detect post-resonance signals with 3G detectors. This allows us to extract more information from GW signals than AT. As pointed out in Ref. \cite{Flanagan+Hinderer+08}, the Love number of non-spinning NS is degenerate with mass ratio $\Xi={M_2}/{M_t}$ at leading order in the adiabatic regime. Only the effective $\lol=\lambda\Xi(11\Xi+1)$ can be constrained by GWs\footnote{We still assume only ${M_1}$ is tidally deformed.}. This degeneracy persists for spinning NSs in AT. In this case, the phase of GW during AT (up to leading tidal order of the Love number) is given by
\begin{align}
&\Psi=2\pi ft_c-\phi_c-\frac{\pi}{4}+\frac{3}{128}\left(\pi\chirp f\right)^{-5/3}\left\{1-\frac{24(\pi f)^{10/3}}{\chirp^{5/3}} \right.\notag \\
&\left.\times\left[\frac{11}{4}\Xi^2\lambda_0+\frac{\lambda_2\Xi}{1-2\Omega_s^2/\omega_2^2}\left(1+\frac{33}{4}\Xi\right)\right]\right\}. \label{phase-AT-spin}
\end{align}
Hence the tidal term is governed by the effective Love number
\begin{align}
\lol_3=\frac{11}{4}\Xi^2\lambda_0+\frac{\lambda_2\Xi}{1-2\Omega_s^2/\omega_2^2}\left(1+\frac{33}{4}\Xi\right). \label{love3}
\end{align}
It is straightforward to see that $\lol_3$ reduces to $\lol$ in the non-spinning limit. Note that our notation of $\lol_3$ differs from Ref.\ \cite{Flanagan+Hinderer+08} by a factor of $\eta=\mu/{M_t}$, since they used total mass ${M_t}$ while we use the chirp mass $\chirp$ here. 
As $\Omega$ increases, the motion of $(2,2)$ mode is resonantly getting excited while $(2,0)$ mode is not, their different reactions to the tidal driving from the orbit lead to distinct effects on GW emission, therefore the degeneracy is broken. To describe this effect, we introduce another parameter 
\begin{align}
\lol_4=\frac{\lambda_2\Xi}{1-2\Omega_s^2/\omega_2^2}\left(1+\frac{33}{4}\Xi\right),
\end{align}
i.e., the second part of Eq.\ (\ref{love3}). Accordingly, the numerical waveforms are determined by a 9-dimensional parameter $\bm{\theta}=\{t_c,\phi_c,D_L,\chirp,\Xi,\Omega_s,\omega_2,\lol_3,\lol_4\}$. Here we ignore $\omega_0$, the mode frequency of $(2,0)$ mode, since this mode does not have DT and its mode frequency is almost degenerate with other parameters.


Let us now turn to parameter estimation, using the Fisher information matrix formalism. Suppose random noise $n(t)$ in observed signal $s(t)$ is stationary and Gaussian, the conditional likelihood function of $s$ given parameters $\bm{\theta}$ can be written as
\begin{align}
p(s|\bm{\theta})\propto e^{-(s-h|s-h)/2},
\end{align}
where $h(\bm{\theta},t)$ stands for the true waveform for parameter $\bm{\theta}$. In the large-SNR approximation, the likelihood function becomes Gaussian,
\begin{align}
p(s|\bm{\theta})\propto e^{-\Gamma_{ij}\Delta\theta^i\Delta\theta^j/2},
\end{align}
where Fisher matrix $\Gamma_{ij}$ is given by
\begin{align}
\Gamma_{ij}=\left(\left.\frac{\partial h}{\partial \theta^i}\right|\frac{\partial h}{\partial \theta^j}\right).
\end{align}
Since waveforms are numerically calculated in our case (from algorithms discussed in the previous subsection), derivatives are computed numerically using the symmetric difference quotient method. The inverse of the Fisher matrix gives the covariance matrix. In particular, the diagonal components are the variances of the estimated parameters
\begin{align}
\Delta\theta_i=\sqrt{(\Gamma^{-1})_{ii}},
\end{align}
which are the projected constraints that we can put on parameters from the observation. 

We still use the H4 and the $\Gamma=2$ polytropic EoS, with ${M_1}={M_2}=1.4M_\odot$. The system is at ${D_L}=100$Mpc and optimally oriented. Projected constraints on several parameters as functions of spin frequency are shown in Figs.\ \ref{fig:h4-fim} and \ref{fig:polytropic-fim}, where the vertical lines stand for values of spins for which resonance takes place right on contact. We can see that the two EoS give similar results. The constraints change with detectors since we have fixed the distance of the source, and 3G detectors can benefit from large SNRs. Among the six detectors, CE provides the best parameter estimations because it is the most sensitive in the high frequency band, where DT takes place.
To quantify the effect of DT, we list the projected constraints on several parameters in Table \ref{table:fim} under two situations: (i) results evaluated with spin frequencies when resonance takes place right on contact and (ii) constraints with spin frequencies 500Hz. The improvement factor, which is the ratio of estimation accuracies between two situations, characterizes the effect of DT. 

Let us discuss each parameter more specifically. First, we can see that for different detectors the relative errors on $\lol_3$ are of order $\sim0.4-20$, which depend most weakly on spins when compared to other parameters. The estimation error even becomes worse when spins are high. This is because this parameter is mainly estimated from AT, and the constraints do not benefit from DT. When spins are high, adiabatic waveforms become relatively short, hence the project constraints become worse. By contrast, estimation error of the other Love number $\lol_4$, which describes the $(2,2)$ mode, improves with spin. This is expected since DT introduces the dependence of waveforms on $\lol_4$. The constraints on this quantity can be improved by a factor of $3-5$, depending on EoS and detectors. In the CE case, the relative error of $\lol_4$ can final decrease to $\sim0.8$ as spins are around 500Hz. However, this parameter is still degenerate with the mass ratio $\Xi$. One need to take into account PN corrections to break such degeneracy. 

DT also helps us put more stringent constraints on the $(2,2)$ mode frequency, since the oscillations of NSs can react back to orbits and influence GW waveforms. As shown in Table \ref{table:fim}, the averaged improvement factors are around $6.6-6.9$ for the polytropic EoS, while $\sim5$ for the H4 EoS. The current detector, like aLIGO, cannot constrain this parameter well, giving relative errors $\sim5$. However, it is improved to 0.2 in the CE case. We have also calculated the effect of DT on constraining spin frequencies. The improvements on spin are the largest among parameters we discuss, since this parameter determines the location of resonance in the time (frequency) domain. The improvements are around $20-27$ for both EoSs. In the CE case, the relative errors are $\sim0.7-1$ when spins reach 500Hz.
\begin{table}
    \centering
    \caption{Projected constraints on $\lol_3$, $\lol_4$, $\omega_2$ and $\Omega_s$ with two EoS for six different detectors. Here we compare two situations: (i) constraints with spins when resonance takes place right on contact (Res) and (ii) constraints with NSs spinning at 500Hz ($\Omega_s^m$). The improvement factor is the ratio of $\Omega_s^m$ to Res, which characterizes the effect of DT.}
    \begin{tabular}{c c c c c c c c c} \hline\hline
   \multirow{13}{*}{H4} & \multicolumn{2}{c}{Detectors} &aVirgo  & KAGRA  & aLIGO & Voyager & ET-D & CE    \\ \cline{2-9}
 &   \multirow{3}{*}{$\frac{\Delta\lol_3}{\lol_3}$} & Res &18.4 & 13.4 & 5.7 & 2.1&0.6 &0.4\\ \cline{3-9}
   & & $\Omega_s^m$ &22.4 &21.0 &14.1 &4.3 &1.5 &0.8 \\  \cline{3-9}
    && Imp &0.8 &0.6 &0.4 &0.5 &0.4 &0.5\\  \cline{2-9}
 &   \multirow{3}{*}{$\frac{\Delta\lol_4}{\lol_4}$} & Res &81.8 & 72.4&41.6 &13.5 &4.3 &2.5\\ \cline{3-9}
   & & $\Omega_s^m$ &23.0 &21.1 &14.1 &4.3 &1.4 &0.8 \\  \cline{3-9}
    && Imp &3.6 &3.4 &3.9 &3.1 &3.0 &3.2\\  \cline{2-9}
 &   \multirow{3}{*}{$\frac{\Delta\omega_2}{\omega_2}$} & Res &43.2 &41.2 &27.0 &8.2 &2.8 &1.4\\ \cline{3-9}
   & & $\Omega_s^m$ &8.6 &7.8 &5.1 &1.6 &0.5 &0.4 \\  \cline{3-9}
    && Imp &5.0 &5.3 &5.2 &5.2 &5.2 &4.0\\  \cline{2-9}
 &   \multirow{3}{*}{$\frac{\Delta\Omega_s}{\Omega_s}$} & Res &575.7 &542.9 &346.6 &106.4 &35.6 &19.4\\ \cline{3-9}
   & & $\Omega_s^m$ &29.9 &27.1 &17.7 &5.4 &1.8 &1.0 \\  \cline{3-9}
    && Imp &19.3 &20.1 &19.6 &19.6 &19.5 &19.9\\  \hline
  \multirow{13}{*}{Poly}&   \multirow{3}{*}{$\frac{\Delta\lol_3}{\lol_3}$} & Res &17.9 &14.0 &6.3 &2.3 &0.7 &0.4\\ \cline{3-9}
   & & $\Omega_s^m$ &18.1 &17.2 &11.4 &3.5 &1.2 &0.6 \\  \cline{3-9}
    && Imp &1.0 &0.8 &0.6 &0.7 &0.6 &0.6\\  \cline{2-9}
 &   \multirow{3}{*}{$\frac{\Delta\lol_4}{\lol_4}$} & Res &95.8 &81.6 &46.8 &15.1 &4.9 &2.8\\ \cline{3-9}
   & & $\Omega_s^m$ &19.5 &18.1 &11.6 &3.6 &1.2 &0.7 \\  \cline{3-9}
    && Imp &4.9 &4.5 &4.0 &4.2 &4.1 &4.2\\  \cline{2-9}
 &   \multirow{3}{*}{$\frac{\Delta\omega_2}{\omega_2}$} & Res & 39.7&36.5 &24.8 &7.3 &2.5 &1.3\\ \cline{3-9}
   & & $\Omega_s^m$ &6.0 &5.6 &3.6 &1.1 &0.4 &0.2 \\  \cline{3-9}
    && Imp &6.6 &6.6 &6.9 &6.6 &6.9 &6.6\\  \cline{2-9}
 &   \multirow{3}{*}{$\frac{\Delta\Omega_s}{\Omega_s}$} & Res &533.4 &496.0 &332.5 &99.2 &33.9 &18.1\\ \cline{3-9}
   & & $\Omega_s^m$ &20.2 &18.7 &12.0 &3.7 &1.2 &0.7 \\  \cline{3-9}
    && Imp &26.4 &26.5 &27.8 &26.7 &27.6 &26.5\\  \hline
     \end{tabular}
     \label{table:fim}
\end{table}

\section{Conclusions and Discussion}
\label{sec:con}
We have systematically studied the $(2,2)$ $f$-mode DT of spinning NSs in coalescencing binaries. In particular, the spin is assumed to be anti-aligned with the orbital angular momentum, in which case the effect of DT is the most pronounced. We began by deriving a complete set of coupled equations for mode oscillation and orbital evolution, with the aid of the phase-space mode expansion method and the Hamiltonian approach. We then extended H+16's model \cite{Steinhoff+Hinderer+16,Hinderer+Taracchini+16} for $f$-mode excitation to spinning NSs and obtained a new approximation which can describe the full dynamics of systems to a high accuracy. One application of this approximation is to study the post-resonance orbital dynamics, where we used the method of osculating orbits and obtained the time evolution of the osculating variables. This framework allowed us to obtain analytic estimations on the orbital information at resonance (e.g. $\dot{r}_r$, $\omd$). We also obtained a simple formula of angular momentum transfer due to DT, which is an extension of L94 \cite{Lai+94} to the spinning case. Based on this result, we derived the averaged post-resonance orbits over tide-oscillation timescale in an alternative way. The result of our averaged treatment turns out to agree with that of FR07 \cite{Flanagan+Racine+07}, to the leading order in angular momentum transfer time $\Delta t$ [Eq.\ (\ref{delta-t-shift})]. By combining the two treatments, we obtained an algebraic equation for $t_r$. We then compared several DT models by computing the mismatches of waveforms. Finally, we carried out a Fisher matrix analysis to estimate the effect of DT on parameter estimation, with current and 3G detectors.

We summarize our main conclusions as follows. (i) The $(2,2)$ $f$-mode in the spinning NS, by defining a new variable $x$ [Eq.\ (\ref{eq-nogw-q2})], can still be treated as a harmonic oscillator, which is oscillating at its eigenfrequency $\zeta$ in the post-resonance regime. (ii) The reason that H+16 \cite{Steinhoff+Hinderer+16,Hinderer+Taracchini+16} cannot describe the post-resonance evolution are two folds. The first is that their phasing $\tha^2$ is not accurate and should be replaced by $\Theta$ [Eq.\ (\ref{Theta})]. Second, their counterterm Eq.\ (\ref{conter-old}) does not contain phase information. (iii) The picture of averaged orbit over the tide-oscillation timescale is accurate: the true pre- and post-resonance orbital motion can be tracked accurately by PP orbits. These PP orbits are related by energy and angular momentum transfers, and hence a jump in the orbital separation at $t_r$. Within the spin range we studied, the match of GW signals between the prediction using the averaged orbit and numerical integration (post-resonance part) is as high as $99\%$. Therefore the additional tidal perturbation is a small effect. However,  such description requires that the post-resonant signal is long enough (i.e. large spin) so that the system can undergo several tidal oscillation cycles. Looking at the full orbit, we found that there is an extra oscillation on top of the averaged trajectory. We also found that the eccentricity of the orbit is induced by the tidal interaction and can grow to $\sim0.08$ at the end of inspiral, the numbers depend weakly on the spin. (iv) The method of effective Love number is not accurate to describe $f$-mode when spin is large and when DT is significant: this method essentially ignores the torque between the orbit and the star. The mismatch of GW signals between this formalism and numerical integrations increases to 0.2 when the spin frequency is larger than 450Hz, even when accurate models for tidal amplitudes $A$ and $B$ are used, therefore, it is the method itself that is inaccurate. (v) We found that DT leads to little improvement on estimating $\lol_3$ in Eq.\ (\ref{love3}), for which constrains are mainly from AT. In our study, they even become worse since the adiabatic part is relatively short when the spin is large. For a system with component masses $(1.4,1.4)M_\odot$ at 100Mpc, the relative errors of $\lol_3$ are around 5 for aLIGO and 0.4 for CE. However, DT does break the degeneracy between $\lol_3$ and $\lol_4$, because the oscillations of $(2,2)$ mode are excited while $(2,0)$ mode are not, hence they contribute differently to GWs. The constraints on $\lol_4$ can be improved by factor of $3\sim4$. In the CE case, the relative errors are $0.7\sim0.8$ when the spin frequency is 500Hz. We also calculated the constraints on the mode frequency $\omega_2$ and the spin $\Omega_s$. We found that they improve by factors of $5\sim6$ and $19\sim27$, respectively. In the CE case, the relative errors of the mode frequency are around $0.2\sim0.4$ while for spin, the numbers are $0.7\sim1.0$. Hence DT potentially provides an alternative channel for people to study the physics of NSs.

Throughout the paper, we have assumed that the NS is in the normal-fluid state, whereas in reality the core of a cold NS is expected to be in the superfluid state~\cite{Yakovlev:99}. Thus a two-fluid formalism should be used to capture the new degree of freedom associated with the superfluidity~\cite{Anderson:01}, and the $f$-mode in particular should split into a doublet~\cite{Prix:02}. However, as shown in Ref.~\cite{Prix:02}, the new $f$-mode due to the superfluid degree of freedom typically have a much higher frequency than the ordinary one (i.e. the $f$-mode we considered here) and consequently we do not expect it to significantly change the results we have here. 

In addition to the ignorance of the superfluidity, there are three caveats we would like to note. First, the H4 EoS has been shown to be less likely based on the observation of GW170817 \cite{LIGOScientific:2019eut}. Second, the spin modifications to mode frequencies through Maclaurin spheroid is merely a toy model and might be too simple for the real situation. Finally, the NS spin frequency should be high enough $(\sim 500$Hz) for DT to have significant effects. Such high frequency is unlikely in astrophysical binaries. However, we here mainly aim to use semi-analytic methods to provide qualitative understandings on DT, different EoS will give similar results. This is because the equations of motion in Eqs.\ (\ref{eq-of-motion}) are generic. EoS only affects the values of $\lambda_{0,2}$ and $\omega_{0,2,3}$. On the other hand, our derivations of tidal excitations $A$ and $B$ [Eq.\ (\ref{new-formulae-tide})] are valid for any systems which couple a harmonic oscillator to a Kepler orbit with a dissipative force in the long timescale. The framework presented in the paper is generic and can be applied to other types of DTs. One possible avenue for future work is to use our discussions to study excitation of $r$-modes with more realistic EoS, since they only require NS to spin at tens of Hz, and are more likely to take place in BNS systems.



All of calculations in this paper are at the Newtonian order, which has allowed us to reveal the insufficiency of the effective Love number approach, and the possibility of gaining further information on neutron stars --- in the regime where the NS has substantial spin, anti-aligned with the orbital angular momentum.  This information must still be complemented by contributions from PN corrections. For instance, at the Newtonian order $\zeta$ and $\Omega_s$ are partially degenerate since they mainly enter equations through the combination $\zeta-\Omega_s$. By introducing PN effect, like spin-orbit and spin-spin coupling, spin will be more constrained, which could break the degeneracy, and consequently, put more stringent constraints on mode frequencies. This is also true for the degeneracy between mass ratio and Love number. In our case, mass ratio is still badly constrained and degenerate with Love number. By including 1PN effect, we could get more accurate estimations on these quantities. 

Secondly, the universal relation for NS is also an important fact to break degeneracy. For example, the universality between Love number and $f$-mode frequencies was observed in Ref.\ \cite{Chan+Sham+14}. With such additional information, constraints on parameters should be improved.

Finally, it is interesting to compare our analytic analyses with recent numerical simulations in Ref. \cite{PhysRevD.99.044008}. To do so, one need to append the tidal Hamiltonian Eq.\ (\ref{Hamiltonian}) to the EOB Hamiltonian, and jointly evolve the orbital motion and the stellar oscillation, to obtain faithful predictions of waveforms.

\begin{acknowledgments}
We thank Jocelyn Read for useful suggestions. The computations presented here were conducted on the Caltech High Performance Cluster, partially supported by a grant from the Gordon and Betty Moore Foundation. H.Y. is supported by the Sherman Fairchild Fundation. Y.C. and S.M. are supported by the Brinson Foundation, the Simons Foundation (Award Number 568762), and the National Science Foundation, through Grants PHY- 1708212 and PHY-1708213.
\end{acknowledgments}

\appendix
\section{The derivation of mode oscillation formulae}
\label{app:mode-oscil}
In this section, we will give a detailed derivation for our new DT formulae of $A$ and $B$, following Ref. \cite{bender2013advanced}

As we have shown in Eq.\ (\ref{eq-nogw-q2}), the stellar oscillation during DT can be described by a harmonic oscillator after a transformation. Its general solution is the sum of a homogeneous solution and the particular solution. Here we assume that there are no free oscillations in the NS initially, hence the solution can be expressed in terms of the retarded Green function and tidal driving
\begin{align}
&x(t)=\frac{1}{\zeta}\int^{t}\frac{3{M_2}\lambda_2\omega_2^2}{2r^{\prime3}}e^{i\Omega_st^\prime+2i\phi(t^\prime)}\sin\zeta(t-t^\prime)dt^\prime.  \label{q2-green} 
\end{align}
By integration by part, we get
\begin{align}
&x(t)=\frac{3{M_2}\lambda_2\omega_2^2}{2\zeta}\left[\frac{\zeta e^{i\Omega_st+2i\phi}}{\zeta^2-(\Omega_s+2\Omega)^2}\frac{1}{r^3} \notag \right.\\
&+e^{-i\zeta t}\int^t\frac{\dot{\Omega}^\prime e^{i\Omega_st^\prime+2i\phi(t^\prime)+i\zeta t^\prime}}{(\Omega_s+2\Omega^\prime+\zeta)^2r^{\prime3}}dt^\prime \notag \\
&\left.-e^{i\zeta t}\int^t\frac{\dot{\Omega}^\prime e^{i\Omega_st^\prime+2i\phi(t^\prime)-i\zeta t^\prime}}{(\Omega_s+2\Omega^\prime-\zeta)^2r^{\prime3}}dt^\prime\right], \label{q2-inte}
\end{align}
where we have ignored $\dot{r}$. However, the method fails once the resonance happens. There is a stationary point within the integration domain. L94 \cite{Lai+94} and H+16 \cite{Steinhoff+Hinderer+16,Hinderer+Taracchini+16} expanded $\phi(t^\prime)$ in Eq.\ (\ref{q2-green}) around $t_r$ and estimated the integral with SPA. Our treatment is slightly different. In order to incorporate both the adiabatic and resonant regimes, we start from Eq. (\ref{q2-inte}) instead of (\ref{q2-green}), where the adiabatic term is separated out initially. At resonance, this adiabatic term goes to infinity. Hence there should be a counterterm arising from the integration, to cancel out such infinity. H+16 \cite{Steinhoff+Hinderer+16,Hinderer+Taracchini+16} chose Eq. (\ref{conter-old}) as the counterterm. Here we derive a better counterterm by studying the integration in Eq.\ (\ref{q2-inte}). 

Since there is no stationary point in the second term on the RHS of Eq.\ (\ref{q2-inte}), it can be ignored. Expanding the integrand of the third term around the resonance point, and neglecting the time derivatives of $\dot{\Omega}$ and $r$, the integration becomes 
\begin{align}
&\int^t\frac{ e^{i\chi_r+i\omd (t^\prime-t_r)^2}}{4r^{3}_r\omd (t^\prime-t_r)^2}dt^\prime=-\frac{e^{i\chi_r+i\tha^2}}{4r_r^3\sqrt{\omd}}\frac{1}{\tha} +\frac{e^{i\chi_r}}{2r_r^3}\sqrt{\frac{\pi}{2\omd}}\notag \\
&\times\left[-\FS\left(\sqrt{\frac{2}{\pi}}\tha\right) +i\FC\left(\sqrt{\frac{2}{\pi}}\tha\right)-\frac{1}{\sqrt{2}}e^{-i\pi/4}\right]. \label{q2-new}
\end{align}
The terms in the bracket are same as H+16 \cite{Steinhoff+Hinderer+16,Hinderer+Taracchini+16}. However, we have a new counterterm
\begin{align}
-\frac{e^{i\chi_r+i\tha^2}}{4r_r^3\sqrt{\omd}}\frac{1}{\tha}, \label{new-conter}
\end{align}
which contains the phase $\chi_r+\tha^2$. As we have discussed in Sec. \ref{sec:new_DT},  the real part of this term gives rise to a contribution to $A$ that is proportional to $\cos(\tha^2-\Theta)/\tha^2$, which reduces to H+16's \cite{Steinhoff+Hinderer+16,Hinderer+Taracchini+16} if we neglect $\cos(\tha^2-\Theta)$. This term cancels the infinity caused by the adiabatic term. On the other hand, the imaginary part of Eq.\ (\ref{new-conter}) does not diverge, since 
\begin{align}
\lim_{t\to t_r}\frac{\sin(\tha^2-\Theta)}{\tha}=0.
\end{align}

Performing the integration by part again on the third term of Eq.\ (\ref{q2-inte}), we get the next order correction
\begin{align}
&\int^t\frac{\dot{\Omega}^\prime e^{i\Omega_st^\prime+2i\phi(t^\prime)-i\zeta t^\prime}}{(\Omega_s+2\Omega^\prime-\zeta)^2r^{\prime3}}dt^\prime=\frac{\dot{\Omega} e^{i\Omega_st+2i\phi-i\zeta t}}{i(\Omega_s+2\Omega-\zeta)^3r^{3}}\notag \\
&+\int^t\frac{6\dot{\Omega}^{\prime2} e^{i\Omega_st^\prime+2i\phi(t^\prime)-i\zeta t^\prime}}{i(\Omega_s+2\Omega^\prime-\zeta)^4r^{\prime3}}dt^\prime.
\end{align}
Follow the same procedure, we obtain a higher order corrections as
\begin{align}
&\Delta A(t)=\frac{{M_2}\lambda_2}{\zeta}\frac{3\omega_2^2}{16r^3_r\omd^{1/2}}\frac{\sin(\tha^2-\Theta)}{\tha^3}.
\end{align}
The correction term contributes a finite value as $t\to t_r$. As shown in Table \ref{table:AB}, this term reduces the error of $A_r$ from tens of percents to $\lesssim$ $4\%$ in the situations we consider.
\bibliography{refer}
\end{document}